\documentclass[aps,pra,reprint,showpacs,showkeys]{revtex4-1}

\newcommand{\Int}{V}
\newcommand{\HIk}[1]{\hat H_{I,\ver{#1}}}
\newcommand{\UIk}[1]{\hat U_{I,\ver{#1}}}
\newcommand{\HI}{\hat H_I}

\newcommand{\HBk}[1]{\hat H_{B,\ver{#1}}}
\newcommand{\UBk}[1]{\hat U_{B,\ver{#1}}}
\newcommand{\HB}{\hat H_B}

\newcommand{\Nm}{N_m}
\newcommand{\neff}{n_{eff}}
\newcommand{\Neff}{N_{eff}}

\newcommand{\myeqdot}{\eqqcolon}
\newcommand{\mydoteq}{\coloneqq} 		
\newcommand{\cancon}{{\dagger}} 			
\newcommand{\eg}{{e.g.\ }} 				
\newcommand{\ie}{{i.e.\ }}					
\renewcommand{\vec}[1]{\left\left\left\left} 	
\newcommand{\rmd}{{\rm{d}}}
\newcommand{\rme}{{\rm{e}}}
\newcommand{\rmi}{{\rm{i}}}

\newcommand{\be}{\begin{equation}}
\newcommand{\ee}{\end{equation}}
\newcommand{\nn}{\nonumber}
\newcommand{\eka}{\left[}
\newcommand{\ekz}{\right]}

\newcommand{\rka}{\left(}
\newcommand{\rkz}{\right)}

\newcommand{\bea}{\begin{eqnarray}}
\newcommand{\eea}{\end{eqnarray}}
\newcommand{\ua}{\uparrow}
\newcommand{\da}{\downarrow}
\newcommand{\ve}[1]{{\boldsymbol{\rm{#1}}}}		
\newcommand{\ver}[1]{{\bf{#1}}}				
\newcommand{\ox}{{(\ve{x})}}					
\newcommand{\val}{{\ve{\alpha}}}					 
\newcommand{\hval}{{\ve{\bar\alpha}}}				
\newcommand{\hvaldot}{{\ve{\dot{\bar{\alpha}}}}}		
\newcommand{\vbet}{{\ve{\beta}}}					
\newcommand{\hvbet}{{\ve{\bar\beta}}}				
\newcommand{\hvbetdot}{{\ve{\dot{\bar{\beta}}}}}		
\newcommand{\bra}[1]{{\langle #1 \rvert}}				
\newcommand{\ket}[1]{{\lvert #1 \rangle}}				
\newcommand{\braket}[2]{{\langle #1 \vert #2 \rangle}}	
\newcommand{\abs}[1]{{\lvert \ver{#1} \rvert}}

\newcommand{\Ele}[1]{{\rm{E}}\!\eka #1 \ekz}
\newcommand{\Elk}[1]{{\rm{K}}\!\eka #1 \ekz}
\newcommand{\Elf}[1]{{\rm{F}}\!\eka #1 \ekz}

\newcommand{\fullpint}
{{\prod\limits_{\ve{k}}
   \mathcal{D}\al[k]      \;
   \mathcal{D}\hal[k]    \;
   \mathcal{D}\bet[k]    \;
   \mathcal{D}\hbet[k]}}

\usepackage{ifthen}

\newcommand{\I}[1][]						
{\ifthenelse{\equal{#1}{}}
{{I}}
{{I_{\ver{#1}}}}}

\newcommand{\hI}[1][]						
{\ifthenelse{\equal{#1}{}}
{{\bar I}}
{{\bar I_{\ver{#1}}}}}

\newcommand{\eps}[1][]						
{\ifthenelse{\equal{#1}{}}
{\varepsilon}
{\varepsilon_{\abs{#1}}}}

\newcommand{\epsw}{\eps}					

\newcommand{\oo}[2][]						
{\ifthenelse{\equal{#1}{}}
{{\mathcal{O}(\eps^{#2})}}
{{\mathcal{O}(#1^{#2})}}}

\newcommand{\op}[2][]						
{\ifthenelse{\equal{#1}{}}
{{\hat{#2}}}
{{\hat{#2}^{\phantom{\dagger}}_{#1}}}}
\newcommand{\hop}[2][]						
{\ifthenelse{\equal{#1}{}}
{{\hat{#2}^{\dagger}_{\phantom{1}}}}
{{\hat{#2}^{\dagger}_{#1}}}}

\newcommand{\E}[1][]						
{\ifthenelse{\equal{#1}{}}
{{E}}
{{E_{\abs{#1}}}}}

\newcommand{\EE}[1][]						
{\ifthenelse{\equal{#1}{}}
{{E^2}}
{{E^2_{\abs{#1}}}}}

\newcommand{\G}[1][]						
{\ifthenelse{\equal{#1}{}}
{{G_0}}
{{G_{0,\abs{#1}}}}}

\newcommand{\GG}[1][]						
{\ifthenelse{\equal{#1}{}}
{{G_1}}
{{G_{1,\abs{#1}}}}}

\newcommand{\hG}[1][]						
{\ifthenelse{\equal{#1}{}}
{{\bar G_0}}
{{\bar G_{0,\abs{#1}}}}}

\newcommand{\hGG}[1][]						
{\ifthenelse{\equal{#1}{}}
{{\bar G_1}}
{{\bar G_{1,\abs{#1}}}}}

\newcommand{\W}[1][]						
{\ifthenelse{\equal{#1}{}}
{{W}}
{{W_{\abs{#1}}}}}

\newcommand{\om}[1][]						
{\ifthenelse{\equal{#1}{}}
{\omega}
{\omega_{\abs{#1}}}}

\newcommand{\ome}[2][]						
{\ifthenelse{\equal{#1}{}}
{{\omega_{#2}}}
{{\omega_{\ver{#1},#2}}}}

\newcommand{\home}[2][]					
{\ifthenelse{\equal{#1}{}}
{{\bar\omega_{#2}}}
{{\bar\omega_{\ver{#1},#2}}}}

\newcommand{\cp}[1][] 						
{\ifthenelse{\equal{#1}{}}
{{\hat{c}^{\phantom{\dagger}}_{\ua}}}
{{\hat{c}^{\phantom{\dagger}}_{\ua,\ver{#1}}}}}
\newcommand{\hcp}[1][]						
{\ifthenelse{\equal{#1}{}}
{{\hat{c}^{\dagger}_{\ua}}}
{{\hat{c}^{\dagger}_{\ua,\ver{#1}}}}}
\newcommand{\cm}[1][]						
{\ifthenelse{\equal{#1}{}}
{{\hat{c}^{\phantom{\dagger}}_{\da}}}
{{\hat{c}^{\phantom{\dagger}}_{\da,\ver{#1}}}}}
\newcommand{\hcm}[1][]						
{\ifthenelse{\equal{#1}{}}
{{\hat{c}^{\dagger}_{\da}}}
{{\hat{c}^{\dagger}_{\da,\ver{#1}}}}}
\newcommand{\cpm}[1][]						
{\ifthenelse{\equal{#1}{}}
{{\hat{c}^{\phantom{\dagger}}_{\ua\da}}}
{{\hat{c}^{\phantom{\dagger}}_{\ua\da,\ver{#1}}}}}
\newcommand{\hcpm}[1][]						
{\ifthenelse{\equal{#1}{}}
{{\hat{c}^{\dagger}_{\ua\da}}}
{{\hat{c}^{\dagger}_{\ua\da,\ver{#1}}}}}

\newcommand{\opa}[1][]						
{\ifthenelse{\equal{#1}{}}
{{\hat{a}}}
{{\hat{a}^{\phantom{\dagger}}_{\ver{#1}}}}}
\newcommand{\hopa}[1][]							
{\ifthenelse{\equal{#1}{}}
{{\hat{a}^{\dagger}_{\phantom{1}}}}
{{\hat{a}^{\dagger}_{\ver{#1}}}}}

\newcommand{\opb}[1][]						
{\ifthenelse{\equal{#1}{}}
{{\hat{b}}}
{{\hat{b}^{\phantom{\dagger}}_{\ver{#1}}}}}
\newcommand{\hopb}[1][]						
{\ifthenelse{\equal{#1}{}}
{{\hat{b}^{\dagger}_{\phantom{1}}}}
{{\hat{b}^{\dagger}_{\ver{#1}}}}}

\newcommand{\opg}[1][]						
{\ifthenelse{\equal{#1}{}}
{{\hat{\Gamma}}}
{{\hat{\Gamma}^{\phantom{\dagger}}_{\ver{#1}}}}}
\newcommand{\hopg}[1][]						
{\ifthenelse{\equal{#1}{}}
{{\hat{\Gamma}^{\dagger}_{\phantom{1}}}}
{{\hat{\Gamma}^{\dagger}_{\ver{#1}}}}}

\newcommand{\al}[1][]						
{\ifthenelse{\equal{#1}{}}
{{\alpha}}
{{\alpha}_{\ver{#1}}}}
\newcommand{\hal}[1][]						
{\ifthenelse{\equal{#1}{}}
{{\bar\alpha}}
{{\bar\alpha}_{\ver{#1}}}}

\newcommand{\ali}[1][]						
{\ifthenelse{\equal{#1}{}}
{{\alpha_i}}
{{\alpha}_{\ver{#1},i}}}
\newcommand{\half}[1][]						
{\ifthenelse{\equal{#1}{}}
{{\bar\alpha_f}}
{{\bar\alpha}_{\ver{#1},f}}}

\newcommand{\bet}[1][]						
{\ifthenelse{\equal{#1}{}}
{{\beta}}
{{\beta}_{\ver{#1}}}}
\newcommand{\hbet}[1][]						
{\ifthenelse{\equal{#1}{}}
{{\bar\beta}}
{{\bar\beta}_{\ver{#1}}}}

\newcommand{\beti}[1][]						
{\ifthenelse{\equal{#1}{}}
{{\beta_i}}
{{\beta}_{\ver{#1},i}}}
\newcommand{\hbetf}[1][]						
{\ifthenelse{\equal{#1}{}}
{{\bar\beta_f}}
{{\bar\beta}_{\ver{#1},f}}}

\newcommand{\gam}[1][]						
{\ifthenelse{\equal{#1}{}}
{{\Gamma}}
{{\Gamma}_{\ver{#1}}}}
\newcommand{\hgam}[1][]					
{\ifthenelse{\equal{#1}{}}
{{\bar\Gamma}}
{{\bar\Gamma}_{\ver{#1}}}}

\newcommand{\gami}[1][]						
{\ifthenelse{\equal{#1}{}}
{{\Gamma_i}}
{{\Gamma}_{\ver{#1},i}}}
\newcommand{\hgamf}[1][]					
{\ifthenelse{\equal{#1}{}}
{{\bar\Gamma_f}}
{{\bar\Gamma}_{\ver{#1},f}}}

\newcommand{\alc}[1][]						
{\ifthenelse{\equal{#1}{}}
{{\alpha_{cl}}}
{{\alpha}_{\ver{#1},cl}}}
\newcommand{\halc}[1][]						
{\ifthenelse{\equal{#1}{}}
{{\bar\alpha_{cl}}}
{{\bar\alpha}_{\ver{#1},cl}}}

\newcommand{\alcdot}[1][]						
{\ifthenelse{\equal{#1}{}}
{{\dot{\alpha}_{cl}}}
{{\dot{\alpha}}_{\ver{#1},cl}}}
\newcommand{\halcdot}[1][]						
{\ifthenelse{\equal{#1}{}}
{{\dot{\bar\alpha}_{cl}}}
{{\dot{\bar\alpha}}_{\ver{#1},cl}}}

\newcommand{\betc}[1][]						
{\ifthenelse{\equal{#1}{}}
{{\beta_{cl}}}
{{\beta}_{\ver{#1},cl}}}
\newcommand{\hbetc}[1][]						
{\ifthenelse{\equal{#1}{}}
{{\bar\beta_{cl}}}
{{\bar\beta}_{\ver{#1},cl}}}

\newcommand{\betcdot}[1][]						
{\ifthenelse{\equal{#1}{}}
{{\dot{\beta}_{cl}}}
{{\dot{\beta}}_{\ver{#1},cl}}}
\newcommand{\hbetcdot}[1][]						
{\ifthenelse{\equal{#1}{}}
{{\dot{\bar\beta}_{cl}}}
{{\dot{\bar\beta}}_{\ver{#1},cl}}}

\newcommand{\vgam}[1][]						
{\ifthenelse{\equal{#1}{}}
{{\ve{\Gamma}}}
{{\ve{\Gamma}_{\ver{#1}}}}}
\newcommand{\hvgam}[1][]					
{\ifthenelse{\equal{#1}{}}
{{\ve{\bar\Gamma}}}
{{\ve{\bar\Gamma}_{\ver{#1}}}}}

\newcommand{\fy}[1][]						
{\ifthenelse{\equal{#1}{}}
{{y}}
{y_\abs{#1}}}
\newcommand{\hfy}[1][]						
{\ifthenelse{\equal{#1}{}}
{{\bar y}}
{{\bar y_\abs{#1}}}}

\newcommand{\fz}[1][]						
{\ifthenelse{\equal{#1}{}}
{{z}}
{z_\abs{#1}}}
\newcommand{\hfz}[1][]						
{\ifthenelse{\equal{#1}{}}
{{\bar z}}
{{\bar z_\abs{#1}}}}

\newcommand{\fY}[1][]						
{\ifthenelse{\equal{#1}{}}
{{Y}}
{Y_\abs{#1}}}
\newcommand{\hfY}[1][]						
{\ifthenelse{\equal{#1}{}}
{{\bar Y}}
{{\bar Y_\abs{#1}}}}

\newcommand{\fZ}[1][]						
{\ifthenelse{\equal{#1}{}}
{{Z}}
{Z_\abs{#1}}}
\newcommand{\hfZ}[1][]						
{\ifthenelse{\equal{#1}{}}
{{\bar Z}}
{{\bar Z_\abs{#1}}}}

\newcommand{\fYdot}[1][]						
{\ifthenelse{\equal{#1}{}}
{{\dot Y}}
{{\dot Y_\abs{#1}}}}
\newcommand{\hfYdot}[1][]					
{\ifthenelse{\equal{#1}{}}
{{\dot{\bar Y}}}
{{\dot{\bar Y}_\abs{#1}}}}

\newcommand{\fZdot}[1][]						
{\ifthenelse{\equal{#1}{}}
{{\dot Z}}
{{\dot Z_\abs{#1}}}}
\newcommand{\hfZdot}[1][]					
{\ifthenelse{\equal{#1}{}}
{{\dot{\bar Z}}}
{{\dot{\bar Z}_\abs{#1}}}}

\newcommand{\opA}[1][]						
{\ifthenelse{\equal{#1}{}}
{{\hat{A}}}
{{\hat{A}^{\phantom{\dagger}}_{\ver{#1}}}}}
\newcommand{\hopA}[1][]								
{\ifthenelse{\equal{#1}{}}
{{\hat{A}^{\dagger}_{\phantom{1}}}}
{{\hat{A}^{\dagger}_{\ver{#1}}}}}

\newcommand{\mfua}[1][]						
{\ifthenelse{\equal{#1}{}}
{{u_{a}}}
{{u_{\ver{#1},a}}}}
\newcommand{\hmfua}[1][]								
{\ifthenelse{\equal{#1}{}}
{{\bar u_{a}}}
{{\bar u_{\ver{#1},a}}}}

\newcommand{\mfva}[1][]						
{\ifthenelse{\equal{#1}{}}
{{v_{a}}}
{{v_{\ver{#1},a}}}}
\newcommand{\hmfva}[1][]							
{\ifthenelse{\equal{#1}{}}
{{\bar v_{a}}}
{{\bar v_{\ver{#1},a}}}}

\newcommand{\mfub}[1][]						
{\ifthenelse{\equal{#1}{}}
{{u_{b}}}
{{u_{\ver{#1},b}}}}
\newcommand{\hmfub}[1][]							
{\ifthenelse{\equal{#1}{}}
{{\bar u_{b}}}
{{\bar u_{\ver{#1},b}}}}

\newcommand{\mfvb}[1][]						
{\ifthenelse{\equal{#1}{}}
{{v_{b}}}
{{v_{\ver{#1},b}}}}
\newcommand{\hmfvb}[1][]							
{\ifthenelse{\equal{#1}{}}
{{\bar v_{b}}}
{{\bar v_{\ver{#1},b}}}}

\newcommand{\mfug}[1][]						
{\ifthenelse{\equal{#1}{}}
{{u}}
{{u_{\ver{#1}}}}}
\newcommand{\hmfug}[1][]							
{\ifthenelse{\equal{#1}{}}
{{\bar u}}
{{\bar u_{\ver{#1}}}}}

\newcommand{\mfvg}[1][]						
{\ifthenelse{\equal{#1}{}}
{{v}}
{{v_{\ver{#1}}}}}
\newcommand{\hmfvg}[1][]						
{\ifthenelse{\equal{#1}{}}
{{\bar v}}
{{\bar v_{\ver{#1}}}}}

\newcommand{\mfu}[2][]						
{\ifthenelse{\equal{#1}{}}
{{u_{#2}}}
{{u_{\ver{#1},#2}}}}
\newcommand{\hmfu}[2][]							
{\ifthenelse{\equal{#1}{}}
{{\bar u_{#2}}}
{{\bar u_{\ver{#1},#2}}}}

\newcommand{\mfv}[2][]						
{\ifthenelse{\equal{#1}{}}
{{v_{#2}}}
{{v_{\ver{#1},#2}}}}
\newcommand{\hmfv}[2][]							
{\ifthenelse{\equal{#1}{}}
{{\bar v_{#2}}}
{{\bar v_{\ver{#1},#2}}}}

\newcommand{\opda}[1][]						
{\ifthenelse{\equal{#1}{}}
{{\hat{d}^{\phantom{\dagger}}_a}}
{{\hat{d}^{\phantom{\dagger}}_{\ver{#1},a}}}}
\newcommand{\hopda}[1][]						
{\ifthenelse{\equal{#1}{}}
{{\hat{d}^{\;\!\dagger}_{a}}}
{{\hat{d}^{\;\!\dagger}_{\ver{#1},a}}}}

\newcommand{\opdb}[1][]						
{\ifthenelse{\equal{#1}{}}
{{\hat{d}^{\phantom{\dagger}}_b}}
{{\hat{d}^{\phantom{\dagger}}_{\ver{#1},b}}}}
\newcommand{\hopdb}[1][]						
{\ifthenelse{\equal{#1}{}}
{{\hat{d}^{\;\!\dagger}_{b}}}
{{\hat{d}^{\;\!\dagger}_{\ver{#1},b}}}}

\newcommand{\opdg}[1][]						
{\ifthenelse{\equal{#1}{}}
{{\hat{d}}}
{{\hat{d}^{\phantom{\dagger}}_{\ver{#1}}}}}
\newcommand{\hopdg}[1][]						
{\ifthenelse{\equal{#1}{}}
{{\hat{d}^{\;\!\dagger}}}
{{\hat{d}^{\;\!\dagger}_{\ver{#1}}}}}

\newcommand{\opd}[2][]						
{\ifthenelse{\equal{#1}{}}
{{\hat{d}^{\phantom{\dagger}}_#2}}
{{\hat{d}^{\phantom{\dagger}}_{\ver{#1}, #2}}}}
\newcommand{\hopd}[2][]						
{\ifthenelse{\equal{#1}{}}
{{\hat{d}^{\;\!\dagger}_{#2}}}
{{\hat{d}^{\;\!\dagger}_{\ver{#1}, #2}}}}

\newcommand{\opDa}[1][]						
{\ifthenelse{\equal{#1}{}}
{{\hat{D}^{\phantom{\cancon}}_a}}
{{\hat{D}^{\phantom{\cancon}}_{\ver{#1},a}}}}
\newcommand{\hopDa}[1][]							
{\ifthenelse{\equal{#1}{}}
{{\hat{D}^{\;\!\cancon}_{a}}}
{{\hat{D}^{\;\!\cancon}_{\ver{#1},a}}}}

\newcommand{\opDb}[1][]						
{\ifthenelse{\equal{#1}{}}
{{\hat{D}^{\phantom{\cancon}}_b}}
{{\hat{D}^{\phantom{\cancon}}_{\ver{#1},b}}}}
\newcommand{\hopDb}[1][]						
{\ifthenelse{\equal{#1}{}}
{{\hat{D}^{\;\!\cancon}_{b}}}
{{\hat{D}^{\;\!\cancon}_{\ver{#1},b}}}}

\newcommand{\opDg}[1][]						
{\ifthenelse{\equal{#1}{}}
{{\hat{D}}}
{{\hat{D}^{\phantom{\cancon}}_{\ver{#1}}}}}
\newcommand{\hopDg}[1][]						
{\ifthenelse{\equal{#1}{}}
{{\hat{D}^{\;\!\cancon}}}
{{\hat{D}^{\;\!\cancon}_{\ver{#1}}}}}

\newcommand{\opD}[2][]							
{\ifthenelse{\equal{#1}{}}
{{\hat{D}^{\phantom{\cancon}}_#2}}
{{\hat{D}^{\phantom{\cancon}}_{\ver{#1}, #2}}}}
\newcommand{\hopD}[2][]						
{\ifthenelse{\equal{#1}{}}
{{\hat{D}^{\;\!\cancon}_{#2}}}
{{\hat{D}^{\;\!\cancon}_{\ver{#1}, #2}}}}

\newcommand{\B}[1][]
{\ifthenelse{\equal{#1}{}}
{{\hat B}}
{{{\hat B}_{\ver{#1}}^{\phantom{\dagger}}}}}

\newcommand{\hB}[1][]
{\ifthenelse{\equal{#1}{}}
{{{\hat B}^{\dagger}}}
{{{\hat B}_{\ver{#1}}^{\dagger}}}}

\newcommand{\nuF}{\nu}
\newcommand{\DeltaF}{\Delta}
\newcommand{\gammaF}{\gamma}
\newcommand{\gF}{g}

\usepackage{graphicx}
\usepackage{color}
\usepackage{amsmath,bbm}	
\usepackage[utf8]{inputenc}	
\usepackage{mathtools}
\usepackage[bookmarks=false]{hyperref}

\begin{document}

\hyphenation{Bo-gol-iu-bov Her-mi-ti-an An-der-son}

\title{Post-adiabatic Hamiltonian for low-energy excitations\\ in a slowly time-dependent BCS-BEC crossover}

\author{Bernhard M. Breid}
\email[]{breidbe@physik.uni-kl.de}
\author{James R. Anglin}
\homepage[]{http://www.physik.uni-kl.de/anglin/home/}
\affiliation{\mbox{Technische Universität Kaiserslautern,} \mbox{D-67653 Kaiserslautern, Germany}}

\pacs{
67.85.Pq,
03.65.Db
}

\keywords{Bardeen--Cooper--Schrieffer--Bose--Einstein condensate crossover; Feshbach resonance; coherent state path integral; classical path; stationary phase approximation; dilute gas approximation; adiabatic approximation; geometric phase; effective Hamiltonian; post-adiabatic; non-adiabatic; Anderson--Bogoliubov modes, BCS, BEC
}

\date{\today}

\begin{abstract}
We develop a Hamiltonian that describes the time-dependent formation of a molecular Bose--Einstein condensate (BEC) from a Bardeen--Cooper--Schrieffer (BCS) state of fermionic atoms as a result of slowly sweeping through a Feshbach resonance. In contrast to many other calculations in the field (see \eg \cite{Holl01,Mils02,Roma06,Mari98}), our Hamiltonian 
includes the leading post-adiabatic effects that arise because the crossover proceeds at a non-zero sweep rate. 
We apply a path integral approach and a stationary phase approximation for the molecular $\ve{k}=\ve{0}$ background, which is a good approximation for narrow resonances (see \eg \cite{Dieh06,Dieh07}). We use two-body adiabatic approximations to solve the atomic evolution within this background. The dynamics of the $\ve{k}\neq\ve{0}$ molecular modes is solved within a dilute gas approximation and by mapping it onto a purely bosonic Hamiltonian. Our main result is a post-adiabatic effective Hamiltonian in terms of the instantaneous bosonic 
(An\-der\-\mbox{son--)Bo}\-gol\-iu\-bov modes, which holds throughout the whole resonance, as long as the Feshbach sweep is slow enough to avoid breaking Cooper pairs. 
\end{abstract}

\maketitle

\section{The BCS-BEC crossover problem}
The ground state of a gas of weakly attractively interacting fermions is a BCS state, in which most of the fermions behave ideally, but some of those near the Fermi surface form something like a Bose condensate of Cooper pairs.  As the extreme example of a gas of protons and electrons shows, however, stronger attractions instead produce a true condensate, of tightly bound composite bosons.  Current laboratory techniques exploit the atomic physics of collisional Feshbach resonances \cite{Chin10} to produce real ultracold gases in which attractions can be varied dramatically, admitting experimental study of the entire BCS-BEC crossover \cite{Chin04,Zwie04,Kohl06,Rega04}.  Several labs can now produce molecular condensates from weakly interacting degenerate Fermi gases, by adiabatically changing a control parameter (typically a magnetic field).  Prospects for more precise and detailed measurements of this process and its products are good.  

Theoretical studies of this problem have so far focused mainly on equilibrium properties of systems with various fixed (instantaneous) values of the interaction parameter \cite{Holl01,Mils02,Roma06,Mari98}. Up-to-date reviews can be found for example in \cite{Kohl06,Zwer12,Ingu08}. Along with wide tunability of equilibrium parameters, however, controllable and observable non-equilibrium dynamics are a major advantage of cold quantum gases as experimental systems.  

The dynamics of cold, dilute quantum gases are of course comparatively simple.  On the one hand, this dynamical simplicity may ultimately allow mesoscopic quantum gas systems to shed light on problems, such as the emergences of irreversibility and classicality, that are even more fundamental than the issue of which effective Hamiltonian is correct.  On the other hand, even for cold dilute gases, quantum many body theory can be difficult enough that understanding will require comparisons among experiments, phenomenological models, and first-principles calculations for simple cases.

In the present work we therefore develop a rigorous theory for weakly non-adiabatic evolution of a dilute quantum gas from BCS to BEC regimes, via a time-dependent Feshbach resonance, in the tractable limit of a narrow resonance.   We show that for a finite effective resonance width (e.g. created by a sufficiently large background scattering length), adiabaticity can in principle be maintained, as far as two-body dynamics are concerned, through the entire evolution. Long-wavelength collective excitations will still be generated, as perfect many-body adiabaticity is not expected to be possible in an infinite system. Our main result is a post-adiabatic Hamiltonian that can describe the excitation of the (An\-der\-\mbox{son-)Bo}\-gol\-iu\-bov modes: Anderson-Bogoliubov excitations on the BCS side and Bogoliubov excitations on the BEC side.

Excitation of these modes in a time-dependent BCS-BEC crossover is an example of the general phenomenon of (quasi-)particle production in a time-dependent background. 
Similar situations can also be found in quite different contexts as, \eg in cosmology, where Hawking radiation \cite{Wald84} is produced in the classical space-time background of a star collapsing into a black hole. 
We will describe the former effect in our context with a post-adiabatic effective Hamiltonian. Post-adiabatic effective Hamiltonians are also common in other physical contexts. Perhaps the best known example describes the effect of ‘geometric magnetism’ \cite{Ahar90, Ahar92, Litt93} and its higher order corrections \cite{Schm96, Litt93, Angl04}. The present paper may thus also be of value in expanding the application of post-adiabatic concepts to experimentally tractable many-body systems.
%
%
\subsection{Structure of this paper}
This paper is organized as follows: 

In Section \ref{kap2} we introduce the BCS-BEC crossover model and explain how we model the background scattering differently from the standard BCS theory approach. In order to simplify our analysis we introduce Fermi units and work in momentum space.

It is important to keep in mind that we will deliberately use path integral language in Section \ref{molfer} and Section \ref{mapping}, but only as a convenient tool to show the equivalence of two systems before we revert to operator language.

In Section \ref{molfer} we consider molecules and atoms as interacting subsystems. The molecular subsystem is treated by a path integral approach whereas the atomic part is treated algebraically by means of a dilute gas approximation. In this context, the choice of initial and final conditions has to be adressed. After solving the atomic dynamics, the results for the molecular dynamics are non-local in time and inconvenient for practical calculations.

Therefore, they are mapped onto a familiar quadratic bosonic system in Section \ref{mapping}. This is achieved by introducing new virtual bosons which mimic the fermionic subsystem. Afterwards we can leave the path integral language for the molecular subsystem in favor of an effective Hamiltonian operator.
  
The newly obtained dummy Hamiltonian is then diagonalized instantaneously in Section \ref{inst} by means of a Bogoliubov transformation. We discuss the properties of mode functions and mode frequencies in general and especially in the case of low-energy excitations, and show some numerical results.

In Section \ref{post} we then use the mode functions and mode frequencies 
to rewrite the dummy Hamiltonian in terms of instantaneous eigenmodes (normal modes). The coupling between these eigenmodes is the leading post-adiabatic effect, and constitutes the main result of this paper.

Finally, we address the validity and consequences of our results in \ref{discussion}.

Appendix \ref{classical} contains a detailed derivation of the classical molecular background that populates the ${\ver{k}=\ver{0}}$ Fourier mode in the adiabatic limit we assume for it.
All our results explicitly depend on this classical molecular background.
\section{A BCS-BEC crossover model}\label{kap2}
\subsection{The model Hamiltonian}
We use here a Hamiltonian similar to those described in \cite{Timm99,Mils02,Holl04,Roma06,Java04,Java05,Lee07,Yi06}, which is based on a delta-like pseudopotential for a collisional interaction in which two fermionic atoms unite into a bosonic molecule (or conversely, in which a molecule splits into two atoms). So our many-body Hamiltonian appears in second-quantized notation as
\begin{align}\label{basicham}
\hat H ={}&\int\limits_V d^3x \biggl[
\frac{\hbar^2}{2\,M} 
\rka{\bf\nabla} \hcp\ox \cdot {\bf\nabla}  \cp\ox + {\bf\nabla} \hcm\ox \cdot {\bf\nabla} \cm\ox \rkz
\nn\\
&-g \, \hcm\ox \hcp\ox \cp\ox \cm\ox
+\frac {\hbar^2}{2(2\,M)}\,{\bf\nabla}\hopa\ox \cdot {\bf\nabla}\opa\ox
\nn\\
& +\frac{\Delta}{2}\rka \hcp\ox\cp\ox+\hcm\ox\cm\ox\rkz 
\nn\\
& +\hbar \, \gamma  \rka \cp\ox\cm\ox\hopa\ox+ \opa\ox\hcm\ox\hcp\ox\rkz\nn\\
&-\frac{\hbar\,\omega}{2}\rka \hcp\ox\cp\ox+\hcm\ox\cm\ox+2\,\hopa\ox\opa\ox \rkz
\biggr]
\end{align}
where $M$ is the fermion mass, $\Delta(t)$ is the external control parameter that is slowly ramped up in time (usually linearly as ${\Delta(t) \mydoteq \hbar\, \nu^2 t}$ for some slow rate $\nu$), and $\gamma$ is an interaction strength, determined by atomic collision physics, and indicating the width of the Feshbach resonance. Finally, $g>0$ is the strength of the scattering among the fermions, defining the BCS ground state as  $\Delta\rightarrow-\,\infty$. It is important to state at this point that it is \textit{not} crucial for our calculations that $\gamma$ and $g$ are time-independent or that $\Delta$ has a linear behavior. Instead, these could all be functions depending adiabatically slowly on time, 
as long as the first two can be approximated as constant, and the last as linearly time-dependent, within the period of significant non-adiabatic excitation.

The annihilation and creation field operators $\opa\ox$ and $\hopa\ox$ are bosonic, whereas $\cpm\ox$ and $\hcpm\ox$ are the field operators of the fermions, with two spin states denoted by $\ua$, $\da$. Although the last term proportional to $\omega(t)$ makes the Hamiltonian look similar to a free energy, we still have the Hamiltonian dynamics of a closed system. This term commutes with the rest of the Hamiltonian at any time and does not affect the dynamics of the initial states considered here, apart from a trivial global phase. 

Note that the Hamiltonian \eqref{basicham} is an example of an effective field theory and as such has to be understood as correctly renormalized. 
As always, the logic behind renormalization is the following: the Hamiltonian is a model, anyway; it must be tuned to yield the correct energy shifts. Since these include higher order corrections, the bare Hamiltonian parameters need counter terms. These sometimes (as in our case) turn out to be infinite.
The only quantity which needs to be renormalized here in order to avoid an ultraviolet divergence is $\Delta$. We always assume that an extra counterterm $\Delta_{ct}$ has been added to $\Delta$, \ie the substitution ${\Delta\rightarrow\Delta+\Delta_{ct}}$ has been done in order to maintain the physical meaning of $\Delta/2$ as the fermionic energy shift compared to a free fermion. This is a standard procedure in field theories. 
In our calculations, we will usually transform the fermionic energy shift resulting from the counterterm into a purely bosonic energy shift 
by adding a term proportional to the conserved total number of particles to the Hamiltonian. 
This is the same as replacing $\omega\rightarrow\omega+\Delta_{ct}/\hbar$.

For large negative $\Delta$, the ground state consists mainly of the standard BCS ground state; for large positive $\Delta$, bosons dominate low energy states.  For small $\Delta$, however, the mediated interactions of either species alone formally diverge in strength.  What this means is that the dynamics involves both species non-trivially; if $\gamma$ is small enough, no truly strong interactions are necessarily involved.  We will therefore consider this small-$\gamma$ limit, which is not typical for experiments so far conducted, but is also attainable with current techniques.  (Many different atomic species are trappable today, and each typically has several collisional Feshbach resonances, some of which are very narrow.)
\subsection{Modeling background scattering}
\subsubsection{The Hubbard–Stratonovich transformation in BCS theory}
The fermionic background scattering is the crucial ingredient in the conventional BCS theory, which is also the $\Delta\rightarrow -\,\infty$ limit of our model. In derivations of the BCS theory, the dynamics is often described within a Grassmann variables path integral. In this framework, the interaction among fermions is mimicked by the interaction with a Hubbard–Stratonovich field, introduced by the Hubbard–Stratonovich transformation.
\subsubsection{An alternative approach: adiabatic elimination}
Rather than reviewing the Hubbard–Stratonovich approach, we emphasize that we take a somewhat different approach here which leads to the same kind of simplifications. We will mimic the fermionic background scattering with interaction with an off-resonant bosonic dummy field which can always be adiabatically eliminated. The fermion-fermion interaction term of the Hamiltonian is then equivalent to:
\begin{align}\label{addres}
&\lim_{\varepsilon\to 0} \int\limits_V  d^3x\, 
\biggl[\frac{\hbar\, \gamma}{\varepsilon}  
\rka \cp\ox\cm\ox\hopb\ox+\opb\ox\hcm\ox\hcp\ox\rkz
\nn\\
&+ \rka \frac{\hbar^2\gamma^2}{g\,\varepsilon^2}-\hbar\, \omega\rkz\hopb\ox\opb\ox
+ \frac {\hbar^2}{2(2M)}\,{\bf\nabla}\hopb\ox \cdot {\bf\nabla}\opb\ox\biggr]	\;.
\end{align}
This approach has an especially nice physical interpretation: the backgroud scattering originates from the interaction of the fermions with a second Feshbach resonance that would create molecules in the $\opb\ox$-modes. This second resonance, however, remains detuned during the whole BCS-BEC crossover. 
For $\varepsilon\rightarrow 0$, the leading order effect of this interaction is just the fermion-fermion interaction in the Hamiltonian \eqref{basicham}. 
Note that this leading order behavior mimicked by just one extra resonance could actually originate from a superposition of  the leading order behavior of many different resonances, which would have the same effect.

Considering only the $\varepsilon\rightarrow 0$ limit, the kinetic term and the finite energy shift ${-\,\hbar\, \omega}$ in equation \eqref{addres} become so small compared to the detuning that they could be neglected completely in the following, making the whole procedure fully equivalent to the usual Hubbard–Stratonovich approach. However, it is useful to maintain the finite energy shift formally, since all terms proportional to $\omega$ reflect the gauge freedom due to the conservation of the total boson number plus one-half the fermion number. In contrast, keeping the kinetic term of the $\opb\ox$-modes would lead to unnecessary complications, so we neglect it henceforth, as being small compared to the ${\hbar^2\gamma^2/(g\,\varepsilon^2)}$ term. 
\subsubsection{Connection of the two approaches}
The connection to the Hubbard–Stratonovich field is seen in the coherent state path integral representation for $\opb\ox$ and $\hopb\ox$: 
Let us introduce new fields that are proportional to the c-number 
fields for $\opb\ox$ and $\hopb\ox$ by a prefactor of order $\varepsilon^{-1}$ and let us perform the limit ${\varepsilon\rightarrow 0}$ afterwards. The resulting action of the path integral would now show the familiar Hubbard–Stratonovich field terms of the BCS action, but  written in terms of our newly defined fields. This rescaling is the reason why the operators $\opb\ox$ and $\hopb\ox$ can be interpreted  as physical bosonic modes whereas the Hubbard–Stratonovich field cannot. 

The Hubbard–Stratonovich field thus corresponds to $\varepsilon^{-1} \opb\ox$, up to a finite prefactor, as $\varepsilon\rightarrow 0$. In this limit, the expectation value of $\hopb\ox\opb\ox$ tends to $0$, whereas that of  $\varepsilon^{-2} \hopb\ox\opb\ox$ remains finite.
\subsection{Momentum space representation}
Since we deal with fermions in the absence of an external spatial potential, it is very convenient to change into momentum space where the description of the fermions simplifies significantly.
\subsubsection{Momentum space operators}   
We transform to $\ver{k}$-space using 
periodic 
boundary conditions in a symmetric cube of volume $V$, 
${\ve{k}\,V^{\frac{1}{3}}= 2\, \pi\, \ve{z}}$ 
with 
${\ve{z}\in\mathbbm{Z}}$ 
by replacing 
\begin{align}
\opa\ox=\sum_{\ve{k}} \frac{\opa[k]}{\sqrt{V}}\,\rme^{\rmi \ve{k}\ve{x}} 
\quad, \quad
\cpm\ox=\sum_{\ve{k}}\frac{\cpm[k]}{\sqrt{V}}\,\rme^{\rmi \ve{k}\ve{x}} 
\end{align}
and by replacing $\opb\ox$ following the same logic. The volume $V$ is assumed to tend to infinity. The resulting momentum space Hamiltonian reads now
\begin{align}\label{Hofk}
\frac{\hat H\bigl(\frac{t}{t_F}\bigr)}{E_{F}}={}&\sum_{\ve{q}}  
\rka\frac{\ve{q}^{2}}{k_{F}^{2}} +\frac{\Delta_F}{2}-\frac{\omega_F}{2}\rkz
\rka \hcp[q]\cp[q]+\hcm[q]\cm[q]\rkz
\nn\\
&+ \frac{\gamma_F}{\sqrt{\Nm}}\sum_{\ve{q}, \ve{k}}
\rka \cp[q]\cm[k-q] \rka\hopa[k]+\frac{\hopb[k]}{\varepsilon_F}\rkz  +\rm{H. c.}\rkz
\nn\\
&+\sum_{\ve{k}}\rka\frac{1}{2}\,\frac{\ve{k}^{2}}{k_{F}^{2}}-\omega_F\rkz \hopa[k]\opa[k] \nn\\
&+\sum_{\ve{k}}\rka\frac{\gamma_F^2}{g_F}\,\frac{1}{\varepsilon_F^2}-\omega_F\rkz \hopb[k]\opb[k]	\;.
\end{align}
We use the Fermi momentum ${k_F\mydoteq(6\,\pi^{2}\Nm/V)^{1/3}}$, the Fermi energy ${E_F\mydoteq\hbar^2 k_F^2/(2\,M)}$ as well as the Fermi time ${t_F\mydoteq\hbar/E_F}$ in order to define ${\gamma_F(t/t_F) \mydoteq \sqrt{\Nm/V}}\, \hbar\, \gamma(t)/E_F$, ${g_F(t/t_F) \mydoteq (\Nm/V)}\, (g(t)/E_F)$, ${\omega_F (t/t_F) \mydoteq t_F\,\omega(t)}$, ${\Delta_F(t/t_F) \mydoteq \Delta(t)/E_F}$ and ${\varepsilon_F (t/t_F) \mydoteq \varepsilon(t)}$. In general, we consider here 
most of these quantities as time-independent.
The conserved quantity $\Nm$ is the total number of bosons, plus one-half the total number of fermions. 
In the case of a linear time dependence as described in the beginning, we have ${\Delta_F (t/t_F) = \nu_F^2 t/t_F}$ with ${\nu_F \mydoteq \hbar\, \nu/E_F}$.
\subsubsection{Fermi units}
The Fermi momentum $k_F$, the Fermi energy $E_F$ and the corresponding Fermi time $t_F$ are natural scales of the system. From now on, we will therefore use the dimensionless quantities ${\ve{k}/k_F}$, ${E/E_F}$, ${\hat H/E_F}$ and ${t/t_F}$, and relabel them as $\ve{k}$, $E$, $\hat H$ and $t$ respectively. Furthermore we will relabel again $\omega_F$ as $\om$, $\Delta_F$ as $\DeltaF$, $\gamma_F$ as $\gammaF$, $\nu_F$ as $\nuF$, $g_F$ as $\gF$ and $\varepsilon_F$ as $\eps$.
\section{Molecules and atoms as interacting subsystems}\label{molfer}
Our Hamiltonian involves two different interacting species, molecules and atoms. This suggests that we consider these as two  subsystems, each of which can be treated with a different approach. This picture is especially useful since we can also adopt it in Section \ref{mapping} where the fermionic subsystem is replaced by a subsystem of virtual bosons.     
\subsection{Molecular subsystem: a path integral approach}
Apart from their interaction, the molecular and the atomic subsystem both have quadratic Hamiltonians. Furthermore, even the interaction term would be quadratic as well, if the molecular operators were replaced with classical fields. Since path integrals treat quantum fields as classical fields, this suggests a convenient hybrid approach to the problem, in which the molecule dynamics is described with a path integral over classical molecular fields, while the quadratic Hamiltonian for the fermionic atoms is treated with canonical operator methods.
\subsubsection{The coherent state path integral}
We will formulate the time evolution of the molecules in terms of a coherent state path integral, which we will only begin to evaluate after solving the atomic problem with canonical operator methods. This will offer us an effective description of the molecular subsystem:
\begin{align}\label{pi}
&\bra{\hval_f,\hvbet_f,\bar F_f}
\hat U(t_f,t_i)
\ket{\val_i,\vbet_i,F_i}=
\nn\\
&\int\limits
^{\substack{
\hval(t_f)=\hval_f \\
\hvbet(t_f)=\hvbet_f}} 
_{\substack{
\val(t_i)=\val_i \\
\vbet(t_i)=\vbet_i}}
\fullpint\,
\biggl[  \rme^{ \hval(t_i)\val_i+\hvbet(t_i)\vbet_i} 
\nn\\
&\cdot\rme^{\int_{t_i}^{t_f} \hvaldot(\tau) \val(\tau)+\hvbetdot(\tau) \vbet(\tau) \rmd \tau} 
\; \rme^{- \rmi \int_{t_i}^{t_f}  \mathcal{H}_{MO}\;\rmd \tau}
\bra{\bar F_f}\hat U_{AT} \ket{F_i}
\biggr]	\;.
\end{align}
With $\hat U$ we always denote the time evolution operator of the corresponding Hamiltonian $\hat H$.
The atomic initial and final states at $t_i$ and $t_f$ are denoted as $\ket{F_i}$ and $\bra{\bar F_f}$, respectively. 
The bar \,$\bar{}$\, denotes complex conjugation.
We use here {\it unnormalized} coherent states
in a notation that highlights the property of the bra and ket states as eigenstates of the creation and annihilation operator, respectively:
\begin{align}\label{cohstate}
\ket{\al}\mydoteq\sum_{n=0}^{\infty}\frac{\al^n}{\sqrt{n!}}\ket{n} 
\quad\text{and}\quad 
\bra{\hal}\mydoteq\sum_{n=0}^{\infty}\frac{\hal^n}{\sqrt{n!}}\bra{n}		\;.
\end{align}
Moreover, this notation makes it possible to read off the amplitudes for a creation and annihilation of excitations directly from an expansion of the transition matrix element in powers of $\al$ and $\hal$.
\subsubsection{The coherent state Hamiltonian}
\label{CSHam}
Within the path integral, the 
Hamiltonian $\hat H$ contains the molecules merely as driving fields $\al[k](t)$, $\hal[k](t)$ and $\bet[k](t)$, $\hbet[k](t)$. The Hamiltonian can then be split into two parts: The atomic part $\hat H_{AT}$, which contains fermionic operators and the molecular part $\mathcal{H}_{MO}$, which is a c-number whose non-trivial effect will be in the path integral:
\begin{align}
\label{HAT}
\hat H_{AT}&\mydoteq \hat H_{TL}+\hat H_{PT}	\; ,\\
\label{HMO}
\mathcal{H}_{MO}&\mydoteq\sum_{\ve{k}}\rka\rka\frac{1}{2}\, \ve{k}^2 - \om\rkz \hal[k]\al[k]+\rka \frac{\gammaF^2}{\gF}\,\frac{1}{\epsw^2}-\om\rkz\hbet[k]\bet[k]\rkz	\!.
\end{align}
Note that the fermionic operators make the construction of the path integral slightly more involved: Unlike for the molecular mode path integration variables, for the fermionic operators the time ordering within the path integral really matters. This makes it necessary to calculate the time evolution operator $\hat U_{AT}$ as seen in equation \nolinebreak \eqref{pi}. The atomic Hamiltonian splits again into two parts of different character: The part $\hat H_{TL}$, which describes atoms that couple just to ${\ve{k}=\ve{0}}$ molecular modes, and the part $\hat H_{PT}$, which describes the coupling between atoms and all ${\ve{k}\neq\ve{0}}$ molecular modes:    
\begin{align}
\label{HTL}
\hat H_{TL}&\mydoteq\sum_{\ve{q}} \hat H_{\ve{q}}		\;, \\
\label{HPT}
\hat H_{PT}&\mydoteq\frac{\gammaF}{\sqrt{\Nm}} 
\,\smash{
\sum_{\substack{\ve{k}\neq \ve{0}\\\ve{q}}}
}
\rka \cp[q]\cm[k-q]\rka\hal[k]+\eps^{-1}\hbet[k] \rkz  +\rm{H. c.}\rkz		.
\end{align}
As the label $PT$ suggests $\hat H_{PT}$ will turn out to be only a relatively small perturbation compared to other terms in $\hat H$. This is understandable if one considers slower and slower sweep rates: The small $\ve{k}$ molecular modes will become more and more dominant, whereas the higher $\ve{k}$-modes will become less and less populated. Thus, the Hamiltonian $\hat H_{TL}$, containing the $\ve{k}=\ve{0}$ molecular modes, is the important part for slow sweep rates. This Hamiltonian $\hat H_{TL}$ has a special feature: It can be decomposed into infinitely many two-level systems $\hat H_{\ve{q}}$, as indicated by the label $TL$. The Hamiltonians of these two-level systems are defined by
\begin{align}
\label{HFQ}
\hat H_{\ve{q}}&\mydoteq
\rka \ve{q}^2 + \frac{\DeltaF}{2}-\frac{\om}{2}\rkz
\rka \hcp[q]\cp[q]+\hcm[-q]\cm[-q]\rkz
\nn\\
&\quad+ \frac{\gammaF}{\sqrt{\Nm}}
\rka \cp[q]\cm[-q]\rka\hal[0]+\eps^{-1}\hbet[0] \rkz  +\rm{H. c.}\rkz
\end{align}
and fullfil the commutation relation
\begin{align}
\label{HFQcom}
\eka \hat H_{\ver{q\vphantom{'}}}(t),\hat H_{\ver{q'}}(t') \ekz=0\quad\text{for}\quad\ver{q}\neq\ver{q'}	\;.
\end{align}
The latter is a sufficient condition for the dynamics to factorize and ensures that our decomposition is meaningful.
\subsection{Initial and final states of the coupled subsystems}\label{boundary}
At this point, we have to address the choice of the initial and final bosonic and fermionic states in \eqref{pi}: We are interested in \textit{transition amplitudes} of the full system between the \textit{ground state} at $t=t_i$ and some \textit{arbitrary low energy state} at ${t=t_f}$. As ${\DeltaF(t_i) \rightarrow -\,\infty}$ the ground state tends to the purely fermionic BCS state as defined in Appendix \ref{classical}. Note that despite $\vbet_i \rightarrow \ver{0}$ as $\eps \rightarrow 0$, the value of $\eps^{-1}\vbet_i$ tends to its self consistent value proportional to the BCS order parameter. 

Up to a global phase, the instantaneous ground state at ${t=t_i}$ follows from the BCS ground state with ${\DeltaF=-\,\infty}$ by an infinitely slow adiabatic $\DeltaF$-sweep up to $t_i$. Therefore, it is sufficient to calculate the transition amplitude from ${t_i=-\,\infty}$ with ${\DeltaF(t_i)=-\,\infty}$ to some arbitrary low energy state at $t_f$. 

As $\DeltaF \rightarrow \infty$, all low energy states turn into purely bosonic free molecules due to the high detuning of the fermions. This means a low energy state at $t_f$ can be evolved adiabatically by an infinitely slow adiabatic $\DeltaF$-sweep into a state without fermions at $t=\infty$. Since the calculation of adiabatic time evolution is straightforward, we choose $t_f=\infty$ with $\DeltaF(\infty)=\infty$ for convenience, such that the low energy states are in fact bosonic.

Note that the coherent states \eqref{cohstate} are an over-complete basis: the coherent states $\ket{\al}=\ket{r\,\rme^{\rmi\, \varphi}}$ with \textit{fixed} ${r>0}$ but arbitrary phase $\varphi$ are sufficient to represent any number state $\ket{n}$ by means of the identity
\begin{align}
\frac{1}{2\, \pi}\, \frac{\sqrt{n!}}{r^n} \int_0^{2\, \pi}\rme^{\rmi\,\varphi\, n}\,\ket{r\,\rme^{-\,\rmi\, \varphi}}\;\rmd\varphi=\ket{n}	\;.
\end{align}
It is therefore sufficient to choose an in principle arbitrary modulus $\lvert \hval_f \rvert$ for our final state. However, we make the choice ${\lvert \hval_f \rvert=\sqrt{\Nm}}$ since this is also the mean field value one would expect if all fermions have been converted into $\Nm$ molecules. 
This choice will let the dominant transition amplitudes be given in a simple semiclassical approximation, rather than as high order corrections after many cancellations of detuned path integral paths. 

In summary we have ${\ket{\val_i,\vbet_i,F_i}=\ket{\ve{0},\ve{0},\textrm{BCS}}}$ and ${\DeltaF(t_i)=-\,\infty}$ at ${t_i=-\,\infty}$ as well as ${\bra{\hval_f,\hvbet_f,\bar F_f}=\bra{\hval_f,\ve{0},vac}}$ with ${\lvert \hval_f \rvert=\sqrt{\Nm}}$ and ${\DeltaF(t_f)=\infty}$ at $t_f=\infty$.
\subsection{Atomic subsystem: an algebraic solution}
Our strategy here is to solve the fermionic dynamics first, before we start to evaluate the bosonic path integral. To this end, the atomic Hamiltonians $\hat H_{TL}$ and $\hat H_{PT}$ deserve different treatments, as explained in the following paragraphs.
\subsubsection{Procedure for $\hat H_{TL}$: fermionic two-level systems in a bosonic background}
\label{HTLProc}
As mentioned before, the general time evolution of a balanced mixture of fermions evolving under $\hat{H}_{TL}$ factorizes into a commuting product of the evolution of two-level systems $\hat U_{\ve{q}}$. If we had no terms involving ${\ve{k}\neq\ve{0}}$ molecules in the Hamiltonian, we would have the same situation as described in \cite{Brei08} apart from our current inclusion of background scattering. In this case the classical path for $\al[0]$, $\hal[0]$ and $\bet[0]$, $\hbet[0]$ shows a total conversion from the BCS state at ${t_i=-\,\infty}$, into molecules at ${t_f=\infty}$ as long as the sweep rate $\nuF^2$ is small enough to avoid leaving fermionic excitations at ${t_f=\infty}$. 
As we will see, this adiabaticity condition requires only that $\nuF$ be small compared to the BCS gap; it is the condition for adiabaticity of the two-body problem of resonant association into molecules. Our focus in this paper, in contrast, is on non-adiabaticity of low-frequency collective excitations in the time-dependent many-body system. We will therefore assume that the two-body dynamics is indeed perfectly adiabatic, so that the final state contains no fermions.

We denote the classical path of the scenario without ${\ve{k}\neq\ve{0}}$ molecular terms in the Hamiltonian by $\alc$, $\halc$ and $\betc$, $\hbetc$. Since the ${\ve{k}=\ve{0}}$ molecular modes are the ones which will mainly be occupied even for the Hamiltonian including the full ${\ve{k}\neq\ve{0}}$ molecular terms, we actually do not evaluate the path integral for them. Instead we evaluate the integrand on $\alc$, $\halc$ and $\betc$, $\hbetc$, as a kind of classical approximation. Although $\alc$, $\halc$ and $\betc$, $\hbetc$ are not the exact saddlepoint paths of the full system (including ${\ve{k}\neq\ve{0}}$ molecular terms), they are a good and self-consistent approximation to it, since the depletion into the other molecular modes turns out to be very small compared to this adiabatic classical background \cite{Brei14}. 

Note that, because of analytic continuation of the path integral variables, for the classical path the \mbox{bar \,$\bar{}$\,} does not \textit{in general} denote complex conjugation any more, but instead denotes independent functions (see Appendix \nolinebreak \ref{classical}). Nevertheless, the classical paths we find in Appendix \nolinebreak \ref{classical} for ${\lvert \half[0] \rvert=\sqrt{\Nm}}$ are indeed still complex conjugate pairs. This means, for the class of classical paths considered in this paper, we can still use the \mbox{bar \,$\bar{}$\,} as complex conjugation. Similar statements hold for all quantities which depend on the classical path. 

Hence, the time evolution of a general balanced fermionic state $\ket{F}$ and of its dual $\bra{\bar F}$ under $\hat H_{TL,cl}$ ($\hat H_{TL}$ on the classical path) are \textit{in general} not the Hermitian conjugate of each other any more --- but they are for our class of classical paths. The general time-evolution along them can be written as
\begin{align}
\label{F1}
\ket{F(t)}\mydoteq&\,\,\hat{U}_{TL, cl}(t,t_i)\ket{F} \nn\\
=&\,\prod_{\ve{q}}\rka \fy[q]\hcm[-q]\hcp[q]+\fz[q]\rkz \ket{vac}
\end{align}
where $\fy[q]$, $\hfy[q]$ and $\fz[q]$, $\hfz[q]$ are general solutions to the fermionic two-level problems $\hat H_{\ve{q},cl}$ ($\hat H_{\ve{q}}$ on the classical path). The factors of the product in brackets are the $\hat U_{\ve{q},cl}$ ($\hat U_{\ve{q}}$ on the classical path) of the single two-level systems. 
The initial conditions at ${t=t_i=-\,\infty}$ of the atomic two-level systems in our specific case follow immediately from the foregoing discussion in Section \ref{boundary}. In order to meet $\ket{F}=\ket{F_i}=\ket{\textrm{BCS}}$, we need
\begin{align}\label{Fdef}
\ket{F}={}&\rme^{\rmi\,\pi\,\Nm}\,\rme^{\rmi\,\theta\,\Nm}\nn\\
&\cdot\prod_{\ve{q}}\rka \fY[q](t_i)\,\hcm[-q]\hcp[q]+\fZ[q](t_i)\rkz \ket{vac}
\end{align}
with the instantaneous eigenstate of the two-state system $\fY[q]$, $\fZ[q]$ \eqref{yzdef} and the phase $\theta$ \eqref{thetadef} of the molecular classical path. These quantities are defined in Appendix \nolinebreak \ref{classical}. 
The fermionic ground state \eqref{Fdef} still holds for finite $t_i$ when imposing also the self-consistent molecular initial conditions
\begin{align}
\ali[0]={}&\alc(t_i)=\sqrt{\Nm}\, r\,  \rme ^{-\rmi\, \theta} 	\;, \\
\beti[0]={}&\betc(t_i)=-\,\frac{\gF\, \epsw\, \om}{\gammaF^2-\gF\, \epsw^2\, \om}\, \sqrt{\Nm}\, r\,  \rme ^{-\rmi\, \theta} 
\end{align}
with $\theta$, $r$ and $\om$ as defined in Appendix \ref{classical} and evaluated at $t_i$. For $t_i \rightarrow -\,\infty$ the values $\ali[0]$ and $\beti[0]$ both tend to zero, while $\eps^{-1}\beti[0]$ becomes proportional to the usual BCS order parameter as explained before. If one would turn off the interaction among the fermions, \ie $\gF \rightarrow 0$, the state $\ket{F}$ \eqref{Fdef} would asymptotically approach a non-interacting Fermi gas.

There is a non-arbitrary parameter $\theta$ appearing in the initial state $\ket{F}$ \eqref{Fdef}, which is the initial phase of the specific classical path considered. This is due to the fact that in general, the number conserving ground state $\ket{\textrm{NCGS}}$ of the system would look like
\begin{align}
&\ket{\textrm{NCGS}}\propto \nn\\
&\int_0^{2\, \pi}\rme^{\rmi\,\theta\, \Nm}\prod_{\ve{q}}\rka \fY[q]\hcm[-q]\hcp[q]+\fZ[q]\rkz \ket{\alc,\betc,vac}
\;\rmd\theta
\end{align}
with $\theta$, $\alc$ and $\betc$ as defined in Appendix \ref{classical}. Thus, the phase of the fermionic state and the phase of the coherent state are necessarily related to insure number conservation.
Consequently, when the stationary phase approximation chooses the phase of the classical path giving the main contribution to the transition amplitude, it also chooses automatically the fermionic phases within the BCS state.

In Appendix \ref{classical} we compute the classical path for $\alc$, $\halc$ and $\betc$, $\hbetc$ as well as the adiabtic solution of the atomic two-level systems on the classical path. The reasoning why the general solutions $\fy[q]$, $\hfy[q]$ and $\fz[q]$, $\hfz[q]$ depend only on the modulus of $\ve{q}$ can also be found there.
\subsubsection{Procedure for $\hat H_{PT}$: a dilute gas approximation}
\label{HPTdlg}
As explained before, we are interested here in the scenario that avoids fermionic excitations at ${t_f=\infty}$. This means $\bra{\bar F_f}\propto\bra{\bar F(t_f)}$, because in the adiabatic solution under $\hat H_{TL,cl}$ no fermions remain. To include the $\ve{k}\neq\ve{0}$ molecular modes, we have to include $\hat H_{PT}$. However, it would be very hard to treat $\hat H_{PT}$ analytically exactly. 
To simplify our calculations, we will perform a dilute gas approximation (DGA). 

The idea behind the DGA is the following: Up to a prefactor (irrelevant for the argumentation), the transition amplitude we calculate can be written as an exponential function. The function in the exponent is a power series in terms of molecular modes with $\ve{k}\neq\ve{0}$. 
Since the occupation of all $\ve{k}\neq\ve{0}$ molecular modes is expected to be small, we can stop this series at the leading order, which is the quadratic one in our case. It is thus the diluteness of the gas and the weak occupation of modes that makes the approximation good. After the approximation, the exponent is equivalent to the term one would obtain by a second order time-dependent perturbation theory. This is just because the powers of an expansion must naturally agree. However, the dilute gas approximation is by far better than the second order time-dependent perturbation theory, since there are no restrictions on the time interval in which it is valid. 
Note that the DGA is especially more than just putting the second order perturbation theory into the exponent: All the extra terms that the DGA contains compared to the time-dependent perturbation theory are actually there — its just that some other terms have been neglected. In this sense the DGA is the resummation of a perturbation series. 

One can also take another point of view in order to see why the DGA works so well: At late times, the Dyson series of the system's time evolution operator might actually involve large contributions from terms of high order in ${\ve{k}\neq\ve{0}}$ moleclular fields. But among these, by far the largest contributions are those in which many different modes are each occupied only slightly — because there are very many ways to distribute a few particles over many modes. Including all of these large contributions, and neglecting all the ones that have the same order in occupation number, but are smaller by a factor of the order of the number of modes, we find we have written nothing but the exponential of the quadratic result. 

For the time evolution operator $\hat U_{AT}$, the identity
\begin{align}
\rmi\, \frac{\rmd}{\rmd t}\eka\hat U^{\dagger}_{TL,cl} \,\hat U_{AT}^{\phantom{-1}}\ekz 
=\eka\hat U_{TL,cl}^{\dagger}\, \hat H_{PT} \,\hat U_{TL,cl}^{\phantom{-1}}\ekz 
\eka\hat U^{\dagger}_{TL,cl} \,\hat U_{AT}^{\phantom{-1}}\ekz
\end{align}
enables us to write a Dyson series for $\hat U^{\dagger}_{TL,cl} \,\hat U_{AT}^{\phantom{-1}}$. The resummation of this series in the way described above leads then to the DGA in $\hat H_{PT}$. Finally, the atomic time evolution under $\hat H_{AT}$ reads
\begin{align}\label{effpi}
&\hspace{-2em}\,
\bra{\bar F_f}
\hat U_{AT}
\ket{F_i}
\nn\\
\overset{\rm{DGA}}{\approx} 
&\,\mathcal{F}\,
\exp\biggr[ -\int_{t_i}^{t_f}\rmd t_2\int_{t_i}^{t_2}\rmd t_1\nn\\
&\bra{ \bar F(t_2)} 
\hat H_{PT}(t_2)\,\hat U_{TL,cl}(t_2,t_1)\,\hat H_{PT}(t_1)
\ket{F(t_1)}\biggl]
\end{align}
where
\begin{align}
\mathcal{F}\mydoteq\braket{\bar F_f}{F(t_f)}=\rme^{\rmi\,\pi\,\Nm}\,\rme^{\rmi\,\theta\,\Nm}
\end{align}
is the semiclassical amplitude of the fermions evolving under the ${\ver{k}=\ver{0}}$ adiabatic, classical molecular background. It is just a phase factor for self-adjoint Hamiltonians, reflecting the creation of $\Nm$ molecules.
The transition amplitude in the exponent is computed with the atomic states given in equation \eqref{F1}: 
\begin{align}\label{exp}
&\bra{F(t_2)}\hat H_{PT}(t_2)\,\hat U_{TL,cl}(t_2,t_1)\,\hat H_{PT}(t_1)\ket{F(t_1)}\nn\\
&\!=\frac{\gammaF^2}{\Nm}\,
\smash{\sum_{\substack{\ve{k}\neq \ve{0}\\\ve{q}}}}\hspace{0.5 em}
\Bigl[ \hfy[q](t_2) \hfy[k-q](t_2)
\Bigl[\al[-k](t_{2})+\eps^{-1}\bet[-k](t_{2}) \Bigr]\nn\\
&\hspace{5em}\,-\hfz[q](t_2) \hfz[k-q](t_2)
\Bigl[\hal[k](t_{2})+\eps^{-1}\hbet[k](t_{2}) \Bigr]\Bigr]\nn\\
&\hspace{5em}\,\Bigl[ \fy[q](t_1) \fy[k-q](t_1)
\Bigl[\hal[-k](t_{1})+\eps^{-1}\hbet[-k](t_{1}) \Bigr]\nn\\
&\hspace{5em}\,-\fz[q](t_1) \fz[k-q](t_1)
\Bigl[\al[k](t_{1})+\eps^{-1}\bet[k](t_{1}) \Bigr]\Bigr]\nn\\
&\hspace{5em}\,\,\rme ^{-\rmi \int_{t_1}^{t_{2}} \ve{q}^2+(\ve{k}-\ve{q})^2 + \DeltaF\;\rmd \tau}	\;.
\end{align}
Later on, we will use the adiabatic solutions for $\fy[q]$, $\hfy[q]$ and $\fz[q]$, $\hfz[q]$  as derived in Appendix \ref{classical}.

Note that effects of back-reaction 
onto the classical path have been ignored here, since the ${\ver{k}=\ver{0}}$ molecular mode has not been included in the DGA. This is justified, since the depletion into ${\ver{k}\neq\ver{0}}$ molecular modes turns out to be small \cite{Brei14}.

In summary, we have formally solved the atomic subsystem's dynamics in this section by means of a DGA. The point of our exercise has been to show that, for slow but not perfectly adiabatic sweeps through the Feshbach resonance, the leading order effect of the fermionic sector on the molecular modes is an effective action that is non-local in time, but \textit{quadratic} in molecular fields. Since equation \eqref{effpi} is non-local in time, it needs to be simplified. The idea is to find a transition amplitude, calculated with a simple Hamiltonian, but matching equation \eqref{effpi}. Since it is only the transition amplitude which matters, we can actually replace the fermionic, atomic subsystem by a different, even bosonic one. The following Section \nolinebreak[4] \ref{mapping} is devoted to this subject: The form of influence functional in \eqref{effpi} is in general equivalent to that induced by an array of harmonic oscillators, linearly coupled. We can therefore use this result to construct an effective \textit{bosonic} \textit{quadratic} Hamiltonian, \ie a linear system that will reproduce the leading non-adiabatic effects of the slow Feshbach sweep on the final molecular modes.

\section{Mapping to a  
familiar system: molecules and virtual bosons as interacting linear subsystems}\label{mapping}
In the previous section we have solved the fermionic subsystem \eqref{effpi}: By perturbing around the non-trivially time-dependent mean field solution for ${\ver{k}=\ver{0}}$, and applying the dilute gas approximation, we have obtained an effective action that describes the quantum dynamics of the molecular field under driving by the associating fermions. The essential feature is that this effective action is quadratic, and its effective quantum dynamics is linear. We will therefore be able to solve this dynamics without assuming equilibrium. The assumptions that have justified our linearization are self-consistent in the limit of a slow crossing of the Feshbach resonance: the variation in time of the molecular path integral variables is either slow, compared to the BCS gap time scale, or else small, compared to the nonlinearity, on all paths that contribute significantly to the path integral.

By not assuming equilibrium, however, we have acquired one significant complication in our linear dynamics. Our effective action is non-local in time, in the sense that it is a double time integral. The saddlepoints of such actions are determined by integro-differential equations, and these bring many technical difficulties. 

We will therefore return to the more familiar context of purely differential equations, by deliberately undoing some steps we have so far done --- except for our linearization. The fermionic subsystem of the previous section will, however, be replaced by a new subsystem, which yields the same effect due to its linear coupling to the actual molecular bosons. 
Maintaining the effective linear dynamics that we have derived, we will restore the eliminated fermionic subsystem, in an alternative but equivalent form, as a subsystem of virtual bosons. In general, we follow the same strategy as before: We consider the system to consist of two subsystems of which one is the molecular subsystem. 

Even though the strategy in this section is the same as in the previous one, the procedure is somewhat reversed: 
In continuation of equation \eqref{effpi} we start out in Section \ref{virbos} within a path integral description for the molecular modes. We show that equation \eqref{effpi} is the very same as calculating a transition amplitude for the virtual bosons under a certain Hamiltonian. 
Having represented our system with a tractable model of coupled oscillators, we will even abandon the path integral of the molecular subsystem in Section \ref{leave}. Thus, we will present our final result as a canonical Hamiltonian operator that incorporates the post-adiabatic effects of the time-dependent BCS-BEC crossover, obtained in Section \ref{molfer}.

\subsection{Virtual bosons as effective atomic subsystem: an algebraic solution}\label{virbos}
The integrand \eqref{exp} in the double time integral of equation \eqref{effpi} shows the creation and annihilation of molecules at $t_1$ and their subsequent annihilation and creation at $t_2$. This is a signature for the elimination of a virtual fermionic  intermediate state between $t_1$ and $t_2$. This insight offers an option to simplify the double time integral: We can consider equation \eqref{effpi} as a transition amplitude for a dummy subsystem under its Hamiltonian. This means putting in bosonic modes $\opg[k,q]$ by hand which serve as intermediate states. Because of their nature as intermediate states we call them virtual bosons. The right-hand side of quation \eqref{effpi} turns out to be  equivalent to
\begin{align}
&\mathcal{F}\,\prod_{\substack{\ve{k}\neq \ve{0}\\\ve{q}}}
\bra{\hgam_{\ver{k},\ver{q},f}}
\,\hat T\,\rme^{-\rmi\int_{t_i}^{t_f} \hat H_{\Int,\ver{k},\ver{q}}^{PI} \;\rmd \tau}
\,\ket{\gam_{\ver{k},\ver{q},i}}\label{test}
\end{align}
with the obligatory boundary conditions ${\hgam_{\ver{k},\ver{q},f}=0}$ and ${\gam_{\ver{k},\ver{q},i}=0}$ (in the notation of unnormalized coherent states \eqref{cohstate}). 

The Hamiltonian $\hat H_{\Int,\ver{k},\ver{q}}^{PI}$ denotes the path integral representation 
(in this case, $\opa[k]$- and $\opb[k]$-modes are just replaced with their path integral variables) 
of $\hat H_{\Int,\ver{k},\ver{q}}$ defined as
\begin{align}\label{hikq}
\hat H_{\Int,\ver{k},\ver{q}}\mydoteq{}&
\Bigl[\W[k-q]+\W[q]\Bigr]
\hopg[k,q]\opg[k,q]\nn\\
&+\opg[k,q] \Bigl[\gammaF\,\hfY[q] \hfY[k-q]\Bigl(\opa[-k]+\eps^{-1}\opb[-k]\Bigr)\Nm^{-1/2}\nn\\
&\quad-\gammaF\,\hfZ[q] \hfZ[k-q]\Bigl(\hopa[k]+\eps^{-1}\hopb[k]\Bigr)\Nm^{-1/2}\Bigr]\nn\\
&+\hopg[k,q]\Bigl[\gammaF\,\fY[q] \fY[k-q]\Bigl(\hopa[-k]+\eps^{-1}\hopb[-k]\Bigr) \Nm^{-1/2}\nn\\
& \quad-\gammaF\,\fZ[q] \fZ[k-q]\Bigl(\opa[k]+\eps^{-1}\opb[k]\Bigr) \Nm^{-1/2}\Bigr]	\;.
\end{align}
The fact that $\hat H_{\Int,\ver{k},\ver{q}}$ contains the interaction with \textit{virtual} excitations, motivates the index \Int. Consequently, the full Hamiltonian of virtual bosons reads:
\begin{align}\label{HV}
\hat H_{\Int}\mydoteq \sum_{\substack{\ve{k}\neq \ve{0}\\\ve{q}}} \hat H_{\Int,\ver{k},\ver{q}}		\;.
\end{align}
The result obtained by the dilute gas approximation is thus represented by means of the transition amplitudes between $\ket{\gam_{\ver{k},\ver{q},i}}$ and $\bra{\hgam_{\ver{k},\ver{q},f}}$ that are calculated with a time evolution under $\hat H_{\Int,\ver{k},\ver{q}}^{PI}$.

The coefficients $\fY[q]$, $\hfY[q]$ and $\fZ[q]$, $\hfZ[q]$ as well as $\W[q]$ are derived in Appendix \ref{classical} in the course of the derivation of an adiabatic solution for $\fy[q]$, $\hfy[q]$ and $\fz[q]$, $\hfz[q]$. They are all explicitly time-dependent via $\DeltaF$ (or $\nuF^2 t$ in the linear case).

Because the operators appear only quadratically in $\hat H_{\Int,\ver{k},\ver{q}}^{PI}$, the transition amplitudes in equation \eqref{test} can for instance be checked by a simple algebraic calculation to agree with equation \eqref{effpi}.
This is especially simple because the dynamics for the subsystem of virtual bosons factorizes within the path integral for ${(\ver{k},\ver{q})\neq(\ver{k'},\ver{q'})}$ by means of the commutation relation
\begin{align}\label{picom}
\eka \hat H_{\Int,\ver{k},\ver{q}}^{PI} (t)\,,
\hat H_{\Int,\ver{k'},\ver{q'}}^{PI}(t')\ekz=0	\;.
\end{align}

Surprisingly, we have thus replaced originally fermionic intermediate states with virtual bosons. Furthermore the Hamiltonian for the virtual bosons is just quadratic. Both features are direct consequences of the fact that we went only up to the second order in molecular modes with $\ve{k}\neq\ve{0}$ when calculating their action in Section \ref{molfer}. 
This is, however, sufficient for the leading effect of non-adiabatic excitations, since for a very slow Feshbach sweep, these excitations will be weak.
 
\subsection{Molecular subsystem: leaving the path integral framework}\label{leave}
It is important to notice that it is only solving the dynamics of the fermionic subsystem or the subsystem of virtual bosons {\it before} the molecular one that produces the double time integral in \eqref{effpi}.  
While it was only by solving the fermions first that we could see that their effect on the molecules was the same as that of virtual bosons, now that this has been established, the two-step procedure itself is not necessary. None of our results for the molecules will be altered if we in fact solve the entire system of molecular and virtual bosons all at once. 
Instead of calculating the dynamics of the $\opg[k,q]$-modes before the one of the $\opa[k]$- and $\opb[k]$-modes, the whole dynamics is then calculated at the same time.
This has the important technical advantage that we will not have to deal with non-local effects in time, for which calculational techniques are less familiar, and can instead use standard linear differential equation techniques, for the enlarged dynamical system.

\subsubsection{Restoring operators}

In fact, since the entire bosonic system that we are now using turns out to be quadratic, it is quite straightforward to solve it all at once, with canonical operator methods. The only reason to use the path integral was in order to solve for the fermions first, and leave the molecules for later; now that we have used the path integral to establish the virtual boson model, we are free to abandon the path integral for the ${\ve{k}\neq\ve{0}}$ molecular modes entirely, and revert to canonical methods.

Leaving the path integral for the ${\ve{k}\neq\ve{0}}$ molecular modes, the c-number fields $\al[k]$, $\hal[k]$, $\bet[k]$, $\hbet[k]$ representing the bosonic modes will now be restored as operators which do not all commute. 
It turns out that the transition amplitude \eqref{pi} considered originally can be rewritten with the help of \eqref{test} as 
\begin{align}
&\bra{\hval_f,\hvbet_f,\bar F_f}
\hat U(t_f,t_i)
\ket{\val_i,\vbet_i, F_i}=  \nn\\
&\mathcal{F}\, \prod_{\substack{\ve{k} \neq \ve{0} \\ k_3 > 0}}
\bra{\hal_{\ver{k},f},\hal_{\ver{-k},f},0,0,\ve{0},\ve{0}}
\UBk{k}(t_f,t_i)
\ket{0,0,0,0,\ve{0},\ve{0}}
\label{t1}
\end{align}
where the full unnormalized coherent state in \eqref{t1} is written as $\ket{\al_{\ver{k}},\al_{\ver{-k}},\bet_{\ver{k}},\bet_{\ver{-k}},\vgam_{\ver{k}},\vgam_{\ver{-k}}}$. The index $\ver{q}$ of $\gam[k,q]$ has been absorbed into the vector notation of $\vgam[k]$. The $k_3>0$ (assuming ${V\to\infty}$) condition in equation \eqref{t1} avoids a double counting in the product as explained below.

Replacing the fermionic subsystem with virtual bosons in \eqref{t1}, we also replace the initial and final fermionic states with the equivalent initial and final states of virtual bosons as obtained in \eqref{test}. 

A new Hamiltonian $\HBk{k}$, creating the time evolution $\UBk{k}$ has been introduced here. Its properties will be discussed in the following section.

\subsubsection{A Hamiltonian for molecules and virtual bosons}\label{newham}
The new Hamiltonian $\HBk{k}$ appearing in $\UBk{k}$ of \eqref{t1} when leaving the path integral for the ${\ve{k}\neq\ve{0}}$ molecular modes is defined as  
\begin{align}\label{dummy}
\HBk{k}\mydoteq{}&
\rka\frac{1}{2}\, \ver{k}^2 -\om\rkz\rka \hopa[k]\opa[k]+\hopa[-k]\opa[-k]\rkz\nn\\
&+\rka\frac{\gammaF^2}{\gF}\frac{1}{\epsw^{2}}-\om\rkz\rka \hopb[k]\opb[k]+\hopb[-k]\opb[-k]\rkz\nn\\
&+\sum_{\ver{q}}\rka\hat H_{\Int,\ver{k},\ver{q}}+\hat H_{\Int,\ver{-k},\ver{q}}\rkz	\;.
\end{align}
The Hamiltonian for the full system reads now
\begin{align}\label{B}
\HB\mydoteq\sum_{\substack{\ve{k} \neq \ve{0} \\ k_3 > 0}}
\HBk{k}	\;.
\end{align}

The Hamiltonian $\HB$ simply restores to $\hat H_{\Int}$ from \eqref{hikq}, \eqref{HV} the self-energy terms of our two molecular modes:
It includes the kinetic energy term of the $\opa[k]$-modes (the actual molecules forming the final condensate) and the total detuning of the $\opb[k]$-modes (the ‘dummy’ molecules that mediate the interactions among the fermions). Previously these were absorbed in the path integral's $\mathcal{H}_{MO}$, but with our return to the canonical operator formalism, they must be made explicit.

The Hamiltonian $\HB$ contains only bosonic operators and is entirely quadratic. Due to the commutation relation
\begin{align}\label{com1}
\eka \HBk{k}(t)\;\!, \HBk{k'}(t')\ekz=0 \quad\text{for}\quad \ver{k'}\neq\ver{\pm k}
\end{align}
the dynamics in \eqref{t1} factorizes into a time evolution under the $\HBk{k}(t)$ which contain as few molecular modes as possible. In contrast, within the path integral (see \eqref{test}), the smallest system factorizing the dynamics was $\hat H_{\Int,\ver{k},\ver{q}}$.

The physical meaning of $\HBk{k}$ is that it describes the complete dynamics of a pair of molecular modes with momenta $\ver{\pm k}$, in terms of \textit{time-dependent}, linear couplings among \textit{three} different types of bosonic fields.

The fact that only opposite momentum molecular modes with $\ver{\pm k}$ need to be coupled in $\HBk{k}$ is due to our assumption that, to leading non-adiabatic order, only the ${\ve{k}=\ve{0}}$ molecular mode will be macroscopically populated in the BCS-BEC crossover. This preserves translational symmetry for the post-adiabatic excitations. 

Note that the symmetry offered by ${\HBk{k}(t)=\HBk{-k}(t)}$ implies also the symmetry ${\UBk{k}(t,t_i)=\UBk{-k}(t,t_i)}$. Due to these symmetries, a double counting in the product over $\ver{k}$ in equation \eqref{t1} is avoided by restricting the third component of $\ver{k}$, $k_3$ to $k_3>0$, assuming that the volume $V$ tends to infinity. A double counting in the sum over $\ver{k}$ in equation \eqref{B} is avoided in the same way. 
\subsection{Initial and final conditions of the coupled subsystems}
The post-adiabatic analysis of the effective Hamiltonian $\HB$ in Section \ref{post} will be the main contribution of this paper. Before proceeding with it, however, we pause here to note that the mapping from the physical fermionic subsystem to the effective subsystem of virtual bosons is not only a mapping between formal Hamiltonians. It also includes the mapping between initial and final quantum states of the physical and effective systems. This has been done in \eqref{t1} for the states at infinitely early and late times: the BCS state at $t=t_i$ and the fermion vacuum at $t=t_f$ are both mapped onto the vacuum of virtual bosons.

Beyond computing the final results of slow but non-adiabatic evolution over infinite times, however, we would like to be able to use our effective model to describe low-frequency excitations at arbitrary intermediate times as well. And indeed we can do this, but the sense in which we can do it is a bit subtle. The necessary technical steps have in principle already been laid out in Section \ref{boundary} and in the mapping \eqref{t1} of initial and final states. 

Our construction of the virtual boson model, as reproducing the effective action from fermions in the molecule path integral, showed that the virtual boson model correctly yields all effects on the molecules, as long as all dynamics is slow compared to the BCS gap. The fact that we succeed in computing the molecular excitations at late times shows that we must be accurately computing the intermediate dynamics that yields them, and this is the full low frequency dynamics of the entire system. What we are really describing, therefore, are the low frequency collective modes of the entire system. Since both the actual fermionic excitations and the effective excitations of our virtual bosons are separated from the low frequency collective modes by the BCS gap, in both cases the high frequency modes adiabatically dress the low frequency excitations. It is thus in principle a straightforward exercise in adiabatic theory, to explicitly determine exactly what virtual fermionic excitations are really dressing the slow molecular excitations at intermediate times, in terms of the virtual boson excitations that dress them in our effective model. 

We will not pursue that adiabatic exercise in this paper, but instead focus on the post-adiabatic collective excitations that arise when the BCS-BEC crossover is effected at any finite speed. We will express the slow but possibly non-adiabatic evolution of the full system in the time-dependent basis of its instantaneous normal modes. And we will represent these instantaneous energy eigenstates within the effective model in which our virtual bosons have replaced the physical fermions. For many practical purposes, the expression in terms of instantaneous normal modes is sufficient in itself. Spectroscopic measurements, for example, measure normal modes directly. In any case, it will only be adiabatic mappings that we leave implicit; our theory will provide the leading post-adiabatic effects explicitly.

\subsection{The mean field approximation}
As explained earlier, the path integral of the ${\ve{k}=\ve{0}}$ modes had been approximated by evaluating it along its classical path. There is still a further step necessary in order to obtain a mean field theory in the common sense which does not depend on specific molecular boundary conditions, \ie specific phases. Looking at the transition amplitude \eqref{t1}, there appears the phase of the classical molecular background $\theta$ in $\mathcal{F}$ as well as in $\UBk{k}$ (through terms in $\HBk{k}$ \eqref{dummy}, \ie $\hat H_{\Int,\ver{k},\ver{q}}$ \eqref{hikq} which create and destroy molecular excitations). This is nothing but the particle number conservation: if we changed from coherent states to number states, the $\theta$ integration would define the allowed particle numbers. 
The approach we take here is to redefine the phases of the fermionic creation and annihilation operators such that they just cancel the phase terms within the Hamiltonian $\HB$ \eqref{B}. The latter depends now only on the mean field values $r$ and $\om$ as defined in Appendix \ref{classical}. This finally makes $\HB$ \eqref{B} a Hamiltonian for perturbations around a classical mean field background.

\section{Instantaneous normal modes}\label{inst}
The Hamiltonians $\HBk{k}$ \eqref{dummy} defined in the previous section let the dynamics factorize. Moreover, they are simply quadratic. As is well known, any quadratic Hamiltonian can be diagonalized (\ie, reduced to the standard $\sum_{\xi}\ome[k]{\xi}\,\hopD[k]{\xi}\,\opD[k]{\xi}$ form for some set of canonical quasiparticle destruction operators $\opD[k]{\xi}$) by means of a Bogoliubov transformation. Unfortunately, the diagonalization of the Hamilton does not in general imply the diagonalization of the time evolution operator. The latter is only true for time-independent Hamiltonians, for which the Bogoliubov transformation identifies the fundamental excitations, which are constant for all times. In the time-dependent case we can still get the fundamental excitations of the system at any instant in time by diagonalizing the Hamiltonian instantaneously. This will however make the diagonalizing Bogoliubov transformation \textit{time-dependent}. This implies a \textit{post-adiabatic} coupling between the instantaneous fundamental excitations in the system's dynamics. We will come back to this issue in Section \ref{post}. 

Instantaneously diagonalizing $\HB$ is still the first step towards our goal of describing these post-adiabatic excitations. To zeroth post-adiabatic order, the system will simply remain in an instantaneous quasiparticle number eigenstate at all times. In this section we will therefore perform the instantaneous diagonalization of the Hamiltonian $\HBk{k}$ \eqref{dummy} in order to get the instantaneous fundamental excitations. The latter excitations are usually called Bogoliubov excitations on the BEC side and Anderson-Bogoliubov excitations on the BCS side. 
On the BCS side of the problem, the lowest energy eigenmodes of the fundamental excitations are also called sound modes due to the nature of their dispersion relation. 

Along with the eigenmodes, we derive the equations for the eigenfrequencies of the fundamental excitations. It will turn out that the lowest energy eigenmodes of $\HBk{k}$ \nolinebreak \eqref{dummy} have eigenfrequencies within the gap of fermionic excitations. 
We focus especially on the low energy excitations in this section.

\subsection{Compact notation}
Since various sums are involved in the course of a Bogoliubov transformation, it is notationally inefficient for this purpose to denote modes by using different characters. Instead, we switch to the following notation involving just one basic character for all modes:
\begin{align}
\opda[k]\mydoteq \opa[k]\text{\quad,\quad}		
\opdb[k]\mydoteq \opb[k]\text{\quad and\quad}	
\opdg[k,q]\mydoteq \opg[k,q]	\;.
\end{align}
The original modes are distinguished now by different subindices. Consequently, the subindex can be either a vector, denoting the old $\opg[k,q]$, or a character, denoting the old $\opa[k]$ and $\opb[k]$. As a joint notation which does not specify either of both possibilities, we will dedicate greek subindices like $\eta$ or $\xi$. Thus, an unspecified operator is given for example by $\opd[k]{\eta}$. Consequently, sums over greek indices are taken over all momenta as well as over $a$ and \nolinebreak $b$.
\subsection{Bogoliubov transformation to $\opD[k]{\xi}$}
As mentioned before it is well known that a quadratic Hamiltonian can be diagonalized by means of a Bogoliubov transformation. We can thus assume that our Hamiltonian $\HBk{k}$ \eqref{dummy} can be rewritten as 
\begin{align}\label{di}
\HBk{k}={}&\sum_{\xi,\pm
}\ome[\pm k]{\xi}\,\hopD[\pm k]{\xi}\,\opD[\pm k]{\xi}
\nn\\
&-\sum_{\xi,\eta,\pm
}\ome[\pm k]{\xi}\,\hmfv[\pm k]{\xi,\eta}\,\mfv[\pm k]{\xi,\eta}
\end{align}
where the transformed operators are defined as
\begin{align}
\label{b}
\opD[k]{\xi}&\mydoteq \sum_{\eta
} \mfu[k]{\xi,\eta}\,		\opd[k]{\eta}+\mfv[k]{\xi,\eta}\, 		\hopd[-k]{\eta}		\;\; , \\
\label{bd}
\hopD[-k]{\xi}&\mydoteq \sum_{\eta
} \hmfu[-k]{\xi,\eta}\, 	\hopd[-k]{\eta}+\hmfv[-k]{\xi,\eta}\, 	\opd[k]{\eta}
\end{align}
and fulfill the canonical commutation relations.
For small $\abs{k}$, we will numerically find real, positive low-frequency solutions for $\ome[k]{\xi}$. We assume that no further, negative or even complex $\ome[k]{\xi}$ exist in the low-frequency regime.

It is easily seen that the c-number term in the Hamiltonian \eqref{di} is necessary to compensate for an operator reordering: If we insert equations \eqref{b} and \eqref{bd} into the Hamiltonian \eqref{di} and reorder the operators as in \eqref{dummy}, we get extra terms due to the canonical commutation relations. 
No c-number additions to the Hamiltonian will affect the probabilities, however, and so all of those may simply be ignored.
\subsubsection{Completeness relations for mode functions}
The above definitions of the Bogoliubov transformation need to imply compliance with the canonical commutation relations. As a consequence of this constraint, the following conditions become part of the definition: 
\begin{align}
\label{condi1}
\sum_{\eta
}\rka\mfu[k]{\xi,\eta}\,\hmfu[k]{\xi',\eta}-\mfv[k]{\xi,\eta}\,\hmfv[k]{\xi',\eta}\rkz=\delta_{\xi, \xi'}
\end{align}
is necessary in order to satisfy the commutation relation $\bigl[ \opD[k]{\xi}\,, \hopD[k']{\xi'}\bigr]=\delta_{\ver{k},\ver{k'}}\,\delta_{\xi, \xi'}$, whereas the commutation relation $\bigl[ \opD[k]{\xi}\,, \opD[k']{\xi'}\bigr]=0$ requires 
\begin{align}
\label{condi2}
\sum_{\eta
}\rka\mfu[k]{\xi,\eta}\,\mfv[-k]{\xi',\eta}-\mfv[k]{\xi,\eta}\,\mfu[-k]{\xi',\eta}\rkz=0
\end{align}
to hold. The commutation relation among the creation operators adds a further condition, which is the barred version of \eqref{condi2}, written here for $\ver{-k}$ instead of $\ver{k}$:
\begin{align}
\label{condi3}
\sum_{\eta
}\rka\hmfu[-k]{\xi,\eta}\,\hmfv[k]{\xi',\eta}-\hmfv[-k]{\xi,\eta}\,\hmfu[k]{\xi',\eta}\rkz=0		\;.
\end{align}
\subsubsection{The inverse Bogoliubov transformation}
In view of a later requirement in Section \ref{post}, we introduce here also the inverse Bogoliubov transformation defined by 
\begin{align}
\label{inv}
\opd[k]{\eta}&=\sum_{\xi'
} 
\hmfu[k]{\xi',\eta}\,\opD[k]{\xi'}-\mfv[-k]{\xi',\eta}\,\hopD[-k]{\xi'}		\;\;, \\
\label{invd}
\hopd[-k]{\eta}&=\sum_{\xi'
} 
\mfu[-k]{\xi',\eta}\,\hopD[-k]{\xi'}-\hmfv[k]{\xi',\eta}\,\opD[k]{\xi'}		\;\;.
\end{align}
The validity of \eqref{inv} and \eqref{invd} is easily checked by inserting it into the definitions \eqref{b} and \eqref{bd} and applying the identities \eqref{condi1}, \eqref{condi2} and \eqref{condi3}.

In order for the $\opd[k]{\eta}$ to obey canonical commutation relations, the necessary and sufficient conditions are a set of orthonormality relations among the $\mfu[k]{\xi,\eta}$, $\mfv[k]{\xi,\eta}$, similar to those obtained for the original transformation \eqref{condi1}, \eqref{condi2}, \eqref{condi3}. It can be shown, however, that the conditions for the original and the inverse transformation imply each other \cite{Walc11}, and so the normalization problem for the $\mfu[k]{\xi,\eta}$, $\mfv[k]{\xi,\eta}$ is not overconstrained.
\subsection{Instantaneous eigenmodes: mode functions and frequencies}\label{ewg}
We want the Bogoliubov transformation \eqref{b}, \eqref{bd} to instantaneously diagonalize the Hamiltonian $\HBk{k}$ \eqref{dummy} as \eqref{di}. To this end we need the appropriate mode functions $\mfu[k]{\xi,\eta}$, $\mfv[k]{\xi,\eta}$ of the eigenmodes and their eigenfrequencies $\ome[k]{\xi}$, respectively. The correct choice of both in order to diagonalize the Hamiltonian \eqref{di} implies that the time-independent Heisenberg equation
\begin{align}\label{comcondi}
\eka\opD[k]{\xi}\,,\HBk{k}\ekz\overset{\boldsymbol{!}}{=}\,\ome[k]{\xi}\,\opD[k]{\xi}
\end{align}
must hold. If we insert here the definitions \eqref{b} and \eqref{bd} of the Bogoliubov transformed modes $\opD[k]{\xi}$ and $\hopD[k]{\xi}$, we can sort the terms by the various $\opd[k]{\xi}$ and $\hopd[k]{\xi}$. Since the coefficient of each of these operators must vanish, we get separate equations for the mode functions. 
The coefficient in front of $\opda[k]$ reads: 
\begin{align}
\label{ua}
0={}&\rka \frac{1}{2}\, \ver{k}^2 -\om-\ome[k]{\xi}\rkz \mfu[k]{\xi,a}\nn\\
&-\sum_{\ver{q}}\gammaF\,\fZ[q]\fZ[k-q]\,\Nm^{-1/2}\,\mfu[k]{\xi,\ver{q}}\nn\\
&-\sum_{\ver{q}}\gammaF\,\hfY[q]\hfY[-k-q]\,\Nm^{-1/2}\,\mfv[k]{\xi,\ver{q}}	\;\,.
\end{align}
The coefficient in front of $\opdb[k]$ reads: 
\begin{align}
\label{ub}
0={}&\rka \frac{\gammaF^2}{\gF}\frac{1}{\epsw^{2}}-\om-\ome[k]{\xi}\rkz \mfu[k]{\xi,b}\nn\\
&-\sum_{\ver{q}}\gammaF\,\fZ[q]\fZ[k-q]\,\Nm^{-1/2}\,\eps^{-1}\,\mfu[k]{\xi,\ver{q}}\nn\\
&-\sum_{\ver{q}}\gammaF\,\hfY[q]\hfY[-k-q]\,\Nm^{-1/2}\,\eps^{-1}\,\mfv[k]{\xi,\ver{q}}	\;\,.
\end{align}
The coefficient in front of $\opdg[k,q]$ reads: 
\begin{align}
\label{uq}
0={}&\rka \W[k-q]+\W[q]-\ome[k]{\xi}\rkz \mfu[k]{\xi,\ver{q}}\nn\\
&-\gammaF\, \hfZ[q]\hfZ[k-q]\,\Nm^{-1/2}\rka\mfu[k]{\xi,a}+\eps^{-1}\,\mfu[k]{\xi,b}\rkz\nn\\
&-\gammaF\, \hfY[q]\hfY[k-q]\,\Nm^{-1/2}\rka\mfv[k]{\xi,a}+\eps^{-1}\,\mfv[k]{\xi,b}\rkz	\;.
\end{align}
The coefficient in front of $\hopda[-k]$ reads: 
\begin{align}
\label{va}
0={}&\rka -\,\frac{1}{2}\,\ve{k}^2+\om-\ome[k]{\xi}\rkz \mfv[k]{\xi,a}\nn\\
&+\sum_{\ver{q}}\gammaF\,\fY[q]\fY[k-q]\,\Nm^{-1/2}\,\mfu[k]{\xi,\ver{q}}\nn\\
&+\sum_{\ver{q}}\gammaF\,\hfZ[q]\hfZ[-k-q]\,\Nm^{-1/2}\,\mfv[k]{\xi,\ver{q}}	\;\,.
\end{align}
The coefficient in front of $\hopdb[-k]$ reads: 
\begin{align}
\label{vb}
0={}&\rka -\,\frac{\gammaF^2}{\gF}\frac{1}{\epsw^{2}}+\om-\ome[k]{\xi}\rkz \mfv[k]{\xi,b}\nn\\
&+\sum_{\ver{q}}\gammaF\,\fY[q]\fY[k-q]\,\Nm^{-1/2}\,\eps^{-1}\,\mfu[k]{\xi,\ver{q}}\nn\\
&+\sum_{\ver{q}}\gammaF\,\hfZ[q]\hfZ[-k-q]\,\Nm^{-1/2}\,\eps^{-1}\,\mfv[k]{\xi,\ver{q}}	\;\,.
\end{align}
The coefficient in front of $\hopdg[-k,q]$ reads: 
\begin{align}
\label{vq}
0={}&\rka -\,\W[-k-q]-\W[q]-\ome[k]{\xi}\rkz \mfv[k]{\xi,\ver{q}}\nn\\
&+\gammaF\, \fY[q]\fY[-k-q]\,\Nm^{-1/2}\rka\mfu[k]{\xi,a}+\eps^{-1}\,\mfu[k]{\xi,b}\rkz\nn\\
&+\gammaF\, \fZ[q]\fZ[-k-q]\,\Nm^{-1/2}\rka\mfv[k]{\xi,a}+\eps^{-1}\,\mfv[k]{\xi,b}\rkz	\;.
\end{align}
With respect to the final index $\eta$, the mode functions $\mfu[k]{\xi,\eta}$ and $\mfv[k]{\xi,\eta}$ are the components of an eigenvector of a linear mapping with the eigenfrequency $\ome[k]{\xi}$. Together with the completeness relation \eqref{condi1}, this defines the mode functions. 

Remember that the mode functions as well as the frequencies defined here are time-dependent via the time dependence of the classical background $\alc$, $\halc$ and $\betc$, $\hbetc$ which appears in $\fY[q]$, $\hfY[q]$ and $\fZ[q]$, $\hfZ[q]$.
\subsection{Symmetry properties of mode frequencies and mode functions}\label{gmf}
As one may check by inspection, the equations \eqref{ua}, \eqref{ub}, \eqref{uq}, \eqref{va}, \eqref{vb} and \eqref{vq} offer certain symmetries, which are especially useful for our calculation. For the following reasoning it is important that we always think of the system's volume as going to infinity. This leads to continuous momenta and we can define an arbitrary rotation $\hat R$ in momentum space. Let us replace $\ver{k}$ by ${\ver{k'}=\hat R \, \ver{k}}$ and $\ver{q}$ by ${\ver{q'}=\hat R \, \ver{q}}$ everywhere in the above equations. Realizing that the sums over $\hat R \,\ver{q}$ and $\ver{q}$ are identical, we find that the new set of equations is the same 
linear mapping
as for the original mode functions and mode frequency. We thus have the identities ${\ome[k']{\xi}=\ome[k]{\xi}}$, ${(\mfu[k]{\xi,a}, \mfu[k]{\xi,b},\mfu[k]{\xi,\ver{q}})}={(\mfu[k']{\xi,a}, \mfu[k']{\xi,b},\mfu[k']{\xi,\ver{q'}})}$ and 
${(\mfv[k]{\xi,a}, \mfv[k]{\xi,b},\mfv[k]{\xi,\ver{q}})}={(\mfv[k']{\xi,a}, \mfv[k']{\xi,b},\mfv[k']{\xi,\ver{q'}})}$. 
As an immediate consequence, the mode functions $\mfu[k]{\xi,a}$, $\mfu[k]{\xi,b}$ and $\mfv[k]{\xi,a}$, $\mfv[k]{\xi,b}$ as well as the mode frequency $\ome[k]{\xi}$ can only depend on the modulus $\abs{k}$ since the rotation $\hat R$ is arbitrary. One rotation which is especially useful is the rotation mapping $\ver{k}$ into $\ver{-k}$ and $\ver{q}$ into $\ver{-q}$.
\subsection{Low-energy (An\-der\-\mbox{son-)Bo}\-gol\-iu\-bov modes}  
In this section, we will reduce the equations for the mode functions and mode frequencies as far as possible. 
We will restrict ourselves to low frequency excitations.
\subsubsection{Elimination of $\mfu[k]{\xi,b}$ and $\mfv[k]{\xi,b}$ for appropriate $\ome[k]{\xi}$}
Comparing the equations \eqref{ua} and \eqref{ub} as well as \eqref{va} and \eqref{vb}, we can express $\mfu[k]{\xi,b}$ and $\mfv[k]{\xi,b}$ in terms of $\mfu[k]{\xi,a}$ and $\mfv[k]{\xi,a}$
\begin{align}
\label{elim1}
\mfu[k]{\xi,b}&=\eps \, f(\ome[k]{\xi}) \, \mfu[k]{\xi,a}		\;\;, \\
\label{elim2}
\mfv[k]{\xi,b}&=\eps \, f(-\,\ome[k]{\xi}) \, \mfv[k]{\xi,a}
\end{align}
by means of the function
\begin{align}\label{ffunction}
f(\ome[k]{\xi})
=\rka \frac{1}{2}\,\ve{k}^2-\om-\ome[k]{\xi}\rkz
\rka \frac{\gammaF^2}{\gF}-\eps^2\rka\om+\ome[k]{\xi}\rkz\rkz^{-1}	\;.
\end{align}
This will help us in the following to eliminate $\mfu[k]{\xi,b}$ and $\mfv[k]{\xi,b}$ from the equations.
Of course, the above denominator should not become zero. Since we consider $\eps$ as going to zero anyway, we restrict our further analysis to cases where the combination of $\xi$ and $\epsw$ fulfills the condition
\begin{align}\label{sat1}
\ome[k]{\xi}<\frac{\gammaF^2}{\gF}\,\frac{1}{\epsw^2}-\om	\;.
\end{align}
For an appropriate choice of $\eps$, this holds for $\xi=a$ or if $\xi$ equals $\ver{q}$ for a small value of $\abs{q}$. However, as $\eps$ tends to zero, more and more $\ver{q}$-values meet this criterion and the upper bound for $\ome[k]{\xi}$ tends to infinity.
\subsubsection{Elimination of $\mfu[k]{\xi,\ver{q}}$ and $\mfv[k]{\xi,\ver{q}}$ for appropriate $\ome[k]{\xi}$}
\label{threshold}
We would also like to eliminate the mode functions $\mfu[k]{\xi,\ver{q}}$ and $\mfv[k]{\xi,\ver{q}}$. This is easily done by rewriting equation \eqref{uq} as
\begin{align}
\label{denomu}
\mfu[k]{\xi,\ver{q}}=\!\biggl[&\gammaF\, \hfZ[q]\hfZ[k-q]\,\Nm^{-1/2}
				\rka 1+f(\ome[k]{\xi})\rkz\mfu[k]{\xi,a}\nn\\
				&+\gammaF\, \hfY[q]\hfY[k-q]\,\Nm^{-1/2}
				\rka 1+f(-\,\ome[k]{\xi})\rkz\mfv[k]{\xi,a}\biggr]\nn\\
				&\Bigl( \W[k-q]+\W[q]-\ome[k]{\xi}\Bigr)^{-1}
\end{align}
and equation \eqref{vq} as
\begin{align}
\label{denomv}
\mfv[k]{\xi,\ver{q}}=\!\biggl[&\gammaF\, \fY[q]\fY[-k-q]\,\Nm^{-1/2}
				\rka 1+f(\ome[k]{\xi})\rkz\mfu[k]{\xi,a}\nn\\
				&+\gammaF\, \fZ[q]\fZ[-k-q]\,\Nm^{-1/2}
				\rka 1+f(-\,\ome[k]{\xi})\rkz\mfv[k]{\xi,a}\biggr]\nn\\
				&\Bigl( \W[-k-q]+\W[q]+\ome[k]{\xi}\Bigr)^{-1}		\;.
\end{align}
Equation \eqref{denomu} is at least well defined as long as its denominator does not become zero. Thus, to avoid unnecessary complications, we restricts us 
to mode frequencies satisfying 
\begin{align}\label{sat2}
{\ome[k]{\xi}<\underset{\ver{q}}{\min}\rka\W[k-q]+\W[q]\rkz}
\end{align} 
with $\W[q]$ as defined in Appendix \ref{classical}.

As long as ${\DeltaF-\om<0}$ holds, not all atoms of the atomic background solution have already gone through the resonance. In this case we have ${\underset{\ver{q}}{\min}\rka\W[k-q]+\W[q]\rkz=2\, \gammaF \sqrt{\neff}}$ for all ${\abs{k}\leq2\,\sqrt{(\om-\DeltaF)/2}}$ (see \cite{Comb06}) with $\neff$ as defined in Appendix \ref{classical}. 
The minimal value is achieved for $\ver{q}$ laying on a circle defined by the intersection of the spheres ${\ver{q}^2=(\om-\DeltaF)/2}$ and ${(\ver{k}-\ver{q})^2=(\om-\DeltaF)/2}$. 
In all other cases, the lower bound is given by ${\underset{\ver{q}}{\min}\rka\W[k-q]+\W[q]\rkz=2\,\W[k/2]}$ (see \cite{Comb06}) and is achieved for ${\ver{q}=\ver{k}/2}$.

$2\,\W[k/2]$ is bounded from below by ${2\, \gammaF \sqrt{\neff}}$. The latter is thus the overall lower bound with respect to $\ver{k}$ on the BCS side, 
where the value $2\, \gammaF \sqrt{\neff}$ is just the minimal width of the energy gap in the two-level systems of the atomic background solution as described in Appendix \ref{classical}. This gap separates 
paired from unpaired atomic (fermionic) 
states. Finally, this means that the above condition for $\ome[k]{\xi}$ is nothing else than the requirement that the excitations considered here must have 
frequencies much lower than the frequencies involved 
in breaking pairs of atoms.

We will focus in the following on excitations which are energetically well separated from the dummy molecular mode (condition \eqref{sat1}) and from the atomic pair breaking threshold (condition \eqref{sat2}). It will turn out that this class of instantaneous normal modes agrees with low-energy free molecular modes $\opda[k]$ at late times. Therefore, we use from now on the index $a$ instead of the general index \nolinebreak $\xi$.
\subsubsection{A simple matrix equation for $\mfu[k]{a,a}$ and $\mfv[k]{a,a}$}
From equations \eqref{ua} and \eqref{va} we get $\mfu[k]{a,a}$ and $\mfv[k]{a,a}$, respectively. In order to close the system of equations, we eliminate $\mfu[k]{a,\ver{q}}$ and $\mfv[k]{a,\ver{q}}$ by means of \eqref{denomu} and \eqref{denomv}. Finally, we end up with a linear mapping
\begin{align}
\label{mat}
\ome[k]{a}
\left(\begin{array}{c}
\mfu[k]{a,a} \\
\mfv[k]{a,a} 
\end{array}\right)
=
\left(\begin{array}{cc}
\mathcal{M}_{11} &
\mathcal{M}_{12} \\
\mathcal{M}_{21} & 
\mathcal{M}_{22} 
\end{array}\right)
\left(\begin{array}{c}
\mfu[k]{a,a} \\
\mfv[k]{a,a} 
\end{array}\right)
\end{align}
where the componenets of the matrix $\mathcal{M}$ are defined as 
\begin{align}
\label{M11}
\mathcal{M}_{11}&\mydoteq \frac{1}{2}\,\ve{k}^2-\om
-\rka 1+f(\ome[k]{a})\rkz A(\ver{k},\ome[k]{a})		\;\;, \\
\label{M12}
\mathcal{M}_{12}&\mydoteq 
-\rka 1+f(-\,\ome[k]{a})\rkz B(\ver{k},\ome[k]{a})	\;\;, \\
\label{M21}
\mathcal{M}_{21}&\mydoteq \rka 1+f(\ome[k]{a})\rkz \bar B(\ver{k},\ome[k]{a})		\;\;,
\\
\label{M22}
\mathcal{M}_{22}&\mydoteq -\,\frac{1}{2}\,\ve{k}^2+\om
+\rka 1+f(-\,\ome[k]{a})\rkz A(\ver{k},-\,\ome[k]{a})	\;\;.
\end{align}
The functions $A(\ver{k},\ome[k]{a})$ and $B(\ver{k},\ome[k]{a})$ are defined after some appropriate relabeling of the summation indices as 
\begin{align}
A(\ver{k},\ome[k]{a})\mydoteq \frac{\gammaF^2}{\Nm}\sum_{\ver{q}}
&\biggl[\frac{\fZ[q]\fZ[k-q]\hfZ[q]\hfZ[k-q]}{\W[k-q]+\W[q]-\ome[k]{a}}\nn\\
&+\frac{\hfY[q]\hfY[k-q]\fY[q]\fY[k-q]}{\W[k-q]+\W[q]+\ome[k]{a}}\biggr]
\end{align}
and
\begin{align}
B(\ver{k},\ome[k]{a})\mydoteq \frac{\gammaF^2}{\Nm}\sum_{\ver{q}}
&\biggl[\frac{\fZ[q]\fZ[k-q]\hfY[q]\hfY[k-q]}{\W[k-q]+\W[q]-\ome[k]{a}}\nn\\
&+\frac{\hfY[q]\hfY[k-q]\fZ[q]\fZ[k-q]}{\W[k-q]+\W[q]+\ome[k]{a}}\biggr]		\;.
\end{align}
Similar functions appear in the time-independent treatment of single-channel and two-channel models as, \eg in \cite{Roma06,Mari98}. 

We have thus reached an eigenvector equation, which will deliver us $\mfu[k]{a,a}$ and $\mfv[k]{a,a}$. The mode functions $\mfu[k]{a,\ver{q}}$ and $\mfv[k]{a,\ver{q}}$ can then be derived trivially from equations \eqref{denomu} and \eqref{denomv}. The mode functions $\mfu[k]{a,b}$ and $\mfv[k]{a,b}$ follow from equations \eqref{elim1} and \eqref{elim2}.
Please keep in mind that after all the manipulations in this section, we assume small $\eps$ or even $\eps \to 0$. The matrix equation \eqref{mat} 
then only depends on 
the  small-$\eps$-limit of $f(\ome[k]{\xi})$ \eqref{ffunction}.

Note that the above calculation could be generalized to excitations which are energetically not well separated from the dummy molecular mode (condition \eqref{sat1}): One could eliminate $\mfu[k]{a,\ver{q}}$ and $\mfv[k]{a,\ver{q}}$ as well, but would end up with a four-by-four matrix instead of $\mathcal{M}$. These four equations would then also depend linearly on $\mfu[k]{a,b}$ and $\mfv[k]{a,b}$. The strategy would just be to solve \eqref{uq} and \eqref{vq} for $\mfu[k]{a,\ver{q}}$ and $\mfv[k]{a,\ver{q}}$ and to apply the result in the other eigenvalue equations \eqref{ua}, \eqref{ub}, \eqref{va} and \eqref{vb}. 
However, we do not follow this approach here, since for a true multi-channel Feshbach resonance with finite $\eps$ it would have been necessary to keep the kinetic energy term of the second resonance, which we neglected right in the beginning. Moreover, it would have been necessary to specify the precise position of the resonance in the Hamiltonian \nolinebreak \eqref{addres}.

\subsection{Frequencies of the low-energy (An\-der\-\mbox{son-)Bo}\-gol\-iu\-bov modes}\label{lowsection}
According to equation \eqref{mat}, the low-energy mode frequency $\ome[k]{a}$ can be found straightforwardly by setting the determinant $\det(\mathcal{M}-\ome[k]{a})$ to zero (compare the single channel model \cite{Mari98}):
\begin{align}\label{deteq}
(\mathcal{M}_{11}-\ome[k]{a})\,(\mathcal{M}_{22}-\ome[k]{a})-\mathcal{M}_{12}\,\mathcal{M}_{21}=0		\;.
\end{align}
Note that this equation involves $\ome[k]{a}$ not just polynomially but in a very non-trivial way, 
because the matrix elements $\mathcal{M}_{ij}$ all depend on $\ome[k]{a}$, as given by \eqref{M11}, \eqref{M12}, \eqref{M21}, \eqref{M22}. Numerical solution is necessary to find the explicit frequencies.

For sufficiently small $\abs{k}$, the possible mode frequencies of $\HBk{k}$ (as the volume tends to infinity) will appear as follows: There exists one isolated solution $\ome[k]{a}$ for each $\HBk{k}$ and a continuum of possible frequencies above some threshold value. The existence of continuous mode frequencies becomes clear by realizing that
\begin{align}
\Bigl[\HBk{k}\,,\hopg[k,q']-\hopg[k,k-q']\Bigr]={}&\Bigl(\W[q']+\W[k-q']\Bigr)\nn\\
&\cdot\Bigl( \hopg[k,q']-\hopg[k,k-q'] \Bigr)	\;.
\end{align}
This means that, when applied onto an eigenstate, the operator $\hopg[k,q']-\hopg[k,k-q']$ creates excitations with frequencies $\W[q']+\W[k-q']$. As the example shows, there exist excitations with frequencies of the uncoupled virtual molecules. For fixed $\ver{k}$, they are bounded from below by the very same minimum discussed in the previous Section \nolinebreak[4] \ref{threshold}.   

On the BCS side ${\DeltaF-{\om}< 0}$, the $\ver{k}$-dependence of $\ome[k]{a}$ for moderate $\abs{k}$-values shows a special feature: It tends to the threshold line associated with the onset of the continuous frequency spectrum. The threshold value is ${2\, \gammaF \sqrt{\neff}}$ for ${\abs{k}\leq2\,\sqrt{(\om-\DeltaF)/2}}$ and $2\,\W[k/2]$ for ${\abs{k}\geq2\,\sqrt{(\om-\DeltaF)/2}}$, where $\neff$ and $\W[k]$ are defined in Appendix \nolinebreak[4] \ref{classical}. This behavior has been stated for a single-channel model in \cite{Comb06} and is shown in Figure \nolinebreak 2 therein. We recover exactly this figure at ${t_i=-\,\infty}$, ${\DeltaF(t_i)=-\,\infty}$, where our two-channel model has turned assymptotically into a single-channel model. This happens as follows:

Since in the above limit (${\DeltaF\to -\,\infty}$) only the quantities $\gF$, ${\rka\DeltaF-\om\rkz/2}$ and ${2\, \gammaF \sqrt{\neff}}$ appear in the equation for $\ome[k]{a}$ as well as in the two equations of motion for the classical background (see Appendix \ref{classical}), the only free undetermined parameter which is left is $\gF$. The parameter $\gammaF$ alone becomes arbitrary since it can be compensated by an appropriate choice of $\neff$. Thus, we are essentially left with a BCS model that involves only the parameter \nolinebreak $\gF$. 
\begin{table}[!h]
\normalsize
\begin{math}
\begin{array}{|c|c|c|c|}
\hline
\text{`$1/k_F a$' in \cite{Comb06}}	&	\gF		&	\rka\DeltaF-\om\rkz/2	&	2\, \gammaF \sqrt{\neff} 	\\
\hline
-\,1/2								&	8/( 3\,\pi )	&	-\,0.849490			&	0.804175				\\
\hline
-\,1\phantom{/2}						&	4/( 3\,\pi )	&	-\,0.953983			&	0.416835				\\
\hline
\end{array}
\end{math}
\caption{\label{tablestringari}Parameters matching Figure 2 in \cite{Comb06} for ${t = t_i}$.}
\end{table}

The parameters for which our two-channel model asymptotically (${\DeltaF\to -\,\infty}$) agrees with the single-channel model of Figure 2 in \cite{Comb06} are given in table \ref{tablestringari}, where the second and third row correspond to the upper and lower thick lines in the cited figure.

\begin{figure}[h]
\includegraphics[width=1.0\linewidth]{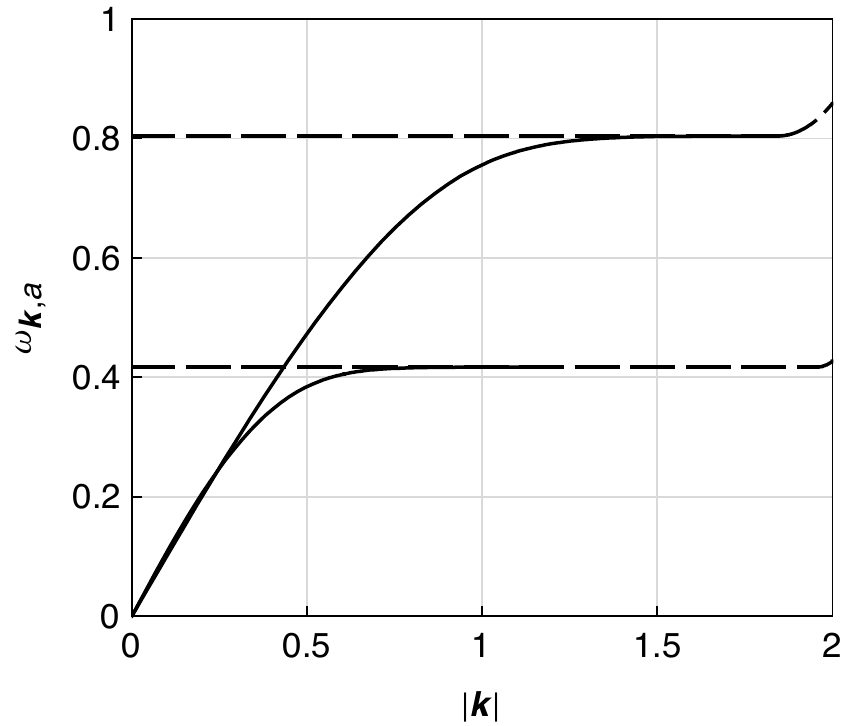}
\caption{\label{d1}Frequency dispersion relation $\ome[k]{a}$ (solid line) and its $\ver{k}$-dependent threshold (dashed line) as ${\DeltaF \to -\,\infty}$. Shown are the two cases given in table \ref{tablestringari}. The ${\DeltaF \to -\,\infty}$ limit recovers a single-channel model as considered in \cite{Comb06}. We chose the parameters such that our results are in agreement with Figure 2 therein. The threshold line is ${2\, \gammaF \sqrt{\neff}}$ for ${\abs{k}\leq2\,\sqrt{(\om-\DeltaF)/2}}$ and $2\,\W[k/2]$ for ${\abs{k}\geq2\,\sqrt{(\om-\DeltaF)/2}}$. To simplify numerics, $\ome[k]{a}$ is only calculated up to values slightly below the threshold line.}
\end{figure}

Figure \ref{d1} shows the $\ver{k}$-dependence of our two-channel model on the BCS side for ${t_i=-\,\infty}$, where ${\DeltaF(t_i)=-\,\infty}$. A similar situation is also found for finite times $t$ on the whole BCS side. 
The approach of the threshold line is the same as in the single-channel case \cite{Comb06}. 

As discussed in \cite{Comb06} for the single-channel case, the behavior of these curves at higher frequencies is non-trivial.
On the BCS side ${\DeltaF-{\om}< 0}$, the dispersion curve for Anderson-Bogoliubov modes 
most probably merges into the continuum for a certain $\abs{k}$ with ${\abs{k}>2\,\sqrt{(\om-\DeltaF)/2}}$ as it does in the single-channel case; the seemingly asymptotic approach to the continuum that one sees in Figure \ref{d1} here and in Figure 2 of \cite{Comb06} changes at high $\abs{k}$. As the BCS-BEC crossover proceeds, 
the dispersion curve is expected to cross into the continuum 
before entering the BEC regime ${\DeltaF-{\om}> 0}$. There, the dispersion curve gradually turns into the parabola of non-interacting bosons. On the BEC side, we expect no special behavior of the dispersion curve at the threshold of the $\opg[k,q]$-continuum. This is simply because the molecular frequency in the extreme BEC limit, $\ver{k}^2/2$, is always smaller than the continuum threshold on the BEC side, ${\underset{\ver{q}}{\min}\rka\W[k-q]+\W[q]\rkz=2\,\W[k/2]}$ (see Appendix \ref{classical} for definition).

In a single-channel model, \cite{Comb06} considers collective modes of all frequencies, for \textit{time-independent} Feshbach detuning $\DeltaF$. 
Our interest here, however, is in the non-adiabatic effects of \textit{time-dependent} $\DeltaF$ in a two-channel model. 
Feshbach sweep rates high enough to induce non-adiabatic behavior at frequencies approaching the BCS gap will also produce more complicated non-adiabatic evolution, including the non-perturbative two-body fermion dynamics that we have here treated adiabatically. 
We thus restrict our attention to Feshbach sweeps slow compared to the gap, so that our treatment of the fermions with a mixture of adiabatic and perturbative approximations is valid. 
We therefore also focus only on collective modes with frequency well below the continuum threshold, since only these will show significant non-adiabatic evolution in the limit of a slow BCS-BEC crossover.
For low frequencies, wave numbers are also low, and the linear dispersion curve indicates that we deal with sound modes.

\begin{figure}[b]
\includegraphics[width=1.0\linewidth]{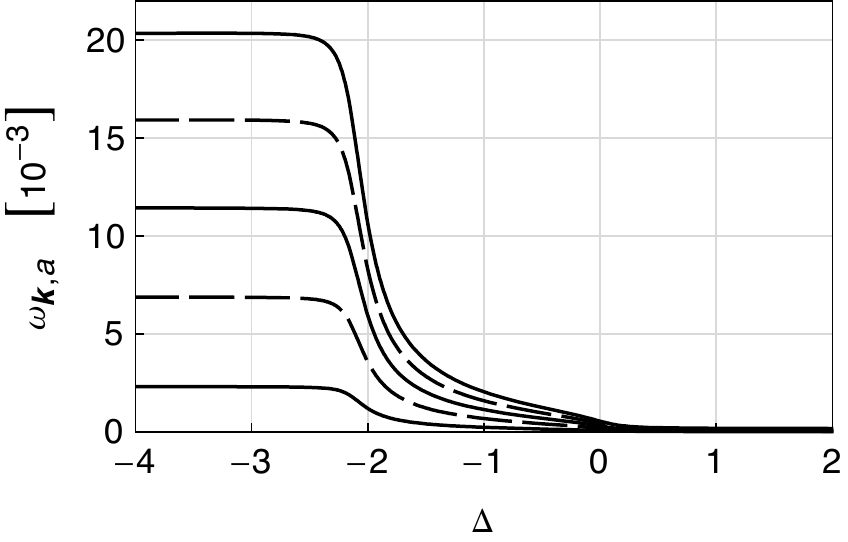}
\caption{\label{d2}Frequency dispersion relation $\ome[k]{a}$ versus $\DeltaF$ for different $\abs{k}$ (${10^{3}\,\abs{k}=2}$, $6$, $10$, $14$, $18$, lower to upper line). The linear $\abs{k}$-dependence for small $\abs{k}$ in the early BCS regime is obvious. The parameters are ${\gammaF=0.1}$ and ${\gF=0.2}$.}
\end{figure}
\begin{figure}[h]
\includegraphics[width=1.0\linewidth]{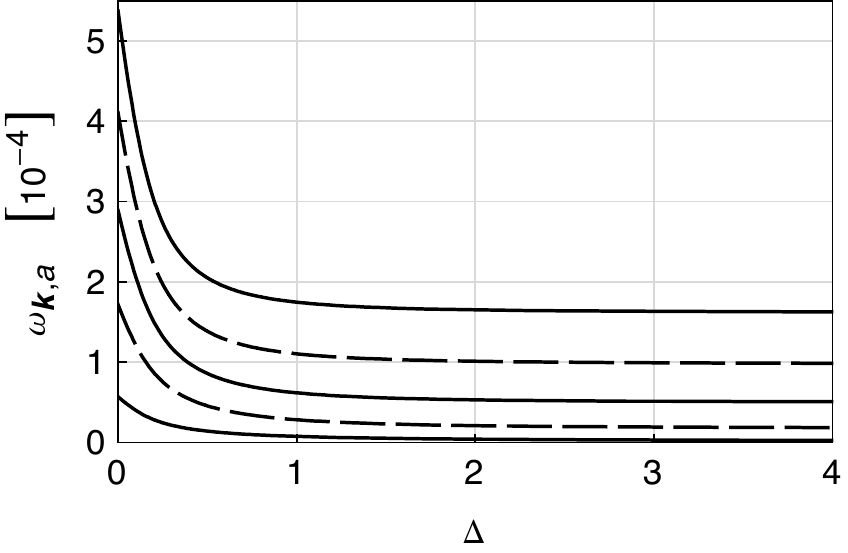}
\caption{\label{d3}Same situation as in Figure \ref{d2}, but focussing onto another region. The free particle dispersion relation $\ver{k}^2/2$ is asymptotically approached in the late BEC regime.}
\end{figure}
The time dependence of instantaneous, low $\ome[k]{a}$ is shown in Figures \ref{d2} and \nolinebreak[4] \ref{d3}. In Figure \ref{d2}, the linear $\abs{k}$-dependence of $\ome[k]{a}$ for small $\abs{k}$ becomes obvious. Since the system does not change very much as long as the atoms are not resonant with the molecules yet (\ie before ${\DeltaF\approx-2}$), the frequencies do not change very much in this region, either. For the parameters chosen, the frequencies drop quite rapidly as soon as the first atoms on the Fermi surface become resonant with the molecules, just before ${\DeltaF\approx-2}$. However, the frequencies plotted here still show a very linear $\abs{k}$-dependence up to $\DeltaF \approx 0$, \ie almost in the full BCS regime ${\DeltaF-\om<0}$.

The mode frequencies are shown in Figure \nolinebreak[4] \ref{d3} for the late BEC regime, \ie ${\DeltaF-\om>0}$. Evidently, the molecules turn into free particles since their frequencies asymptotically approach $\ver{k}^2/2$. 
In the transition region between BCS and BEC, \ie  for ${\DeltaF-\om\approx0}$, the frequency dispersion changes smoothly from linear into quadratic.

\section{Post-adiabatic normal modes Hamiltonian}\label{post}
In this section we will derive the central result of this paper, namely a post-adiabatic Hamiltonian $\HI$ 
for which the subscript indicates that it is expressed in terms of the instantaneous normal modes. The post-adiabatic effects will thus be expressed as non-diagonal couplings among these modes. Within the approximations and constraints made in the previous sections and in the following paragraph, this Hamiltonian is equivalent to our original problem. 

We will derive the coupling among normal modes, which is of order ${\dot{\Delta}}$ and originates from time dependence of the Bogoliubov transformation $\B[k]$.
Note that in general, there are also corrections to $\fY[q]$ and $\fZ[q]$ of the order ${\dot{\Delta}}$ contained in $\fy[q]$ and $\fz[q]$ due to the adiabatic time evolution of the fermions in the molecular background (see Appendix \ref{classical}). This contribution could be expressed in terms of 
normal modes 
as well, leading to additional couplings among these. However, we neglect these couplings, since we are mainly interested in the dynamics of low energy normal modes located at the 
lower end 
within the fermionic gap. For these low energy normal modes, the situation is as follows: compared with the coupling terms of order ${\dot{\Delta}}$ which result from the time dependence of the Bogoliubov transformation $\B[k]$, the coupling terms resulting from the time dependence of the fermionic subsystem are roughly speaking \textit{at least} smaller by the ratio of $\ome[k]{a}$ and the fermionic gap. Since we will finally consider momenta $\ver{k}$ for which this ratio tends to be small, we will neglect these coupling terms.
\subsection{Time evolution of the smallest factorizing part}
Our procedure was exactly the same as the familiar Bogoliubov diagonalization of a quadratic Hamiltonian, but since the Hamiltonian in question was time-dependent, the resulting quasiparticle operators were also time-dependent, and not only in the trivial sense of having time-dependent phase pre-factors. Our Bogoliubov transformation was itself non-trivially time-dependent, with each of the instantaneously diagonalizing operators being \textit{time-dependent} combination of \textit{time-\textbf{in}dependent} creation and destruction operators. 

In this sense our time-dependent Bogliubov transformation simply has not yet gone quite far enough. For \textit{any given time} $t$, we have a convenient set of instantaneously diagonalizing operators; but this is still a \textit{different} set of operators for every $t$. We do not yet have a \textit{single} set of operators that provides a simple description of excitations \textit{at all times}.

The easiest way to explain exactly what we still need to do is probably just to do it. We introduce a time-dependent change of basis in the many-body Hilbert subspace of modes with fixed $\ver{k}$, such that for all states $\ket{\Psi_\ver{k}(t)}$, the new-basis representation $\ket{\tilde{\Psi}_\ver{k}(t)}$ is given by
\begin{align}
\ket{\Psi_\ver{k}(t)} \myeqdot \B[k](t) \ket{\tilde{\Psi}_\ver{k}(t)} \;,
\end{align}
where $\B[k](t)$ is the unitary operator that effects the time-dependent Bogoliubov transformation which instantaneously diagonalizes $\HBk{k}$:

\begin{align}\label{Dd}
\opD[k]{\xi}\mydoteq\B[k]\,\opd[k]{\xi}\,\hB[k]		\;.	
\end{align}
We usually omit the argument of $\B[k](t)$ whenever it is $t$. In general $\B[k]$ is the exponential of a bilinear mapping in terms of the old creation and annihilation operators appearing in $\HBk{k}$ \eqref{dummy}. 

The effect of this change of basis is simple, and very convenient: it re-labels states so that they count as the same, at different times, if they are the same modulo adiabatic time evolution. So, for example, the instantaneous ground state of the $\ver{k}$-subspace at any time $t$ was defined in our original basis by
\begin{align}
\opD[k]{\xi}(t)\ket{0(t)} = 0		\;.
\end{align}
This implies that, in the new basis, the ground state is determined by
\begin{align}
\opD[k]{\xi}(t)\, \B[k](t) \ket{\tilde{0}} 
\equiv \B[k](t)\, \opd[k]{\xi} \ket{\tilde{0}}&
= 0 \nn\\
\Longleftrightarrow \opd[k]{\xi} \ket{\tilde{0}}&
= 0 		\;.
\end{align}
Since all the operators $\opd[k]{\xi}$ are \textit{time-independent}, the ground state $\ket{\tilde{0}}$ is now also time-independent. A similar property holds for all instantaneous eigenstates of $\HBk{k}$: their eigenvalues are time-dependent, but the states themselves are now time-independent, in this new time-dependent basis, because the adiabatic time evolution (up to phases) has been built into the basis itself.

Exactly this change of many-body Hilbert space basis is normally made implicitly at the same time as one diagonalizes a quadratic Hamiltonian with a Bogoliubov transformation. In the case of a time-independent Bogoliubov transformation, diagonalizing a time-independent Hamiltonian, the basis transformation is trivial enough that it needs no explicit attention. In the general case of a time-dependent transformation, however, the basis change is time-dependent, and some non-trivial effects are involved. For where the time evolution in the original basis was given by
\begin{align}
\ket{\Psi_\ver{k}(t)} = \UBk{k}(t,t_i) \ket{\Psi_\ver{k}(t_i)} 		\;,
\end{align}
now in the new basis the time evolution is
\begin{align}
\ket{\tilde{\Psi}_\ver{k}(t)} ={}& \UIk{k}(t,t_i) \ket{\tilde{\Psi}_\ver{k}(t_i)} 		\;, \\
\label{ubofui}
\UIk{k}(t,t_i)\mydoteq{}&\hB[k](t)\,\UBk{k}(t,t_i)\,\B[k](t_i)		\;,
\end{align}
with
\begin{align}\label{hiklong}
\HIk{k}=\hB[k]\,\HBk{k}\,\B[k]-\rmi \, \hB[k]\rka\frac{\rmd}{\rmd t}\,\B[k]\rkz
\end{align}
where we have also used the identity ${\hB[k](t)\, \B[k](t) \equiv 1}$.

The first term in $\HIk{k}$ represents strictly adiabatic evolution. It is a sum of rather trivial terms ${\ome[k]{\xi}(t)\,\hopd[k]{\xi}\, \opd[k]{\xi}}$ where the time-dependent instantaneous eigenfrequencies $\ome[k]{\xi}(t)$ are those of Section \ref{inst} above, and the operators $\opd[k]{\xi}$ and $\hopd[k]{\xi}$ are \textit{time-independent}. The second term in $\HIk{k}$ is less trivial, but because the Bogoliubov transformation $\B[k]$ only varies on the slow time scale ${\DeltaF}$, it will be a post-adiabatic correction that is small if the crossover sweep is slow. It can be considered a second-quantized analog to first post-adiabatic correction terms in single-particle quantum mechanics, such as so-called geometric magnetism and other artificial gauge fields.

We would also like this second term to be expressed in terms of the \textit{time-independent} operators $\opd[k]{\xi}$.
As we will see, this term couples different eigenmodes of the instantaneous Hamiltonian. Its representation at hand is still not very practical, however, since it contains $\hB[k]$ and the time derivative of $\B[k]$. To compute the coupling term directly would thus require an explicit representation of $\B[k]$. We will take a simpler approach in the following section.   

Keep in mind that due to the transformation into the normal modes basis, the physical meaning of $\opd[k]{\xi}$ has changed; instead of describing the basic excitations of molecules and virtual bosons, it now describes the excitation \textit{in terms of} normal modes. Here `\textit{in terms of}' means that $\opd[k]{\xi}$ is still \textit{not time-dependent} itself, since the time-dependence (up to phases) of the normal modes has been absorbed in $\hB[k](t)$ within the definition of $\UIk{k}$ \nolinebreak\eqref{ubofui}.

%
\subsection{Coupling of instantaneous eigenmodes}
In order to derive the coupling term among normal modes in $\HIk{k}$ \eqref{hiklong} without using an explicit representation for $\B[k]$, we take the following approach. We calculate the commutator of the coupling term, with all  the operators it contains, in terms of the \textit{time-independent} operators $\opd[k]{\xi}$. It is then possible to reconstruct an operator leading to the same commutation relations. The coupling term reconstructed by this method is just defined up to 
c-number terms, 
which we freely choose to be zero.

Following this strategy, we first of all calculate the commutator 
of the coupling term (omitting the prefactor) with $\opd[k]{\xi}$
\begin{align}\label{dcom}
\biggl[ \opd[k]{\xi} &\, , \hB[k]\rka\frac{\rmd}{\rmd t}\,\B[k]\rkz \biggr]  \nn\\
&=\opd[k]{\xi} \, \hB[k]\rka\frac{\rmd}{\rmd t}\,\B[k]\rkz + \rka \frac{\rmd}{\rmd t}\,\hB[k] \rkz \B[k] \, \opd[k]{\xi}  \nn\\
&=\hB[k]\, \opD[k]{\xi}\rka\frac{\rmd}{\rmd t}\,\B[k]\rkz + \rka \frac{\rmd}{\rmd t}\,\hB[k] \rkz \opD[k]{\xi} \, \B[k]  \nn\\
&=\frac{\rmd}{\rmd t}\rka \hB[k]\, \opD[k]{\xi}\, \B[k] \rkz - \hB[k] \rka \frac{\rmd}{\rmd t}\,\opD[k]{\xi} \rkz \B[k]  \nn\\
&=-\, \hB[k] \rka \frac{\rmd}{\rmd t}\,\opD[k]{\xi} \rkz \B[k] 		\;.
\end{align}
In the first step of this derivation we use the identity 
\begin{align}\label{identderiv}
\frac{\rmd}{\rmd t}\rka\hB[k]\,\B[k]\rkz
=\hB[k]\rka \frac{\rmd}{\rmd t}\,\B[k] \rkz+\rka \frac{\rmd}{\rmd t}\,\hB[k] \rkz \B[k]=0
\end{align}
and in its last step we use the inverse Bogoliubov transformation
\begin{align}
\opd[k]{\xi}=\hB[k]\,\opD[k]{\xi}\,\B[k]		\;.
\end{align}
The expression for the commutation relation \eqref{dcom} 
\begin{align}
-\, \hB[k] \rka \frac{\rmd}{\rmd t}\,\opD[k]{\xi} \rkz \B[k] 
\end{align}
might not look very useful at first sight, but in fact it is. If we had the time derivative of $\opD[k]{\xi}$ in terms of $\opD[k]{\xi}$-modes, the inverse Bogoliubov transformation in the above equation would map them into $\opd[k]{\xi}$-modes, giving the desired result.

Thus we calculate the time derivative of $\opD[k]{\xi}$ in terms of $\opD[k]{\xi}$-modes by using its explicit definition \eqref{b}
\begin{align}
\frac{\rmd}{\rmd t}\,\opD[k]{\xi}
=\sum_{\eta
}
\rka
\opd[k]{\eta}		\,\frac{\rmd}{\rmd t}\,\mfu[k]{\xi,\eta}
+\hopd[-k]{\eta}	\,\frac{\rmd}{\rmd t}\,\mfv[k]{\xi,\eta}
\rkz
\end{align}
and by replacing the $\opd[k]{\xi}$-modes therein with the \textit{explicit} inverse Bogoliubov transformation \eqref{inv}, \eqref{invd}. 
We obtain an equation which contains only the various operators $\opD[k]{\xi'}$ and $\hopD[-k]{\xi'}$. 
The desired result is then
\begin{align}\label{dcomd}
\eka \opd[k]{\xi} \, , \hB[k]\rka\frac{\rmd}{\rmd t}\,\B[k]\rkz \ekz ={}& \dot{\DeltaF}\sum_{\xi'}
G_{\ver{k},\xi,\xi'} \, \opd[k]{\xi'} \nn\\
&-\dot{\DeltaF}\sum_{\xi'}
\I_{\ver{k},\xi,\xi'} \, \hopd[-k]{\xi'}  
\end{align}
giving the commutator of $\opd[k]{\xi}$ with the coupling term in terms of the \textit{time-independent} modes $\opd[k]{\xi'}$. The commutator of the coupling term with $\hopd[k]{\xi}$ can be found straightforwardly by taking the Hermitian conjugate of this equation \eqref{dcomd} and using the identity \eqref{identderiv}.

In the above equation \eqref{dcomd}, the coupling between the operators involves the newly defined functions
\begin{align}
G_{\ver{k},\xi,\xi'}&\mydoteq\sum_{\eta
}
\Big(
\hmfu[k]{\xi',\eta}\,\frac{\rmd}{\rmd \DeltaF}\,\mfu[k]{\xi,\eta}
-\hmfv[k]{\xi',\eta}\,\frac{\rmd}{\rmd \DeltaF}\,\mfv[k]{\xi,\eta}
\Big)\label{Gdef}		\;, \\
I_{\ver{k},\xi,\xi'}&\mydoteq\sum_{\eta
}
\Big(
\mfv[-k]{\xi',\eta}\,\frac{\rmd}{\rmd \DeltaF}\,\mfu[k]{\xi,\eta}
-\mfu[-k]{\xi',\eta}\,\frac{\rmd}{\rmd \DeltaF}\,\mfv[k]{\xi,\eta}
\Big)\label{Idef}
\end{align}
where we should keep in mind the properties of the mode functions as discussed in Section \ref{gmf}. Since the mode functions only depend on time via $\DeltaF$ in our case, the above definitions immediately imply that the same also holds for $G_{\ver{k},\xi,\xi'}$ and $I_{\ver{k},\xi,\xi'}$.

The functions $G_{\ver{k},\xi,\xi'}$ and $I_{\ver{k},\xi,\xi'}$ offer some symmetries: If we take the time derivative of the completeness relation \nolinebreak \eqref{condi1},  
we find (note the position of the prime)
\begin{align}\label{G}
G_{\ver{k},\xi,\xi'}=-\,\bar{G}_{\ver{k},\xi',\xi}		\;.
\end{align}
This means that $G_{\ver{k},\xi,\xi'}$ is a purely imaginary function.
By taking the time derivative of the orthogonality relation \eqref{condi2},  
we find (note the position of the prime)
\begin{align}\label{I}
I_{\ver{k},\xi,\xi'}=I_{-\ver{k},\xi',\xi}		\;.
\end{align}

However, one might already guess from the rotational symmetry of our system as the volume tends to infinity, that $G_{\ver{k},\xi,\xi'}$ and $I_{\ver{k},\xi,\xi'}$ depend on $\ver{k}$ just via the modulus $\abs{k}$.
More rigorously, this can also be seen as follows:
If the system's volume tends to infinity, we can replace $\ver{k}$ by rotated ${\ver{k'}=\hat R \, \ver{k}}$ everywhere in \eqref{Gdef} and \eqref{Idef}. We also replace the part of the $\eta$-sum over $\ver{q}$ by a sum over the rotated ${\ver{q'}=\hat R \, \ver{q}}$. Using the identities derived in Section \nolinebreak[4] \ref{gmf} and realizing that the sums over $\hat R \,\ver{q}$ and $\ver{q}$ are identical, we find that $G_{\ver{k},\xi,\xi'}=G_{\ver{k'},\xi,\xi'}$ and $I_{\ver{k},\xi,\xi'}=I_{\ver{k'},\xi,\xi'}$. Since the rotation $\hat R$ is arbitrary, both functions depend only on the modulus $\abs{k}$.
\subsection{The Hamiltonian in instantaneous eigenmodes}
In this section, we will calculate $\HIk{k}$, the Hamiltonian \textit{in terms of} instantaneous eigenmodes (equivalently denoted as normal modes). 
Since the first part of $\HIk{k}$ \eqref{hiklong} is already diagonal in the \textit{time-independent} $\opd[k]{\xi}$-modes, all we need to do is to reconstruct the second remaining part. This is achived by stating an operator which has the correct commutation relation with $\opd[k]{\xi}$ \eqref{dcomd}, as well as with $\hopd[k]{\xi}$, $\opd[-k]{\xi}$ and $\hopd[-k]{\xi}$.
This is done in general up to a c-number term which we choose to be zero, since it leads only to a global phase in the dynamics.

For {\it fixed} $\ver{k}$ the following Hamiltonian satiafies the required conditions for $\opd[k]{\xi}$, $\hopd[k]{\xi}$, $\opd[-k]{\xi}$ and $\hopd[-k]{\xi}$:

\begin{align}\label{result}
\HIk{k}={}
&\sum_{\zeta}			
\ome[k]{\zeta} \rka \hopd[k]{\zeta}\, \opd[k]{\zeta} + \hopd[-k]{\zeta}\, \opd[-k]{\zeta} \rkz \nn\\
&-\rmi\,\dot{\DeltaF}\sum_{\zeta,\xi'}
G_{\ver{k},\zeta,\xi'} \rka \hopd[k]{\zeta}\, \opd[k]{\xi'} + \hopd[-k]{\zeta}\, \opd[-k]{\xi'} \rkz \nn\\
&+\rmi\,\dot{\DeltaF}\sum_{\zeta,\xi'}
\rka \I_{\ver{k},\zeta,\xi'}\, \hopd[k]{\zeta}\, \hopd[-k]{\xi'} - \hI_{\ver{k},\zeta,\xi'}\, \opd[k]{\zeta}\, \opd[-k]{\xi'} \rkz 		\!.
\end{align}
This is checked by inspection using the properties \eqref{G} and \eqref{I} of $G_{\ver{k},\xi,\xi'}$ and $I_{\ver{k},\xi,\xi'}$ as well as the 
$\ver{k} \leftrightarrow \ver{-k}$ symmetry
of all c-number coefficients in \eqref{result}. Furthermore, the latter properties ensure that the Hamiltonian is Hermitian (assuming the frequencies are real) and satisfies ${\HIk{k}=\HIk{-k}}$. 
The full Hamiltonian in terms of instantaneous eigenmodes reads then
\begin{align}\label{t2}
\HI=\sum_{\substack{\ve{k} \neq \ve{0} \\ k_3 > 0}}
\HIk{k}
\end{align}
where the summands $\HIk{k}$ can be used to factorize the dynamics by means of the commutation relation 
\begin{align}
\eka \HIk{k}(t)\; \!, \HIk{k'}(t')\ekz=0 \quad\text{for}\quad \ver{k'}\neq\ver{\pm k}	\;.
\end{align}
The restriction to $k_3>0$ (assuming ${V\to\infty}$) in \eqref{t2} avoids a double counting in $\ver{k}$.

Let us consider the post-adiabatic effects in $\HIk{k}$ \eqref{result} created 
by the time-dependent functions $G_{\ver{k},\zeta,\xi'}$ and $\I_{\ver{k},\zeta,\xi'}$.
The effect of the diagonal elements of $G_{\ver{k},\zeta,\xi'}$, $G_{\ver{k},\zeta,\zeta}$ is to cause corrections to the instantaneous eigenfrequencies.
More interestingly, the terms in $\HIk{k}$ \eqref{result} containing the couplings $\I_{\ver{k},\zeta,\xi'}$ and $\hI_{\ver{k},\zeta,\xi'}$ create and destroy pairs of quasiparticles.
This is an example of quantum field theoretical pair production in a time-dependent background, such as typically occurs in cosmology.

The Hamiltonians $\HIk{k}$ \eqref{result} and $\HI$ \eqref{t2} as well as the definitions of their coefficients are the main result of this paper. We will focus on the low energy excitations of this system in a subsequent publication \cite{Brei14}. 
\subsection{Numerical results}
For the present we show some numerical results illustrating the qualitative character of $I_{\ver{k},a,a}$ over the course of a crossover sweep.

\begin{figure}[h]
\includegraphics[width=\linewidth]{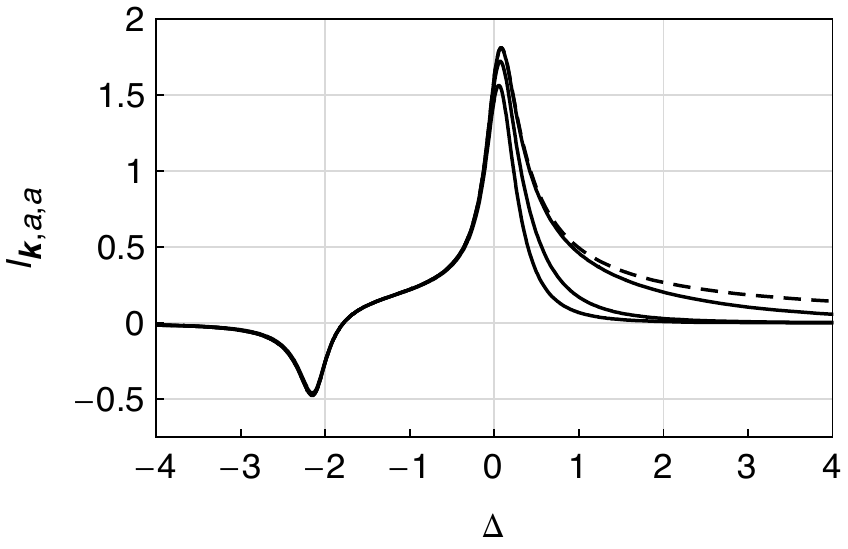}
\caption{\label{d4}Matrix element $I_{\ver{k},a,a}$ versus $\DeltaF$ for different $\abs{k}$ 
(${10^{3}\,\abs{k}=2}$, $10$, $18$, solid lines with decreasing peak size and limit ${\abs{k} \rightarrow 0}$, dashed line). 
Characteristic peaks appear when the first atoms become resonant for ${\DeltaF \lesssim -2}$ and when the last atoms leave the resonance for ${\DeltaF \approx 0}$. The parameters are $\gammaF=0.1$ and $\gF=0.2$.}
\end{figure}
As Figure \ref{d4} shows, there are two regions where the matrix element $I_{\ver{k},a,a}$ shows a characteristic behavior for the given parameters. One characteristic peak appears when the first atoms become resonant around ${\DeltaF \lesssim -2}$ and the other characteristic peak appears when the last atoms leave the resonance around ${\DeltaF \approx 0}$. Between these peaks, the matrix element crosses zero, meaning that there is one instance in time without molecular pair production for any slow crossing speed.

Figure \ref{d4} also shows the $\ver{k}$-dependence of the matrix element $I_{\ver{k},a,a}$ for small $\abs{k}$. It is important to notice that the matrix element considered is bounded even in the limit ${\abs{k} \rightarrow 0}$. The peak around ${\DeltaF \lesssim -2}$ is almost constant in the considered $\ver{k}$-regime, whereas the peak at ${\DeltaF \approx 0}$ is lowered for increasing $\abs{k}$, especially for ${\DeltaF \gtrsim 0}$, \ie mainly on the BEC side. Consequently, the latter peak becomes sharper with increasing $\abs{k}$.

In summary, Figure \ref{d4} immediately implies that the non-adiabatic quasiparticle production can be expected to occur most strongly at the beginning and at the end of the Fermi sphere's sweep through the Feshbach resonance. 

\begin{figure}[h]
\includegraphics[width=\linewidth]{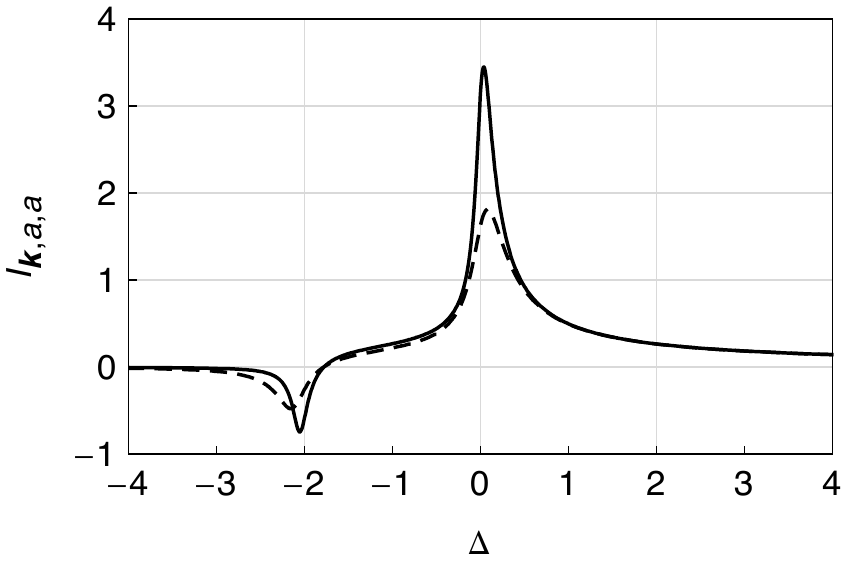}
\caption{\label{d5}Matrix element $I_{\ver{k},a,a}$ versus $\DeltaF$ in the limit ${\abs{k} \to 0}$ for $\gF=0.2$ and different $\gammaF$. The solid, the dashed and the dotted line show ${\gammaF=0.05}$, $0.1$ and $0.2$. Characteristic peaks appear when the first atoms become resonant for ${\DeltaF \lesssim -2}$ and when the last atoms leave the resonance for ${\DeltaF\approx 0}$.}
\end{figure}
Figure \ref{d5} shows the matrix element $I_{\ver{k},a,a}$ in the limit ${\abs{k} \to 0}$ for different values of $\gammaF$. With decreasing $\gammaF$, the peaks at ${\DeltaF \lesssim -2}$ and ${\DeltaF \approx 0}$ become both continuously amplified, whereas their maxima are shifted continuously. For increasing $\gammaF$, the peak at ${\DeltaF \lesssim -2}$ is shifted to smaller $\DeltaF$-values while the peak at ${\DeltaF \approx 0}$ is shifted to higher \mbox{$\DeltaF$-values}.

\begin{figure}[h]
\includegraphics[width=\linewidth]{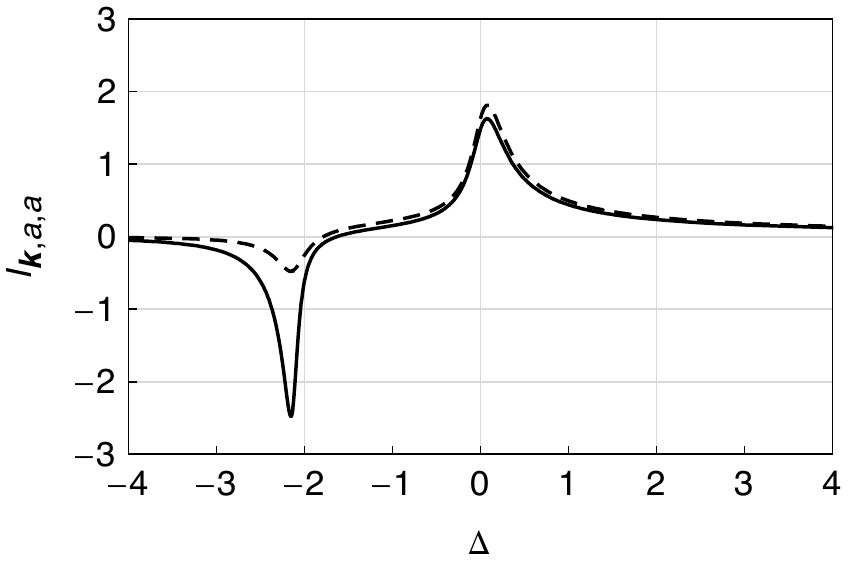}
\caption{\label{d6}Matrix element $I_{\ver{k},a,a}$ versus $\DeltaF$ in the limit $\abs{k} \to 0$ for $\gammaF=0.1$ and different $\gF$. The solid, the dashed and the dotted line show ${\gF=0.1}$, $0.2$ and $0.4$. Characteristic peaks appear when the first atoms become resonant for ${\DeltaF \lesssim -2}$ and when the last atoms leave the resonance for ${\DeltaF \approx 0}$.}
\end{figure}
In contrast, Figure \ref{d6} shows the matrix element $I_{\ver{k},a,a}$ in the limit ${\abs{k} \to 0}$ for different values of $\gF$. Here, the situation is different from the $\gammaF$-behavior in Figure \ref{d5}. The peak around ${\DeltaF \lesssim -2}$ increases with decreasing $\gF$. Its minimum gets shifted to the right for increasing $\gF$. With increasing $\gF$, the maximum around ${\DeltaF \approx 0}$ is also slightly shifted to the right while the peak size increases.

\section{Discussion}\label{discussion}
In this paper, we have addressed the post-adiabatic dynamics of a two-channel model, describing the transition from a BCS ground state of fermionic atoms through a Feshbach resonance into a possibly excited state of bosonic molecules.
\subsection{Summary of results}
Throughout this paper, we have mimicked the effect of background scattering by the adiabatic elimination of off-resonant bound states. In comparison with the standard Hubbard–Stratonovich approach, this offers an especially physical interpretation of the mathematical simplifications.

The basic intermediate result continuously used in the calculations is the classical path for ${\ver{k}=\ver{0}}$ molecular modes as derived in Appendix \ref{classical} and shown in Figure \nolinebreak[4] \ref{classicplot}. It serves as a background field within all our calculations. 

In the course of this paper we have introduced virtual bosonic modes in order to mimic the effects derived before by a dilute gas approximation. The common idea of both descriptions is that the system can be interpreted as consisting of two subsystems. The order of solving them, one after another, or both at once, is the key step in simplifying the systems description.  
We have finally mimicked the leading effects of the subsystem of fermionic atoms by means of a subsystem of virtual bosons, while the subsystem of molecular bosons is treated in terms of a coherent state path integral. Instead of solving the dynamics of one subsystem within the path integral, we left the path integral in order to solve the dynamics of both subsystems simultaneously. This led us to a purely bosonic, quadratic Hamiltonian, which should be an excellent approximation to the original problem in the parameter range we are looking at. The latter Hamiltonian $\HBk{k}$ \eqref{dummy} is an important intermediate result of this paper since all our following main results are based on it.
\subsubsection{Instantaneous diagonalization}
As a main result, we have diagonalized the quadratic Hamiltonian $\HBk{k}$ \eqref{dummy} instantaneously and thus obtained equations for the spectrum of instantaneous eigenfrequencies $\ome[k]{\zeta}$ as well as for the instantaneous mode functions $\mfu[k]{\xi,\eta}$ and $\mfv[k]{\xi,\eta}$. 
The equations obtained can be used for a numerical analysis as shown in the Figures \ref{d1}, \ref{d2} and \ref{d3}. Thus, we have described the instantaneous (An\-der\-\mbox{son-)Bo}\-gol\-iu\-bov excitations completely. 
\subsubsection{The post-adiabatic Hamiltonian}\label{discpostadi}
The most important result of this paper is the post-adiabatic Hamiltonian $\HIk{k}$ \eqref{result}. It describes the Hamiltonian $\HBk{k}$ \eqref{dummy} in terms of its instantaneous, time-dependent normal modes. As an effect of the time dependence, there appear various coupling terms, quadratic in the normal mode's creation and annihilation operators with prefactors $G_{\ver{k},\zeta,\xi'}$, $\I_{\ver{k},\zeta,\xi'}$ and $\hI_{\ver{k},\zeta,\xi'}$. These prefactors which introduce new couplings among the modes are the main post-adiabatic effect. The different prefactors can be obtained numerically as the examples in Figures \nolinebreak[4] \ref{d4}, \ref{d5} and \ref{d6} show for various, 
significant
cases.
\subsubsection{Validity of results}
The main restriction on the validity of our result is the constraint on $\nuF$. The validity of almost all other assumptions are immediate self-consistent consequences of the latter one: the smaller $\nuF$ is, the fewer modes are expected to become excited. Thus, the total depletion out of the system's adiabatic ground state is expected to decrease. As the total depletion decreases, the quality of the dilute gas approximation (DGA) is expected to increase. Furthermore, a small depletion justifies also the neglect of back-reaction on the classical background.

There is another restriction on $\nu$, originating from the derivation of the underlying classical background: the adiabaticity condition for the atomic two-level systems $\nu$ as well as the slow time-dependence of the molecular classical path are both self-consistent. Both hold as long as $\nu$ is small compared to the energy gaps of all atomic two-level systems (\ie the BCS gap on the BCS side). This condition ensures that we end up in a state without remaining 
fermionic
excitations. 

Small sweep rates $\nuF^2$ also suggest low energy modes to be involved. Focussing on the dynamics of low energy normal modes located at the lower end within the fermionic gap, we can neglect higher order adiabatic corrections to the time evolution of the fermions in the molecular background.

Apart from the restrictions on $\nuF$, the classical approximation for the ${\ver{k}=\ver{0}}$ molecular modes has to hold. 
This is ensured by the choice of a small $\gammaF$ (see \cite{Dieh06,Dieh07}): as $\gammaF$ decreases, the coupling between the fermionic and the bosonic subsystem decreases and the eigenstates of the whole system approach product states out of either Hilbert space more closely. Since the classical approximation approximates the system's ground state by a product of an (unnormalized) coherent state and a fermionic squeezed state it improves as $\gammaF$ decreases. 
Note that according to its definition, $\gammaF$ decreases with increasing $\Nm$.

Several assumptions have been made in the derivation of the classical path which turn out to be fulfilled self-consistently. Despite this, these assumptions might in principle have ruled out some possible further classical paths. However, we consider this extremely unlikely, since the obtained classical path shows the expected bahaviour.      
\subsection{Relation to experiments}
In comparison with real experiments, our results have surely to be adapted in so far as all experiments are conducted in traps, whereas we deal with a potential-free Hamiltonian. However, potentials with nice momentum space representations might be tractable within our calculations as well. Numerical results which can be compared with experimental results at least qualitatively will be presented in a future publication \cite{Brei14}.
\subsection{Squeezing and quasiparticle production}
The basic implication of our analysis is that the leading post-adiabatic effect of
a finitely slow BCS-BEC crossover, through a narrow Feshbach resonance, is the excitation of
opposite-momenta squeezed states of the low-frequency collective modes of the molecular quantum gas. This is a typical form of non-adiabatic excitation in systems described by quantum fields in time-dependent backgrounds. It occurs, for example, in cosmology, where excitations of long-wavelength modes in cosmic fields during expansion of the universe have been considered as sources of non-uniformity in the spatial distribution of energy and matter.

As in cosmology, one may expect that subsequent nonlinear interactions will drive the non-adiabatically excited squeezed state into a more typical thermal distribution. In this sense the coherent non-adiabatic excitations are a source of thermal fluctuations, and represent an essential initial stage of entropy production. Our results in this paper can thus be considered as an analysis of the first roots of entropy in chemical reactions, as they appear in molecule production under the most tightly controlled conditions of full quantum coherence.

\appendix{
\section{Adiabatic, classical molecular and atomic background solution}
\label{classical}
Our main strategy in dealing with the path integral is to consider a slow but possibly large classical background and small but possibly fast perturbations around it. For small enough sweep rates, the depletion into other modes will be small in general, and we can restrict completely on the ${\ver{k}=\ver{0}}$ classical molecular background, which we treat non-perturbatively. For the ${\ver{k}=\ver{0}}$ molecular modes the decomposition into classical path $\alc$, $\betc$ and fluctuations ${\delta \al[0]}$, ${\delta \bet[0]}$ is done formally by writing ${\al[0]=\alc+\delta \al[0]}$ with ${\delta \al[0] \mydoteq \al[0]-\alc}$ and ${\bet[0]=\betc+\delta \bet[0]}$ with ${\delta \bet[0] \mydoteq \bet[0]-\betc}$.

Consequently, we consider in this appendix the system \nolinebreak \eqref{Hofk} in the case where only ${\ver{k}=\ver{0}}$ molecular modes are present as discussed in Section \ref{HTLProc}. This immediately implies ${\hat H_{AT}=\hat H_{TL}}$, meaning that the atomic Hamiltonian consists out of many fermionic two-level systems. We derive the adiabatic classical background solution, \ie the classical path $\alc$, $\halc$ and $\betc$, $\hbetc$ for ${\ver{k}=\ver{0}}$ molecules and small $\dot{\DeltaF}$. Simultaneously, we derive the adiabatic classical background solution for atoms $ \ket{F(t)}$, $\bra{\bar F(t)}$, \ie the evolution of the fermionic atoms under $\hat H_{TL,cl}$ ($\hat H_{TL}$ on the classical path). 

As for all classical path calculations, we need to extremize the molecular action by linearizing it for small fluctuations and setting the first variation to zero. 
To this end, we formally set up the general classical atomic problem for the time evolution under $\hat H_{TL,cl}$. 
Subsequently, we can construct the first variation of the system's action, set it to zero and get the molecular equation of motion for the molecular classical path. Together with the time-dependent Schrödinger equation for the atomic evolution under $\hat H_{TL,cl}$, the equations at hand are a sufficient set to determine the classical molecular and atomic background solution. Their solution is found by an appropriate ansatz for the adiabatic limit, \ie for small $\dot{\DeltaF}$ or, strictly speaking, for ${\dot{\DeltaF}\rightarrow0}$. 

Note that making the action stationary means in general to go into the complex plane with both, the real and imaginary parts of the path integral variables. Consequently, the bar \,$\bar{}$\, does in general not mean complex conjugation any more but denotes independent functions. 
However, we will only consider classical path for which the bar in fact means complex conjugation, since we expect the latter to give the most important contribution to the path integral.
\subsection{The classically driven fermionic problem $\hat H_{TL,cl}$}
As described in Section \ref{CSHam} and Section \ref{HTLProc}, the Hamiltonian $\hat H_{TL}$ \eqref{HTL} 
\begin{align}
\hat H_{TL}&\mydoteq\sum_{\ve{q}} \hat H_{\ve{q}}
\end{align}
can be decomposed into a sum over $\hat H_{\ve{q}}$ \eqref{HFQ}
\begin{align}
\hat H_{\ve{q}}&\mydoteq
\rka \ve{q}^2 + \frac{\DeltaF}{2}-\frac{\om}{2}\rkz
\rka \hcp[q]\cp[q]+\hcm[-q]\cm[-q]\rkz
\nn\\
&\quad+ \frac{\gammaF}{\sqrt{\Nm}}
\rka \cp[q]\cm[-q]\rka\hal[0]+\eps^{-1}\hbet[0] \rkz  +\rm{H. c.}\rkz		\,.
\end{align}
The latter can be used in order to factorize the dynamics created by $\hat H_{TL}$ since they fulfill the commutation relation \eqref{HFQcom}. The time-dependent Schrödinger equation for the whole atomic state $\ket{F(t)}$ \eqref{F1} on the classical path,
\begin{align}
\rmi \,\frac{\rmd}{\rmd t}\, \ket{F(t)}=\hat H_{TL,cl}\, \ket{F(t)}		\;,
\end{align}
can therefore also be separated into time-dependent Schrö\-din\-ger equations for the two-level systems  
\begin{align}\label{HQCL}
\rmi \,\frac{\rmd}{\rmd t}
\left(\begin{array}{c}
\fy[q] \vspace{0.2em}\\
\fz[q]
\end{array}\right)
=
\mathcal{H}_{\ver{q},cl}
\left(\begin{array}{c}
\fy[q] \vspace{0.2em}\\
\fz[q] 
\end{array}\right)
\end{align}
where the atomic Hamiltonian matrix $\mathcal{H}_{\ver{q},cl}$ is given by 
\begin{align}\label{HQCLmatrix}
\mathcal{H}_{\ver{q},cl}
=
\left(\begin{array}{cc}
2\, \ve{q}^2 + \DeltaF - \om &
\frac{\gammaF}{\sqrt{\Nm}} \rka\alc+\eps^{-1}\betc \rkz \vspace{0.2em}\\
\frac{\gammaF}{\sqrt{\Nm}} \rka\halc+\eps^{-1}\hbetc \rkz& 
0
\end{array}\right)		\;.
\end{align}
These two-level systems have to fulfill the initial conditions stated in Section \ref{HTLProc}. 
The initial conditions as well as $\mathcal{H}_{\ver{q},cl}$ only depend on $\ver{q}$ by means of $\abs{q}$, as anticipated by our notation. 
\subsection{The general atomic problem $\hat H_{TL}$ and the first variation}

While fully performing the path integral is impossible, since the atomic problem cannot be solved explicitly for arbitrary molecular fields, it is possible to construct the saddlepoint approximation to the path integral, by considering linear fluctuations around a sufficiently slow classical molecular path.
We \textit{could} compute the classical action explicitly, but do not, since it is not necessary for quantities computed within this paper.

The Hamiltonian $\hat H_{TL}$ can be rewritten as ${\hat H_{TL,cl}+\delta\hat H_{TL}}$, where the term ${\delta\hat H_{TL}\mydoteq \hat H_{TL}-\hat H_{TL,cl}}$ is linear in the deviations from the molecular classical path ${\delta \al[0]}$, ${\delta \bet[0]}$ and its barred counterparts. The Hamiltonian $\hat H_{TL,cl}$ is independent of the latter deviations.    
We calculate the first variation of the part of the molecular action which is caused by fermionic time evolution. This reads
\begin{align}\label{vari}
\delta \ln(
\bra{\bar F_f}
\hat U_{TL}
\ket{F_i})
=\frac
{\delta
\bra{\bar F_f}
\hat U_{TL}
\ket{F_i}}
{\bra{\bar F_f}
\hat U_{TL,cl}
\ket{F_i}} 
=\frac
{\bra{\bar F_f}
\delta \hat U_{TL}
\ket{F_i}}
{\mathcal{F}}
\end{align}
where we have used the definition
\begin{align}
\mathcal{F}\mydoteq\braket{\bar F_f}{F(t_f)}
=\bra{\bar F_f} \hat U_{TL,cl} \ket{F_i}		\;.
\end{align}
We assume ${\abs{\mathcal{F}}\equiv 1}$, meaning ${\ket{F_f}}$ is proportional to ${\hat U_{TL,cl}\ket{F_i}}$ up to a phase factor. Since we chose ${\bra{\bar F_f}}$ as the state without remaining fermions, any fermionic dynamics along a slowly growing molecular path 
will fulfill this condition. Using this fact, we can simplify the above equation \eqref{vari} by means of
\begin{align}\label{dgaclassical}
\bra{\bar F_f}
\delta\hat U_{TL}
\ket{F_i} 
\equiv 
-\,\rmi\,\mathcal{F}\int_{t_i}^{t_f}\rmd t\,
\bra{ \bar F(t)} 
\delta\hat H_{TL}(t)
\ket{F(t)}			\;.
\end{align}
The matrix element in the variation \eqref{dgaclassical} is given by
\begin{align}
\bra{ \bar F(t)}&		\delta\hat H_{TL}(t)		\ket{F(t)}\nn\\
={}&\sum_{\ve{q}}\frac{\gammaF}{\sqrt{\Nm}}
\Bigl[ \hfy[q] \fz[q] \rka \delta \al[0]+\eps^{-1}\delta \bet[0] \rkz\nn\\
&\quad\,\,\;+ \fy[q] \hfz[q] \rka\delta \hal[0]+\eps^{-1}\delta \hbet[0] \rkz\Bigr]		\;.
\end{align}
By linearizing the other terms of the action as well around $\alc$, $\halc$, $\betc$, $\hbetc$ and by setting the first variation to zero, we obtain the equations of motion for the molecular classical path as
\begin{align}
\label{emotionalc}
\frac{\rmd}{\rmd t}\,\alc 
&=\rmi \,\om\,  \alc
-\rmi \,\sum_{\ve{q}}\frac{\gammaF}{\sqrt{\Nm}}\,\fy[q] \hfz[q]		\;\;,\\
\label{emotionbetc}
\frac{\rmd}{\rmd t}\,\betc 
&=\rmi \,\om\,  \betc -\rmi \,\frac{\gammaF^2}{\gF}\,\epsw^{-2}\betc
-\rmi \,\eps^{-1}\sum_{\ve{q}}\frac{\gammaF}{\sqrt{\Nm}}\,\fy[q] \hfz[q] 
\end{align}
and
\begin{align}
\label{emotionhalc}
\frac{\rmd}{\rmd t}\,\halc 
&=-\,\rmi \,\om\,  \halc
+\rmi \,\sum_{\ve{q}}\frac{\gammaF}{\sqrt{\Nm}}\,\hfy[q] \fz[q]		\;\;,\\
\label{emotionhbetc}
\frac{\rmd}{\rmd t}\,\hbetc 
&=-\,\rmi \,\om\,  \hbetc + \rmi\, \frac{\gammaF^2}{\gF}\,\epsw^{-2}\hbetc
+\rmi \,\eps^{-1}\sum_{\ve{q}}\frac{\gammaF}{\sqrt{\Nm}}\,\hfy[q] \fz[q] 
\end{align}
for the corresponding barred quantities. As discussed in Section \ref{boundary}, the boundary conditions to the above differential equations are ${\alc(t_i)}=0$, ${\betc(t_i)=0}$ and ${\halc(t_f)=\sqrt{N_m}\,\rme^{\rmi\,\varphi_f}}$, ${\hbetc(t_f)=0}$. 
Together with the differential equations \eqref{HQCL}, their barred versions and the initial conditions for both (see Section \ref{HPTdlg}), these differential equations determine the classical background. 

Since all the equations of motion have the structure of complex conjugate pairs, they lead to complex conjugate $\alc$, $\halc$, as soon as the boundary conditions allow this. 
We assume this and will find self-consistently that the assumption is true.

In the following, the equations \eqref{emotionalc}, \eqref{emotionbetc}, \eqref{emotionhalc} and \eqref{emotionhbetc} will be mapped onto another set of four equations by a superposition. The new equations are equivalent to the old ones as long as the determinant of the linear mapping is not zero. This condition is fulfilled as long as $\Neff\neq0$ holds ($\Neff$ as defined below in \eqref{Neff}). 
\subsection{The molecular equations of motion}
\subsubsection{Number conservation}
The first new equation 
\begin{align}
	&\frac{\rmd}{\rmd t} \rka \halc \, \alc + \hbetc \, \betc \rkz 
	\nn\\ 
={}	&\rmi \,\sum_{\ve{q}}\frac{\gammaF}{\sqrt{\Nm}}\,\hfy[q] \fz[q] \rka\alc+\eps^{-1}\betc \rkz 
\nn\\
	&-\rmi \,\sum_{\ve{q}}\frac{\gammaF}{\sqrt{\Nm}}\,\fy[q] \hfz[q] \rka\halc+\eps^{-1}\hbetc \rkz 
\end{align}
we find is a first integral of motion and in general exact. We rewrite it with the help of the fermionic equations of motion \eqref{HQCL} as
\begin{align}\label{1v4}
\frac{\rmd}{\rmd t}
\rka \halc \, \alc + \hbetc \, \betc +\sum_{\ve{q}}\, \hfy[q] \fy[q]\rkz 
 =0		\;.
\end{align}
Integration of this equation leads to
\begin{align}\label{1v4int}
\rka \halc \, \alc + \hbetc \, \betc +\sum_{\ve{q}}\, \hfy[q] \fy[q]\rkz = \Nm		\;.
\end{align}
This is nothing but the conservation of the number of bosons plus twice the number of fermions.
We determine the integration constant from the fact that as ${t \to t_i}$ and ${\eps\to0}$, the number of bosons tends to zero whereas the expectation value for the number of atom pairs tends to $\Nm$ in the pure BCS state. On the other hand, as ${t \to t_f}$, the left hand side of the equation tends to ${\halc \, \alc=N_m}$, the mean field value at $t_f$. 
\subsubsection{Choice of the free parameter $\om$}
The second new equation reads
\begin{align}\label{4v4}
&\frac{2\,\eps^2\sqrt{\neff}}{1+\eps^2}\,(\om+\dot{\theta})
-\sum_{\ve{q}}\frac{\gammaF}{\Nm}\rka\rme^{-\,\rmi\,\theta}\,\hfy[q]\fz[q]+\rme^{\rmi\,\theta}\,\fy[q]\hfz[q]\rkz\nn\\
&\quad-\frac{\gammaF^2\sqrt{\neff}}{g\,(1+\eps^2)}
\rka 1-\frac{\halc\alc}{\Neff}+\frac{\hbetc\betc}{\eps^2\,\Neff} \rkz=0
\end{align}
where $\theta$ is the phase of ${\halc+\eps^{-1}\hbetc}$ and with
\begin{align}
\label{Neff}
\Neff&\mydoteq \rka \halc+\eps^{-1}\hbetc \rkz \rka \alc+\eps^{-1}\betc \rkz		\;,\\
\label{neff}
\neff&\mydoteq \frac{\Neff}{\Nm}
\end{align}
where $\Nm$ is the exact constant of motion. Next we examine the evolution of the variable $\theta$ which is the phase corresponding to this conserved number.
Our basic Hamiltonian \eqref{Hofk} involves the arbitrary parameter $\om \in \mathbbm{R}$. 
We will use this freedom and make a choice which we expect to simplify our calculations.
Looking at the equations of motion for molecules \eqref{emotionalc}, \eqref{emotionbetc}, \eqref{emotionhalc}, \eqref{emotionhbetc} and for fermionic atoms \eqref{HQCL}, \eqref{HQCLmatrix}, we realize that the solution for $\om \not\equiv 0$ can be obtained out of the solution for $\om \equiv 0$ by the transformation
\begin{align}
(\alc,\betc,\fy[q])&=(\alc,\betc,\fy[q])\big\vert_{\om\, \equiv\, 0}\,\rme^{\rmi \int_{t_i}^t \om \;\rmd\tau}	\;\,,\\
(\halc,\hbetc,\hfy[q])&=(\halc,\hbetc,\hfy[q])\big\vert_{\om\, \equiv\, 0}\,\rme^{-\,\rmi \int_{t_i}^t \om \;\rmd\tau}		\;\,,\\
\fz[q]=\fz[q]\big\vert_{\om\, \equiv\, 0}
\quad&\text{and}\quad
\hfz[q]=\hfz[q]\big\vert_{\om\, \equiv\, 0}		\;\,.
\end{align}
As this transformation shows, we can and will choose $\om$ such that the phase $\theta$ of ${\halc+\eps^{-1}\hbetc}$ is a constant, simplifying equation \eqref{4v4} a lot. 
The motivation for this choice is the following: since we would like to perform an adiabatic approximation to the system \eqref{HQCL} in the following Section \nolinebreak[4] \ref{adiabatic}, we would prefer for this purpose an ${\alc+\eps^{-1}\betc}$ without phases that vary rapidly in time. More generally, we also want the \textit{complete} function ${\alc+\eps^{-1}\betc}$ to depend only adiabatically slowly on time through $\dot{\DeltaF}$ as explained in the next section. 
At this point it is not clear whether the above choice of $\om$ will lead to the desired adiabatic behavior of the \textit{complete} function ${\alc+\eps^{-1}\betc}$ in time. This has to be checked later self-consistently.

Keep in mind that the choice of the in principle arbitrary ${\om \in \mathbbm{R}}$ is only determined by our goal of simplifying the adiabatic approximation for the atomic two-level system \eqref{HQCL} and is not a further assumption or approximation.
\subsubsection{Adiabatic dragging of $\betc$}
The third and forth new equation are
\begin{align}
\label{2v4}
\frac{\rmd}{\rmd t}\,\betc-\rmi\rka\om-\frac{\gammaF^2}{\gF}\,\epsw^{-2}\rkz\betc 
&=\eps^{-1}\rka\frac{\rmd}{\rmd t}\, \alc - \rmi \, \om \, \alc \rkz
\end{align}
and its complex conjugate. This equation relates $\alc$ and $\betc$ and can be formally solved for $\betc$ as
\begin{align}
\label{2v4int}
\betc&=
\int_{t_i}^t\rme^{\rmi \int^t_\tau 
(\om\,-\,\gammaF^2 \gF^{-1}\epsw^{-2})
\,\rmd \tau'}
\eps^{-1}\rka\frac{\rmd}{\rmd \tau}\,\alc - \rmi \, \om \, \alc \rkz\rmd \tau		\;.
\end{align}
However this does not help very much in solving all equations of motion. Further assumptios are necessary before we proceed:  
As long as the sweep rate $\dot{\DeltaF}$ is small enough, we expect the Hamiltonian \eqref{HQCLmatrix} to depend only adiabatically slow on time by means of $\DeltaF$. 
From now on, we assume that $\alc$ and $\betc$ as well as $\om$ and consequently $\mathcal{H}_{\ver{q},cl}$ \eqref{HQCLmatrix} are a power series in terms of ${\dot{\DeltaF}=\nuF^2}$ the coefficients of which depend on time \textit{only} via ${\DeltaF=\nuF^2 t}$. We will determine these quantities up to corrections of the order $\mathcal{O}(\dot{\DeltaF}^2)$.
The advantage of this notation is that we can make clear statements about the scaling for asymptotically small $\dot{\DeltaF}$ (\ie $\nuF^2$) independent of the size of $\DeltaF$ (\ie ${\nuF^2 t}$).

Equation \eqref{2v4int} can then be adiabatically approximated by an asymptotic expansion for ${\dot{\DeltaF}\to 0}$ (\ie ${\nuF^2\to0}$). This expansion strategy is called {\it integration by parts} (see \eg \cite{Bend78}) which is systematic in orders of $\dot{\DeltaF}$ (\ie $\nuF^2$). The smaller $\eps$ is, the better it works, since this will increase the frequency in the integrand more and more. The result is
\begin{align}\label{betcadi}
\betc=&-\frac{\gF\, \epsw\, \om}{\gammaF^2-\gF\, \epsw^2\, \om}\, \alc 
-\rmi\, \frac{\gF\, \epsw\, \gammaF^2}{\rka\gammaF^2-\gF\, \epsw^2\, \om\rkz^2}\,\alcdot\nn\\
&-\rmi\, \frac{\gF^2\,\epsw^3\,\gammaF^2}{\rka\gammaF^2-\gF\, \epsw^2\, \om\rkz^3}\,\dot{\om}\,\alc+\mathcal{O}(\dot{\DeltaF}^2)		\;.
\end{align}
This equation by construction solves \eqref{2v4} up to terms of the kind $\mathcal{O}(\dot{\DeltaF}^2)$ or $\nuF^4$.
Obviously, the assumption made before that $\betc$ is a power series in $\dot{\DeltaF}$ is an immediate consequence of $\alc$ having this property and is not an extra condition. 

Using the ansatz
\begin{align}\label{alcadi}
\alc&=\sqrt{\Nm} \, r\, \rme^{-\, \rmi\, \varphi}
\end{align}
we can now calculate the leading orders of the phase $\theta$ of ${\halc+\eps^{-1}\hbetc}$, which has to be constant by construction
\begin{align}\label{thetadef}
\theta=\rmi \ln \frac{\alc+\eps^{-1}\betc}{\sqrt{\Neff}}\equiv const		\;.
\end{align}
Together with the assumption (see also \eqref{omegacond} and following discussion)
\begin{align}
\om<\frac{\gammaF^2}{\gF\rka 1 + \epsw^2\rkz} 
\end{align}
this leads us immediately to ${\theta\equiv\varphi_f}$ and to an equation for $\varphi$
\begin{align}
\varphi={}&\varphi_f
-\frac{\gF\,\gammaF^2\,\dot{r}}{r\rka\gammaF^2-\gF\, \epsw^2\, \om\rkz \rka\gammaF^2-\gF\rka 1 + \epsw^2\rkz \om\rkz}
\nn\\
&
-\frac{\gF^2\,\gammaF^2\,\eps^2\,\dot{\om}}{\rka\gammaF^2-\gF\, \epsw^2\, \om\rkz^2 \rka\gammaF^2-\gF\rka 1 + \epsw^2\rkz \om\rkz} 
+\mathcal{O}(\dot{\DeltaF}^2)		\;.
\end{align}
Furthermore, we are now able to rewrite $\Neff$ as
\begin{align}\label{naqa}
\Neff={}&\frac{\rka\gammaF^2-\gF\rka 1 + \epsw^2\rkz \om\rkz^2}{\rka\gammaF^2-\gF\, \epsw^2\, \om\rkz^2}\,\halc\alc\nn\\
&-\frac{2\, \gF\, \gammaF^2\rka\gammaF^2-\gF\rka 1 + \epsw^2\rkz \om\rkz}{\rka\gammaF^2-\gF\, \epsw^2\, \om\rkz^3}\,\halc\alc\, \dot{\theta} +\mathcal{O}(\dot{\DeltaF}^2)
\end{align}
and $\hbetc\betc$ as
\begin{align}\label{bqbaqa}
\hbetc\betc={}&\frac{\gF^2\, \epsw^2\, \om^2}{\rka\gammaF^2-\gF\, \epsw^2\, \om\rkz^2}\,\halc\alc\nn\\
&+\frac{2\,\gF^2\,\gammaF^2\,\epsw^2\,\om}{\rka\gammaF^2-\gF\, \epsw^2\, \om\rkz^3}\,\halc\alc\,\dot{\theta}+\mathcal{O}(\dot{\DeltaF}^2)
\end{align}
where in both equations \eqref{naqa} and \eqref{bqbaqa}
\begin{align}
\dot{\theta}={}&\frac{\rmi}{2}\rka\frac{\alcdot+\eps^{-1}\betcdot}{\alc+\eps^{-1}\betc}-\frac{\halcdot+\eps^{-1}\hbetcdot}{\halc+\eps^{-1}\hbetc}\rkz \equiv 0
\end{align}
has to be zero by construction.
\subsection{The fermionic equations of motion}\label{adiabatic}
We need to find a solution which satisfies the differential equations for $\fy[q]$, $\fz[q]$, $\alc$ and $\betc$ simultaneously. This is a priori a difficult task, but can be simplified in general by seeking for the solutions $\fy[q]$ and $\fz[q]$ under the assumption that $\alc$, $\betc$, $\om$ and thus $\mathcal{H}_{\ver{q},cl}$ \eqref{HQCLmatrix} depend on time in the special adiabatic way discussed before: These quantities are power series in terms of ${\dot{\DeltaF}=\nu^2}$ with coefficients that depend on time only through ${\DeltaF=\nuF^2t}$. Thus, a general adiabatic asymptotic series reads
\begin{align}
f(\DeltaF)=f(\nuF^2t)=\sum_{n=0}^{\infty}\nuF^{2n}f_{n}(\nuF^2 t)=\sum_{n=0}^{\infty} \dot{\DeltaF}^n f_{n}(\DeltaF)		\;.
\end{align}
Consequently, we can solve the fermionic equations of motion adiabatically in the following.

Given the solutions for $\fy[q]$ and $\fz[q]$ we can solve for $\alc$ and $\betc$ in the following sections. Finally we need to check whether the latter molecular classical path $\alc$ and $\betc$ fulfills self-consistently the assumptions made when deriving $\fy[q]$ and $\fz[q]$. This can of course not be done to all orders but is usually done up to terms of the kind $\mathcal{O}(\dot{\DeltaF}^2)$ in our case. If the self-consistency criterion holds, we obtained the desired solution for the case of slow time dependence, \ie small $\dot{\DeltaF}$ or $\nuF^2$. As will become clear at the end of our calculations, the adiabaticity criterion indeed holds for the solution we will find. 
\subsubsection{Adiabatic atomic two-level systems}\label{adiabatic2l}
Under the assumptions made in the previous section, the adiabatic approximation holds and the general solution to the fermionic two-level problem \eqref{HQCL} can be replaced by the adiabatic one
\begin{align}\label{adsol}
\Biggl(\begin{array}{c}
\fy[q] \vspace{0.2em}\\
\fz[q] 
\end{array}\Biggr)
&\mydoteq
\Biggl[
\Biggl(\begin{array}{c}
\fY[q] \vspace{0.2em}\\
\fZ[q] 
\end{array}\Biggr)
+\rmi\,\frac{\fZ[q]\fYdot[q]-\fY[q]\fZdot[q]}{2\,\W[q]}
\Biggl(\begin{array}{c}
\hfZ[q] \vspace{0.2em}\\
-\,\hfY[q] 
\end{array}\Biggr) \nn\\ 
&\quad\;\,+\mathcal{O}(\dot{\DeltaF}^2)\Biggr]
\rme ^{-\,\rmi \int_{t_i}^t \E[q](\tau)\,\rmd \tau\,+\,\rmi\,\G[q]+\,\rmi\,\GG[q]}
\end{align}
where the instantaneous lowest energy eigenvector of the matrix \eqref{HQCLmatrix} is given by
\begin{align}\label{yzdef}
\Biggl(\begin{array}{c}
\fY[q] \vspace{0.2em}\\
\fZ[q] 
\end{array}\Biggr)
\mydoteq{}&
\frac{h_{\abs{q}}}{\sqrt{\EE[q]+\gammaF^2\, \neff}}
\Biggl(\begin{array}{c}
\rme^{-\,\rmi\,\varphi_f}\,\E[q] \vspace{0.2em}\\
\gammaF\, \sqrt{\neff}
\end{array}\Biggr)
+\mathcal{O}(\dot{\DeltaF}^2)
\nn\\={}&
\frac{h_{\abs{q}}}{\sqrt{2\,\W[q]}}
\Biggl(\begin{array}{c}
-\,\rme^{-\,\rmi\,\varphi_f}\sqrt{-\,\E[q]} \vspace{0.2em}\\
\sqrt{\E[q]+2\,\W[q]}
\end{array}\Biggr)
+\mathcal{O}(\dot{\DeltaF}^2)		\;.
\end{align}
The respective lowest instantaneous energy $\E[q]$ is given by
\begin{align}\label{Edef}
\E[q]&\mydoteq\rka \ver{q}^2+\frac{\DeltaF-{\om}}{2}\rkz-\W[q]+\mathcal{O}(\dot{\DeltaF}^2)		\;,\\
\W[q]&\mydoteq\sqrt{\rka \ver{q}^2+\frac{\DeltaF-{\om}}{2}\rkz^2+\gammaF^2\, \neff}+\mathcal{O}(\dot{\DeltaF}^2)	\;.
\end{align}
The prefactor $h_{\abs{q}}$ satisfying ${\bar{h}_{\abs{q}} h_{\abs{q}}=1}$ has been chosen as 
\begin{align}\label{hdef}
h_{\abs{q}}&=
\begin{cases}
-\,\rme^{\rmi\,\varphi_f}	&\text{for } \abs{q}\le1\\
1					&\text{for } \abs{q}>1
\end{cases}
\end{align}
such that the instantaneous eigenvector at $t=t_i$ would tend to a Fermi gas
\begin{align}
\left(\begin{array}{c}
\fY[q] \vspace{0.2em}\\
\fZ[q]
\end{array}\right)
=
\begin{cases}
\Bigl(\begin{array}{c} 1 \vspace{-0.5em}\\ 0 \end{array}\Bigr)&\text{for } |\ve{q}|\le1\\
\Bigl(\begin{array}{c} 0 \vspace{-0.5em}\\ 1 \end{array}\Bigr)&\text{for } |\ve{q}|>1
\end{cases}
\end{align}
if the interaction among the fermions would be tuned down.

Note that the choice of the lower energy branch of the fermionic two-level system will automatically lead to a self-consistent molecular and atomic BCS state solution, making it unnecessary to worry about the initial boundary conditions. 

The functions $\G[q]$, $\GG[q]$ are geometric phases starting with powers of $\dot{\DeltaF}^0$ and $\dot{\DeltaF}^1$ respectively. They explicitly read
\begin{align}
\G[q]\mydoteq{}&\rmi \int_{t_i}^t\rka\hfY[q]\fYdot[q]+\hfZ[q]\fZdot[q]\rkz\,\rmd \tau +\mathcal{O}(\dot{\DeltaF}^2)	\;,\\
\GG[q]\mydoteq{}&\int_{t_i}^t \frac{\fZ[q]\fYdot[q]-\fY[q]\fZdot[q]}{2\,\W[q]}\nn\\
			&\;\;\quad\cdot\rka\hfZ[q]\hfYdot[q]-\hfY[q]\hfZdot[q]\rkz \,\rmd \tau+\mathcal{O}(\dot{\DeltaF}^2)
\end{align}
and have the properties ${\G[q]=\hG[q]}$ and ${\GG[q]=\hGG[q]}$. The former of these identities results from the conservation of the norm ${\hfY[q]\fY[q]+\hfZ[q]\fZ[q]=1}$. Note that since we chose $\om$ such that ${\dot{\theta}\equiv 0}$, we have ${\G[q]=0}$ along the classical path. Furthermore, the knowledge of $\GG[q]$ is never necessary in the derivation of the main result of this paper and we will not compute it explicitly. Also the time integral over the energy $\E[q]$ is never needed explicitly, such that it is enough to compute $\E[q]$ up to $\mathcal{O}(\dot{\DeltaF}^2)$ instead of $\mathcal{O}(\dot{\DeltaF}^3)$. Consequently, the adiabatic solution \eqref{adsol} simplifies to
\begin{align}
\Biggl(\begin{array}{c}
\fy[q] \vspace{0.2em}\\
\fz[q] 
\end{array}\Biggr)
\mydoteq{}&
\Biggl[
\Biggl(\begin{array}{c}
\fY[q] \vspace{0.2em}\\
\fZ[q] 
\end{array}\Biggr)
+\frac{\rmi}{2\,\W[q]}
\Biggl(\begin{array}{c}
\fYdot[q] \vspace{0.2em}\\
\fZdot[q] 
\end{array}\Biggr)
+\mathcal{O}(\dot{\DeltaF}^2)\Biggr]\nn\\
&\cdot\rme ^{-\,\rmi \int_{t_i}^t \E[q](\tau)\,\rmd \tau\,+\,\mathcal{O}(\dot{\DeltaF}^1)}
\end{align}
where we have also simplified the equation using norm conservation.
\subsection{A self-consistent adiabatic solution}
Inserting the adiabatic solution $\fy[q]$ \eqref{adsol} into the integral of motion \eqref{1v4int} we find that the notation can be shortened by writing
\begin{align}\label{1von2ad}
\frac{\halc \, \alc}{\Nm} + \frac{\hbetc \, \betc}{\Nm} - \frac{\partial f(\om+\dot{\theta},\neff)}{\partial \, \om} = 1+\mathcal{O}(\dot{\DeltaF}^2)
\end{align}
and then using ${\dot{\theta}\equiv 0}$. 
The function $f(\om,\neff)$ is defined by 
\begin{align}
f(\om,\neff)\mydoteq
\sum_{\ve{q}}\frac{\E[q]}{\Nm}
\end{align}
where it is assumed that $\W[q]$ has been substituted into $\E[q]$ in order to let the latter depend on $\neff$.

In the same way, we can insert the adiabatic solutions $\fy[q]$, $\fz[q]$ \eqref{adsol} into the molecular equation of motion \eqref{4v4}. It turns out that we can also shorten the notation here by writing
\begin{align}\label{2von2ad}
&\frac{2\,\eps^2\sqrt{\neff}}{1+\eps^2}\,(\om+\dot{\theta})-2\,\sqrt{\neff}\,
\frac{\partial f(\om+\dot{\theta},\neff)}{\partial \, \neff}
\nn\\
&\quad-\frac{\gammaF^2 \sqrt{\neff}}{g\,(1+\eps^2)}
\rka 1-\frac{\halc\alc}{\Neff}+\frac{\hbetc\betc}{\eps^2\,\Neff} \rkz=\mathcal{O}(\dot{\DeltaF}^2)
\end{align}
and then applying ${\dot{\theta}\equiv 0}$. 
\subsubsection{The continuum limit}
The sum appearing in the definition of $f(\om,\neff)$ can be approximated by means of the integrals
\begin{align}
&\sum_{\ve{q}}b(\abs{q}) \approx \frac{Vk_{F}^3}{(2\pi)^3}\int_{\mathbbm{R}^3} \rmd^3 q\, b(\abs{q}) = 3\, \Nm\int_0^\infty \rmd q\; q^2\, b(q)
\end{align}
where the approximate sign turns into an equal sign as ${V \to \infty}$. 
We can thus rewrite $f(\om,\neff)$ as
\begin{align}
f(\om,\neff)=3\, \gammaF^{5/2}\,\neff^{5/4}\, \Phi\biggl(\frac{\DeltaF-{\om}}{2\,\gammaF\sqrt{\neff}}\biggr)
\end{align}
with the integral
\begin{align}
\Phi(x)=\int_0^\infty \rmd q\; q^2(q^2+x)-q^2\sqrt{1+(q^2+x)^2}+\frac{1}{2}		\;\,.
\end{align}
In order to make the integral converge, an extra term is added. This is a standard renormalization procedure when dealing with an effective Hamiltonian, as we do.
Finally, we express the partial derivatives of $f(\om,\neff)$ as
\begin{align}
\label{fw}
\frac{\partial f(\om,\neff)}{\partial\, \om}=-&\,\frac{3}{2}\, \gammaF^{3/2}\, \neff^{3/4}\,\Phi'		\;\;,\\
\label{fn}
\frac{\partial f(\om,\neff)}{\partial \, \neff}=-&\,\frac{3}{4}\rka\DeltaF-{\om}\rkz \gammaF^{3/2}\,\neff^{-1/4}\,\Phi'\nn\\
+&\,\frac{15}{4}\, \gammaF^{5/2}\, \neff^{1/4}\,\Phi 
\end{align}
where $\Phi$ and $\Phi'$ are evaluated at ${(\DeltaF-{\om})/(2\,\gammaF\sqrt{\neff})}$.
\subsubsection{Adiabatic solution for $\alc$}

In order to make further simplifications, we assume that 
\begin{align}\label{omegacond}
\om<\frac{\gammaF^2}{\gF\rka 1 + \epsw^2\rkz} 		\;.
\end{align}
An assumption which has also to be checked in the end for self-consistency. This assumption implies that 
\begin{align}
0<\frac{\gammaF^2-\gF\rka 1 + \epsw^2\rkz \om}{\gammaF^2-\gF\, \epsw^2\, \om}
\end{align}
and enables us to rewrite $\sqrt{\Neff}$ from \eqref{naqa}. Especially for $\sqrt{\neff}$, we get 
\begin{align}\label{wnrg}
\sqrt{\neff}=\frac{\gammaF^2-\gF\rka 1 + \epsw^2\rkz \om}{\gammaF^2-\gF\, \epsw^2\, \om}\,r+\mathcal{O}(\dot{\DeltaF}^2)		\;.
\end{align}
Now we replace the partial derivatives in the integral of motion \eqref{1von2ad} and in the second remaining molecular equation of motion \eqref{2von2ad} by \eqref{fw} and \eqref{fn}. Moreover, we express $\hbetc\betc$ and $\Neff$ as well as some $\neff$ in terms of ${\halc\alc=r^2}$ by means of \eqref{bqbaqa} and \eqref{naqa}. Applying ${\dot{\theta}\equiv 0}$, we finally obtain 
\begin{align}\label{emo1rg}
r^2+\frac{\gF^2\, \epsw^2\, \om^2\,r^2}{\rka\gammaF^2-g\, \epsw^2\, \om\rkz^2}
+\frac{3}{2}\, \gammaF^{3/2}\, \neff^{3/4}\,\Phi' = 1+\mathcal{O}(\dot{\DeltaF}^2)
\end{align}
and
\begin{align}\label{emo2rg}
&2\,r\, \om+\frac{3}{2}\rka\DeltaF-{\om}\rkz \gammaF^{3/2}\,\neff^{1/4}\,\Phi'\nn\\
&
\quad
-\frac{15}{2}\, \gammaF^{5/2}\, \neff^{3/4}\,\Phi
= \mathcal{O}(\dot{\DeltaF}^2)
\end{align}
where $\Phi$ and $\Phi'$ are evaluated at ${(\DeltaF-{\om})/(2\,\gammaF\sqrt{\neff})}$. The last three equations \eqref{wnrg}, \eqref{emo1rg}, \eqref{emo2rg} define $r$ and $\om$. If we find a solution satisfying all assumptions made so far, we finally have the desired classical path. Note that, since the latter equations do not depend on $\dot{\DeltaF}$, \ie $\nuF^2$ explicitly, we will get ${r(\DeltaF)+\mathcal{O}(\dot{\DeltaF}^2)}$ and ${\om(\DeltaF)+\mathcal{O}(\dot{\DeltaF}^2)}$. This is only true provided that the functional determinant of the two equations of motion (definition of $\neff$ plugged in) with respect to $r$ and $\om$ is not zero. 
\subsubsection{The function $\Phi$}
For real $x$, the integrand
\begin{align}
\int_0^q \rmd \tilde{q}\; \tilde{q}^2(\tilde{q}^2+x)-\tilde{q}^2\sqrt{1+(\tilde{q}^2+x)^2}+\frac{1}{2}
\end{align}
reads
\begin{align}\label{integrand}
&\frac{1}{2}\,q+\frac{x}{3}\,q^3+\frac{1}{5}\,q^5-\frac{1}{15}\,q \left(3 q^2+2 x\right) \sqrt{1+\left(q^2+x\right)^2}\nn
\\&+\frac{2\, q \left(3-x^2\right) \sqrt{1+\left(q^2+x\right)^2} \left(q^2-\sqrt{1+x^2}\right)}{15 \left(1-q^4+x^2\right)}\nn
\\&+\frac{2}{15} \left(3-x^2\right) \left(1+x^2\right)^{1/4} 
\Biggl(2\, \Ele{\frac{1}{2}-\frac{x}{2 \sqrt{1+x^2}}}\nn
\\&\quad-\Ele{\arccos\Biggl(\frac{q^2-\sqrt{1+x^2}}{q^2+\sqrt{1+x^2}}\Biggr),
\frac{1}{2}-\frac{x}{2 \sqrt{1+x^2}}}\Biggr)\nn
\\&+\frac{1}{15} \left(1+x^2\right)^{1/4} \left(3-x \left(x+\sqrt{1+x^2}\right)\right) \nn
\\&\quad\Biggl(\Elf{\arccos\Biggl(\frac{q^2-\sqrt{1+x^2}}{q^2+\sqrt{1+x^2}}\Biggr),\frac{1}{2}-\frac{x}{2 \sqrt{1+x^2}}}\nn
\\&\quad-2\, \Elk{\frac{1}{2}-\frac{x}{2 \sqrt{1+x^2}}}\Biggr)
\end{align}
where $\neff>0$ ensures that we always have
\begin{align}
0<\frac{1}{2}-\frac{x}{2 \sqrt{1+x^2}}<1
\end{align}
as long as ${\DeltaF-{\om}\neq\pm\,\infty}$. 
With considerable effort, the integrand \eqref{integrand} can be found with the help of similar integrals (see, \eg \cite{Byrd71}). 
The functions $\Elf{\varphi, m}$ and $\Elk{m}$ are the incomplete and complete elliptic integral of the first kind, respectively. They are defined for ${0<m<1}$ as
\begin{align}
\Elf{\varphi, m}\mydoteq{}&\int_0^\varphi \rmd\varphi\, \rka 1- m \sin^2(\varphi)\rkz^{-1/2}		\;, \\
\Elk{m}\mydoteq{}&\Elf{\frac{\pi}{2}, m}		\;.
\end{align}
The functions $\Ele{\varphi, m}$ and $\Ele{m}$ are the incomplete and complete elliptic integral of the second kind, respectively. They are defined for ${0<m<1}$ as
\begin{align}
\Ele{\varphi, m}\mydoteq{}&\int_0^\varphi \rmd\varphi\, \rka 1- m \sin^2(\varphi)\rkz^{1/2}		\;, \\
\Ele{m}\mydoteq{}&\Ele{\frac{\pi}{2}, m}		\;.
\end{align}
Taking the $q \to \infty$ limit of the integrand \eqref{integrand} leads to
\begin{align}
\Phi(x)={}&\frac{4}{15} \left(3-x^2\right) \left(1+x^2\right)^{1/4} \Ele{\frac{1}{2}-\frac{x}{2 \sqrt{1+x^2}}}\nn
\\&-\frac{2}{15}\left(3-x \left(x+\sqrt{1+x^2}\right)\right) \left(1+x^2\right)^{1/4}\nn
\\&\quad\cdot\Elk{\frac{1}{2}-\frac{x}{2 \sqrt{1+x^2}}}		\;.
\end{align}
At this point the asymptotic expansions of the elliptic integrals might be of interest in order to get an idea of the behavior of $\Phi$. The asymptotic expansions for ${m \to 0}$ from above and for ${m \to 1}$ from below are especially helpful and comparably simple as well. They read for the complete elliptic integral of the first kind
\begin{align}
\Elk{m}\approx{}& \frac{\pi }{2}+\frac{\pi }{8}\,m+\mathcal{O}(m^2)		\;, \\
\label{ellekasymp}
\Elk{m}\approx{}& -\frac{1}{2} \ln(1-m)+2 \ln(2)\nn\\
&-\rka\frac{1}{4}-\frac{\ln(2)}{2}+\frac{1}{8} \ln(1-m)\rkz(1-m)\nn\\
&+\mathcal{O}((1-m)^2)
\end{align}
and 
\begin{align}
\Ele{m}\approx{}& \frac{\pi }{2}-\frac{\pi }{8}\,m+\mathcal{O}(m^2)		\;, \\
\label{elleasymp}
\Ele{m}\approx{}& 1-\rka\frac{1}{4}-\ln(2)+\frac{1}{4} \ln(1-m)\rkz(1-m)\nn\\
&+\mathcal{O}((1-m)^2)
\end{align}
for the complete elliptic integral of the second kind. One may check with these expansions that the condition for the upper bound of $\om$ \eqref{omegacond} holds at late times where it is not obvious from just plotting the numerical solution. 
\subsection{Results for ${\eps \to 0}$}
We are mainly interested in the limit ${\eps \to 0}$ which we use in the main part of this paper. Thus, we take this limit for $\alc$ \eqref{alcadi}, $\betc$ \eqref{betcadi} and $\sqrt{\neff}$ \eqref{wnrg} leading to
\begin{align}
\alc&=\sqrt{\Nm} \rka r-\rmi \, \frac{\gF\,\dot{r}}{\gF\,\om-\gammaF^2}\rkz \rme^{-\, \rmi\, \varphi_f}
+\mathcal{O}(\dot{\DeltaF}^2)		\;,
\end{align}
\begin{align}
\betc&=
\epsw\,
\sqrt{\Nm}\rka 
-\,
 \frac{\gF\,\om}{\gammaF^2}\, r
+\rmi\, 
\frac{\gF\, \dot{r}}{\gF\, \om - \gammaF^2} 
\rkz \rme^{-\, \rmi\, \varphi_f}
+\mathcal{O}(\dot{\DeltaF}^2)
\end{align}
and
\begin{align}
\label{wnr}
\sqrt{\neff}&=\rka1-\frac{\gF\, \om}{\gammaF^2}\rkz r+\mathcal{O}(\dot{\DeltaF}^2)
\end{align}
where the imaginary parts of $\alc$ and $\betc$ are just the expansion of the slow phase terms. 
The leading linear $\eps$-order of $\betc$ is kept, since $\betc$ appears as ${\betc/\eps}$ in many equations. 

The ${\eps \to 0}$ limit of the molecular equations of motion \eqref{emo1rg} and \eqref{emo2rg} simplifies to
\begin{align}\label{emo1r}
r^2+\frac{3}{2}\, \gammaF^{3/2}\rka1-\frac{\gF\, \om}{\gammaF^2}\rkz^{3/2} r^{3/2}\,\Phi' = 1+\mathcal{O}(\dot{\DeltaF}^2)		\;,
\end{align}
\begin{align}\label{emo2r}
&2\,r\, \om+\frac{3}{2}\rka\DeltaF-{\om}\rkz \gammaF^{3/2}\rka1-\frac{\gF\, \om}{\gammaF^2}\rkz^{1/2} r^{1/2}\,\Phi'\nn\\
&\quad-\frac{15}{2}\,\gammaF^{5/2}\rka1-\frac{\gF\, \om}{\gammaF^2}\rkz^{3/2} r^{3/2}\,\Phi
= \mathcal{O}(\dot{\DeltaF}^2)		\;.
\end{align}
Here, ${\gammaF\,(\DeltaF-{\om})/(2\,(\gammaF^2-\gF\,\om)\,r)}$ is the argument of $\Phi$ and of $\Phi'$. The number conservation \eqref{emo1r} and the molecular equation of motion \eqref{emo2r} together finally define ${r(\DeltaF)+\mathcal{O}(\dot{\DeltaF}^2)}$ and ${\om(\DeltaF)+\mathcal{O}(\dot{\DeltaF}^2)}$. 
It is important to notice, that both, $r$ and $\om$ have \textit{no} terms of \textit{first order} in \nolinebreak ${\dot{\DeltaF}}$.
\subsubsection{Numerical results}
\begin{figure}[h]
\includegraphics{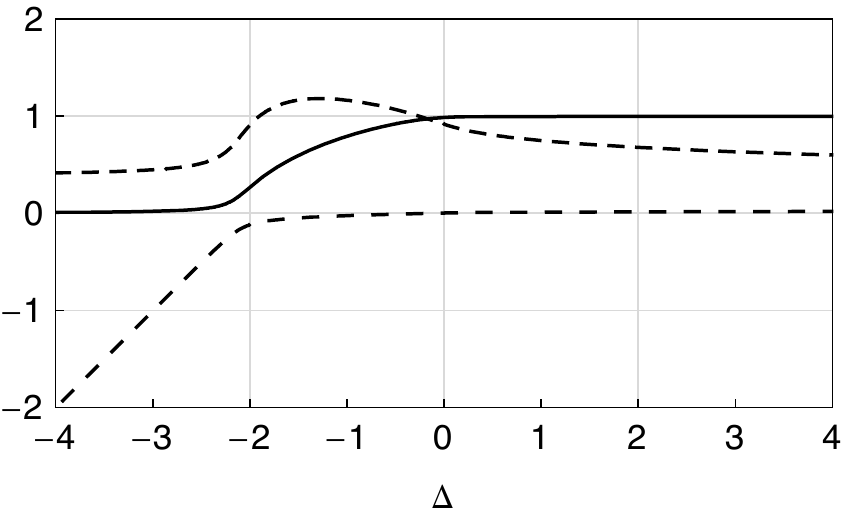}
\caption{\label{classicplot}Square root of the classical relative molecule number (and density), $r$ (solid line), square root of the relative effective density, $\sqrt{\neff}$ (dashed line), frequency $\om$ (dash-dotted line) and ${\rka\DeltaF-{\om}\rkz/2}$ (dotted line) as functions of $\DeltaF$. At late times $\om$ approaches ${\gammaF^2/\gF}$. Parameters are ${\gammaF=0.1}$ and ${\gF=0.2}$.}
\end{figure}
Numerical results for ${\gammaF=0.1}$ and ${\gF=0.2}$ are given in Figure \ref{classicplot}. 
The square root of the relative molecule number (and density), $r$, starts to grow significantly as soon as the atomic part of the system becomes resonant with the molecular part around ${\DeltaF \approx -2}$. Around this point $\om$ changes its behavior and approaches asymptotically the value $\gammaF^2/\gF$. The value $\rka\DeltaF-{\om}\rkz/2$ can be interpreted as half the negative chemical potential of an atom pair. Its almost constant behavior at early times reflects the deep BCS regime. The square root of the relative effective density, $\sqrt{\neff}$, is also almost constant within the deep BCS regime, ensuring that the instantaneous energy gaps of the atomic two-level systems have a lower bound. As discussed before, this makes an adiabatic approximation for these two-level systems possible.
The results obtained here agree with our previous results presented in \cite{Brei08} and generalize them for background scattering ${\gF \neq 0}$.
\subsubsection{Early and late time limits}
As ${t \to t_f}$ we get ${r\to1}$ already very early. In contrast, the asymptotic values ${\om\to\gammaF^2/\gF}$ and ${\sqrt{\neff}\to 0}$ are approached more slowly.

As ${t \to t_i}$, the number conservation \eqref{emo1r} and the molecular equation of motion \eqref{emo2r} tend to
\begin{align}\label{emo1rearly}
\frac{3}{2}\rka\gammaF \sqrt{\neff}\rkz^{3/2}\Phi' = 1+\mathcal{O}(\dot{\DeltaF}^2)
\end{align}
and
\begin{align}\label{emo2rearly}
&\frac{3}{2}\, \gF \rka\frac{\DeltaF-{\om}}{2}\rkz \rka\gammaF \sqrt{\neff}\rkz^{-1/2} \Phi'\nn\\
&
\quad
-\frac{15}{4}\, \gF \rka\gammaF \sqrt{\neff}\rkz^{1/2} \Phi
= 1+\mathcal{O}(\dot{\DeltaF}^2)
\end{align}
both written in terms of $\sqrt{\neff}$ \eqref{wnr} instead in terms of $r$ since the latter really tends to zero at early times. 
The functions $\Phi$ as well as $\Phi'$ are evaluated at ${(\DeltaF-{\om})/(2\,\gammaF\sqrt{\neff})}$. These equations \eqref{emo1rearly}, \eqref{emo2rearly} are obtained by taking the leading order when ${\om \to -\,\infty}$.

At $t_i$, the system of equations depends only on ${(\DeltaF-\om)/2}$ and ${\gammaF \sqrt{\neff}}$, which are the new variables to solve for in dependence of the background scattering parameter $\gF$. Using the asymptotic expansions of the complete elliptic integrals \eqref{ellekasymp}, \eqref{elleasymp}, we find the leading order solution to be
\begin{align}
\frac{\DeltaF-{\om}}{2}&\approx -\,1 		\;\,,\\
\gammaF \sqrt{\neff}&\approx \frac{8}{\rme^2}\,\rme^{-\,2/(3\, \gF)}
\end{align}
for sufficiently small $\gF$. This is the familiar scaling of the BCS order parameter.
}
\begin{acknowledgments}
Support from the Deutsche Forschungsgemeinschaft for a part of this work via the Graduiertenkolleg 792 ‘Nichtlineare Optik und Ultrakurzzeitphysik’ is gratefully acknowledged.
\end{acknowledgments}

\begin{thebibliography}{31}%
\makeatletter
\providecommand \@ifxundefined [1]{%
 \@ifx{#1\undefined}
}%
\providecommand \@ifnum [1]{%
 \ifnum #1\expandafter \@firstoftwo
 \else \expandafter \@secondoftwo
 \fi
}%
\providecommand \@ifx [1]{%
 \ifx #1\expandafter \@firstoftwo
 \else \expandafter \@secondoftwo
 \fi
}%
\providecommand \natexlab [1]{#1}%
\providecommand \enquote  [1]{``#1''}%
\providecommand \bibnamefont  [1]{#1}%
\providecommand \bibfnamefont [1]{#1}%
\providecommand \citenamefont [1]{#1}%
\providecommand \href@noop [0]{\@secondoftwo}%
\providecommand \href [0]{\begingroup \@sanitize@url \@href}%
\providecommand \@href[1]{\@@startlink{#1}\@@href}%
\providecommand \@@href[1]{\endgroup#1\@@endlink}%
\providecommand \@sanitize@url [0]{\catcode `\\12\catcode `\$12\catcode
  `\&12\catcode `\#12\catcode `\^12\catcode `\_12\catcode `\%12\relax}%
\providecommand \@@startlink[1]{}%
\providecommand \@@endlink[0]{}%
\providecommand \url  [0]{\begingroup\@sanitize@url \@url }%
\providecommand \@url [1]{\endgroup\@href {#1}{\urlprefix }}%
\providecommand \urlprefix  [0]{URL }%
\providecommand \Eprint [0]{\href }%
\providecommand \doibase [0]{http://dx.doi.org/}%
\providecommand \selectlanguage [0]{\@gobble}%
\providecommand \bibinfo  [0]{\@secondoftwo}%
\providecommand \bibfield  [0]{\@secondoftwo}%
\providecommand \translation [1]{[#1]}%
\providecommand \BibitemOpen [0]{}%
\providecommand \bibitemStop [0]{}%
\providecommand \bibitemNoStop [0]{.\EOS\space}%
\providecommand \EOS [0]{\spacefactor3000\relax}%
\providecommand \BibitemShut  [1]{\csname bibitem#1\endcsname}%
\let\auto@bib@innerbib\@empty
\bibitem [{\citenamefont {Holland}\ \emph {et~al.}(2001)\citenamefont
  {Holland}, \citenamefont {Kokkelmans}, \citenamefont {Chiofalo},\ and\
  \citenamefont {Walser}}]{Holl01}%
  \BibitemOpen
  \bibfield  {author} {\bibinfo {author} {\bibfnamefont {M.}~\bibnamefont
  {Holland}}, \bibinfo {author} {\bibfnamefont {S.~J. J. M.~F.}\ \bibnamefont
  {Kokkelmans}}, \bibinfo {author} {\bibfnamefont {M.~L.}\ \bibnamefont
  {Chiofalo}}, \ and\ \bibinfo {author} {\bibfnamefont {R.}~\bibnamefont
  {Walser}},\ }\href@noop {} {\bibfield  {journal} {\bibinfo  {journal} {Phys.
  Rev. Lett.}\ }\textbf {\bibinfo {volume} {87}},\ \bibinfo {pages} {120406}
  (\bibinfo {year} {2001})}\BibitemShut {NoStop}%
\bibitem [{\citenamefont {Milstein}\ \emph {et~al.}(2002)\citenamefont
  {Milstein}, \citenamefont {Kokkelmans},\ and\ \citenamefont
  {Holland}}]{Mils02}%
  \BibitemOpen
  \bibfield  {author} {\bibinfo {author} {\bibfnamefont {J.~N.}\ \bibnamefont
  {Milstein}}, \bibinfo {author} {\bibfnamefont {S.~J. J. M.~F.}\ \bibnamefont
  {Kokkelmans}}, \ and\ \bibinfo {author} {\bibfnamefont {M.~J.}\ \bibnamefont
  {Holland}},\ }\href@noop {} {\bibfield  {journal} {\bibinfo  {journal} {Phys.
  Rev. A}\ }\textbf {\bibinfo {volume} {66}},\ \bibinfo {pages} {043604}
  (\bibinfo {year} {2002})}\BibitemShut {NoStop}%
\bibitem [{\citenamefont {Romans}\ and\ \citenamefont {Stoof}(2006)}]{Roma06}%
  \BibitemOpen
  \bibfield  {author} {\bibinfo {author} {\bibfnamefont {M.~W.~J.}\
  \bibnamefont {Romans}}\ and\ \bibinfo {author} {\bibfnamefont {H.~T.~C.}\
  \bibnamefont {Stoof}},\ }\href {\doibase 10.1103/PhysRevA.74.053618}
  {\bibfield  {journal} {\bibinfo  {journal} {Phys. Rev. A}\ }\textbf {\bibinfo
  {volume} {74}},\ \bibinfo {pages} {053618} (\bibinfo {year}
  {2006})}\BibitemShut {NoStop}%
\bibitem [{\citenamefont {Marini}\ \emph {et~al.}(1998)\citenamefont {Marini},
  \citenamefont {Pistolesi},\ and\ \citenamefont {Strinati}}]{Mari98}%
  \BibitemOpen
  \bibfield  {author} {\bibinfo {author} {\bibfnamefont {M.}~\bibnamefont
  {Marini}}, \bibinfo {author} {\bibfnamefont {F.}~\bibnamefont {Pistolesi}}, \
  and\ \bibinfo {author} {\bibfnamefont {G.~C.}\ \bibnamefont {Strinati}},\
  }\href@noop {} {\bibfield  {journal} {\bibinfo  {journal} {Eur. Phys. J. B}\
  }\textbf {\bibinfo {volume} {1}},\ \bibinfo {pages} {151} (\bibinfo {year}
  {1998})}\BibitemShut {NoStop}%
\bibitem [{\citenamefont {Diehl}\ and\ \citenamefont
  {Wetterich}(2006)}]{Dieh06}%
  \BibitemOpen
  \bibfield  {author} {\bibinfo {author} {\bibfnamefont {S.}~\bibnamefont
  {Diehl}}\ and\ \bibinfo {author} {\bibfnamefont {C.}~\bibnamefont
  {Wetterich}},\ }\href@noop {} {\bibfield  {journal} {\bibinfo  {journal}
  {Phys. Rev. A}\ }\textbf {\bibinfo {volume} {73}},\ \bibinfo {pages} {033615}
  (\bibinfo {year} {2006})}\BibitemShut {NoStop}%
\bibitem [{\citenamefont {Diehl}\ \emph {et~al.}(2007)\citenamefont {Diehl},
  \citenamefont {Gies}, \citenamefont {Pawlowski},\ and\ \citenamefont
  {Wetterich}}]{Dieh07}%
  \BibitemOpen
  \bibfield  {author} {\bibinfo {author} {\bibfnamefont {S.}~\bibnamefont
  {Diehl}}, \bibinfo {author} {\bibfnamefont {H.}~\bibnamefont {Gies}},
  \bibinfo {author} {\bibfnamefont {J.~M.}\ \bibnamefont {Pawlowski}}, \ and\
  \bibinfo {author} {\bibfnamefont {C.}~\bibnamefont {Wetterich}},\ }\href
  {\doibase 10.1103/PhysRevA.76.053627} {\bibfield  {journal} {\bibinfo
  {journal} {Phys. Rev. A}\ }\textbf {\bibinfo {volume} {76}},\ \bibinfo
  {pages} {053627} (\bibinfo {year} {2007})}\BibitemShut {NoStop}%
\bibitem [{\citenamefont {Chin}\ \emph {et~al.}(2010)\citenamefont {Chin},
  \citenamefont {Grimm}, \citenamefont {Julienne},\ and\ \citenamefont
  {Tiesinga}}]{Chin10}%
  \BibitemOpen
  \bibfield  {author} {\bibinfo {author} {\bibfnamefont {C.}~\bibnamefont
  {Chin}}, \bibinfo {author} {\bibfnamefont {R.}~\bibnamefont {Grimm}},
  \bibinfo {author} {\bibfnamefont {P.}~\bibnamefont {Julienne}}, \ and\
  \bibinfo {author} {\bibfnamefont {E.}~\bibnamefont {Tiesinga}},\ }\href
  {\doibase 10.1103/RevModPhys.82.1225} {\bibfield  {journal} {\bibinfo
  {journal} {Rev. Mod. Phys.}\ }\textbf {\bibinfo {volume} {82}},\ \bibinfo
  {pages} {1225} (\bibinfo {year} {2010})}\BibitemShut {NoStop}%
\bibitem [{\citenamefont {Chin}\ \emph {et~al.}(2004)\citenamefont {Chin},
  \citenamefont {Bartenstein}, \citenamefont {Altmeyer}, \citenamefont {Riedl},
  \citenamefont {Jochim}, \citenamefont {Denschlag},\ and\ \citenamefont
  {Grimm}}]{Chin04}%
  \BibitemOpen
  \bibfield  {author} {\bibinfo {author} {\bibfnamefont {C.}~\bibnamefont
  {Chin}}, \bibinfo {author} {\bibfnamefont {M.}~\bibnamefont {Bartenstein}},
  \bibinfo {author} {\bibfnamefont {A.}~\bibnamefont {Altmeyer}}, \bibinfo
  {author} {\bibfnamefont {S.}~\bibnamefont {Riedl}}, \bibinfo {author}
  {\bibfnamefont {S.}~\bibnamefont {Jochim}}, \bibinfo {author} {\bibfnamefont
  {J.~H.}\ \bibnamefont {Denschlag}}, \ and\ \bibinfo {author} {\bibfnamefont
  {R.}~\bibnamefont {Grimm}},\ }\href@noop {} {\bibfield  {journal} {\bibinfo
  {journal} {Science}\ }\textbf {\bibinfo {volume} {305}},\ \bibinfo {pages}
  {1128} (\bibinfo {year} {2004})}\BibitemShut {NoStop}%
\bibitem [{\citenamefont {Zwierlein}\ \emph {et~al.}(2004)\citenamefont
  {Zwierlein}, \citenamefont {Stan}, \citenamefont {Schunck}, \citenamefont
  {Raupach}, \citenamefont {Kerman},\ and\ \citenamefont {Ketterle}}]{Zwie04}%
  \BibitemOpen
  \bibfield  {author} {\bibinfo {author} {\bibfnamefont {M.~W.}\ \bibnamefont
  {Zwierlein}}, \bibinfo {author} {\bibfnamefont {C.~A.}\ \bibnamefont {Stan}},
  \bibinfo {author} {\bibfnamefont {C.~H.}\ \bibnamefont {Schunck}}, \bibinfo
  {author} {\bibfnamefont {S.~M.~F.}\ \bibnamefont {Raupach}}, \bibinfo
  {author} {\bibfnamefont {A.~J.}\ \bibnamefont {Kerman}}, \ and\ \bibinfo
  {author} {\bibfnamefont {W.}~\bibnamefont {Ketterle}},\ }\href@noop {}
  {\bibfield  {journal} {\bibinfo  {journal} {Phys. Rev. Lett.}\ }\textbf
  {\bibinfo {volume} {92}},\ \bibinfo {pages} {120403} (\bibinfo {year}
  {2004})}\BibitemShut {NoStop}%
\bibitem [{\citenamefont {K\"ohler}\ \emph {et~al.}(2006)\citenamefont
  {K\"ohler}, \citenamefont {G\'oral},\ and\ \citenamefont
  {Julienne}}]{Kohl06}%
  \BibitemOpen
  \bibfield  {author} {\bibinfo {author} {\bibfnamefont {T.}~\bibnamefont
  {K\"ohler}}, \bibinfo {author} {\bibfnamefont {K.}~\bibnamefont {G\'oral}}, \
  and\ \bibinfo {author} {\bibfnamefont {P.~S.}\ \bibnamefont {Julienne}},\
  }\href@noop {} {\bibfield  {journal} {\bibinfo  {journal} {Rev. Mod. Phys.}\
  }\textbf {\bibinfo {volume} {78}},\ \bibinfo {pages} {1311} (\bibinfo {year}
  {2006})}\BibitemShut {NoStop}%
\bibitem [{\citenamefont {Regal}\ \emph {et~al.}(2004)\citenamefont {Regal},
  \citenamefont {Greiner},\ and\ \citenamefont {Jin}}]{Rega04}%
  \BibitemOpen
  \bibfield  {author} {\bibinfo {author} {\bibfnamefont {C.~A.}\ \bibnamefont
  {Regal}}, \bibinfo {author} {\bibfnamefont {M.}~\bibnamefont {Greiner}}, \
  and\ \bibinfo {author} {\bibfnamefont {D.~S.}\ \bibnamefont {Jin}},\
  }\href@noop {} {\bibfield  {journal} {\bibinfo  {journal} {Phys. Rev. Lett.}\
  }\textbf {\bibinfo {volume} {92}},\ \bibinfo {pages} {040403} (\bibinfo {year}
  {2004})}\BibitemShut {NoStop}%
\bibitem [{\citenamefont {Zwerger}(2012)}]{Zwer12}%
  \BibitemOpen
  \bibinfo {editor} {\bibfnamefont {W.}~\bibnamefont {Zwerger}},\ ed.,\
  \href@noop {} {\emph {\bibinfo {title} {The BCS-BEC Crossover and the Unitary
  Fermi Gas}}},\ \bibinfo {series} {Lecture Notes in Physics}, Vol.\ \bibinfo
  {volume} {836}\ (\bibinfo  {publisher} {Springer},\ \bibinfo {address}
  {Berlin},\ \bibinfo {year} {2012})\BibitemShut {NoStop}%
\bibitem [{\citenamefont {Inguscio}\ \emph {et~al.}(2008)\citenamefont
  {Inguscio}, \citenamefont {Ketterle},\ and\ \citenamefont
  {Salomon}}]{Ingu08}%
  \BibitemOpen
  \bibinfo {editor} {\bibfnamefont {M.}~\bibnamefont {Inguscio}}, \bibinfo
  {editor} {\bibfnamefont {W.}~\bibnamefont {Ketterle}}, \ and\ \bibinfo
  {editor} {\bibfnamefont {C.}~\bibnamefont {Salomon}},\ eds.,\ \href@noop {}
  {\emph {\bibinfo {title} {Ultra-cold Fermi Gases}}},\ Proceedings of the
  International School of Physics “Enrico Fermi”, Course CLXIV, Varenna,
  2006\ (\bibinfo  {publisher} {ISO},\ \bibinfo {address} {Amsterdam},\
  \bibinfo {year} {2008})\BibitemShut {NoStop}%
\bibitem [{\citenamefont {Wald}(1984)}]{Wald84}%
  \BibitemOpen
  \bibfield  {author} {\bibinfo {author} {\bibfnamefont {R.~M.}\ \bibnamefont
  {Wald}},\ }\href@noop {} {\emph {\bibinfo {title} {General Relativity}}}\
  (\bibinfo  {publisher} {The University of Chicago Press},\ \bibinfo {address}
  {Chicago},\ \bibinfo {year} {1984})\BibitemShut {NoStop}%
\bibitem [{\citenamefont {Aharonov}\ \emph {et~al.}(1990)\citenamefont
  {Aharonov}, \citenamefont {Ben-Reuven}, \citenamefont {Popescu},\ and\
  \citenamefont {Rohrlich}}]{Ahar90}%
  \BibitemOpen
  \bibfield  {author} {\bibinfo {author} {\bibfnamefont {Y.}~\bibnamefont
  {Aharonov}}, \bibinfo {author} {\bibfnamefont {E.}~\bibnamefont
  {Ben-Reuven}}, \bibinfo {author} {\bibfnamefont {S.}~\bibnamefont {Popescu}},
  \ and\ \bibinfo {author} {\bibfnamefont {D.}~\bibnamefont {Rohrlich}},\
  }\href {\doibase 10.1103/PhysRevLett.65.3065} {\bibfield  {journal} {\bibinfo
   {journal} {Phys. Rev. Lett.}\ }\textbf {\bibinfo {volume} {65}},\ \bibinfo
  {pages} {3065} (\bibinfo {year} {1990})}\BibitemShut {NoStop}%
\bibitem [{\citenamefont {Aharonov}\ and\ \citenamefont
  {Stern}(1992)}]{Ahar92}%
  \BibitemOpen
  \bibfield  {author} {\bibinfo {author} {\bibfnamefont {Y.}~\bibnamefont
  {Aharonov}}\ and\ \bibinfo {author} {\bibfnamefont {A.}~\bibnamefont
  {Stern}},\ }\href {\doibase 10.1103/PhysRevLett.69.3593} {\bibfield
  {journal} {\bibinfo  {journal} {Phys. Rev. Lett.}\ }\textbf {\bibinfo
  {volume} {69}},\ \bibinfo {pages} {3593} (\bibinfo {year}
  {1992})}\BibitemShut {NoStop}%
\bibitem [{\citenamefont {Littlejohn}\ and\ \citenamefont
  {Weigert}(1993)}]{Litt93}%
  \BibitemOpen
  \bibfield  {author} {\bibinfo {author} {\bibfnamefont {R.~G.}\ \bibnamefont
  {Littlejohn}}\ and\ \bibinfo {author} {\bibfnamefont {S.}~\bibnamefont
  {Weigert}},\ }\href {\doibase 10.1103/PhysRevA.48.924} {\bibfield  {journal}
  {\bibinfo  {journal} {Phys. Rev. A}\ }\textbf {\bibinfo {volume} {48}},\
  \bibinfo {pages} {924} (\bibinfo {year} {1993})}\BibitemShut {NoStop}%
\bibitem [{\citenamefont {Schmiedmayer}\ and\ \citenamefont
  {Scrinzi}(1996)}]{Schm96}%
  \BibitemOpen
  \bibfield  {author} {\bibinfo {author} {\bibfnamefont {J.}~\bibnamefont
  {Schmiedmayer}}\ and\ \bibinfo {author} {\bibfnamefont {A.}~\bibnamefont
  {Scrinzi}},\ }\href@noop {} {\bibfield  {journal} {\bibinfo  {journal}
  {Quantum Semiclass. Opt.}\ }\textbf {\bibinfo {volume} {8}},\ \bibinfo
  {pages} {693} (\bibinfo {year} {1996})}\BibitemShut {NoStop}%
\bibitem [{\citenamefont {Anglin}\ and\ \citenamefont
  {Schmiedmayer}(2004)}]{Angl04}%
  \BibitemOpen
  \bibfield  {author} {\bibinfo {author} {\bibfnamefont {J.~R.}\ \bibnamefont
  {Anglin}}\ and\ \bibinfo {author} {\bibfnamefont {J.}~\bibnamefont
  {Schmiedmayer}},\ }\href {\doibase 10.1103/PhysRevA.69.022111} {\bibfield
  {journal} {\bibinfo  {journal} {Phys. Rev. A}\ }\textbf {\bibinfo {volume}
  {69}},\ \bibinfo {pages} {022111} (\bibinfo {year} {2004})}\BibitemShut
  {NoStop}%
\bibitem [{\citenamefont {Timmermans}\ \emph {et~al.}(1999)\citenamefont
  {Timmermans}, \citenamefont {Tommasini}, \citenamefont {Hussein},\ and\
  \citenamefont {Kerman}}]{Timm99}%
  \BibitemOpen
  \bibfield  {author} {\bibinfo {author} {\bibfnamefont {E.}~\bibnamefont
  {Timmermans}}, \bibinfo {author} {\bibfnamefont {P.}~\bibnamefont
  {Tommasini}}, \bibinfo {author} {\bibfnamefont {M.}~\bibnamefont {Hussein}},
  \ and\ \bibinfo {author} {\bibfnamefont {A.}~\bibnamefont {Kerman}},\
  }\href@noop {} {\bibfield  {journal} {\bibinfo  {journal} {Phys. Rep.}\
  }\textbf {\bibinfo {volume} {315}},\ \bibinfo {pages} {199} (\bibinfo {year}
  {1999})}\BibitemShut {NoStop}%
\bibitem [{\citenamefont {Holland}\ \emph {et~al.}(2004)\citenamefont
  {Holland}, \citenamefont {Menotti},\ and\ \citenamefont {Viverit}}]{Holl04}%
  \BibitemOpen
  \bibfield  {author} {\bibinfo {author} {\bibfnamefont {M.~J.}\ \bibnamefont
  {Holland}}, \bibinfo {author} {\bibfnamefont {C.}~\bibnamefont {Menotti}}, \
  and\ \bibinfo {author} {\bibfnamefont {L.}~\bibnamefont {Viverit}},\
  }\href@noop {} {\bibfield  {journal} {\bibinfo  {journal}
  {arXiv:cond-mat/0404234 [cond-mat.soft]}\ } (\bibinfo {year}
  {2004})}\BibitemShut {NoStop}%
\bibitem [{\citenamefont {Javanainen}\ \emph {et~al.}(2004)\citenamefont
  {Javanainen}, \citenamefont {Ko{\v{s}}trun}, \citenamefont {Zheng},
  \citenamefont {Carmichael}, \citenamefont {Shrestha}, \citenamefont {Meinel},
  \citenamefont {Mackie}, \citenamefont {Dannenberg},\ and\ \citenamefont
  {Suominen}}]{Java04}%
  \BibitemOpen
  \bibfield  {author} {\bibinfo {author} {\bibfnamefont {J.}~\bibnamefont
  {Javanainen}}, \bibinfo {author} {\bibfnamefont {M.}~\bibnamefont
  {Ko{\v{s}}trun}}, \bibinfo {author} {\bibfnamefont {Y.}~\bibnamefont
  {Zheng}}, \bibinfo {author} {\bibfnamefont {A.}~\bibnamefont {Carmichael}},
  \bibinfo {author} {\bibfnamefont {U.}~\bibnamefont {Shrestha}}, \bibinfo
  {author} {\bibfnamefont {P.~J.}\ \bibnamefont {Meinel}}, \bibinfo {author}
  {\bibfnamefont {M.}~\bibnamefont {Mackie}}, \bibinfo {author} {\bibfnamefont
  {O.}~\bibnamefont {Dannenberg}}, \ and\ \bibinfo {author} {\bibfnamefont
  {K.-A.}\ \bibnamefont {Suominen}},\ }\href@noop {} {\bibfield  {journal}
  {\bibinfo  {journal} {Phys. Rev. Lett.}\ }\textbf {\bibinfo {volume} {92}},\
  \bibinfo {pages} {200402} (\bibinfo {year} {2004})}\BibitemShut {NoStop}%
\bibitem [{\citenamefont {Javanainen}\ \emph {et~al.}(2005)\citenamefont
  {Javanainen}, \citenamefont {Ko{\v{s}}trun}, \citenamefont {Mackie},\ and\
  \citenamefont {Carmichael}}]{Java05}%
  \BibitemOpen
  \bibfield  {author} {\bibinfo {author} {\bibfnamefont {J.}~\bibnamefont
  {Javanainen}}, \bibinfo {author} {\bibfnamefont {M.}~\bibnamefont
  {Ko{\v{s}}trun}}, \bibinfo {author} {\bibfnamefont {M.}~\bibnamefont
  {Mackie}}, \ and\ \bibinfo {author} {\bibfnamefont {A.}~\bibnamefont
  {Carmichael}},\ }\href@noop {} {\bibfield  {journal} {\bibinfo  {journal}
  {Phys. Rev. Lett.}\ }\textbf {\bibinfo {volume} {95}},\ \bibinfo {pages}
  {110408} (\bibinfo {year} {2005})}\BibitemShut {NoStop}%
\bibitem [{\citenamefont {Lee}\ \emph {et~al.}(2007)\citenamefont {Lee},
  \citenamefont {Lin},\ and\ \citenamefont {Rivers}}]{Lee07}%
  \BibitemOpen
  \bibfield  {author} {\bibinfo {author} {\bibfnamefont {D.-S.}\ \bibnamefont
  {Lee}}, \bibinfo {author} {\bibfnamefont {C.-Y.}\ \bibnamefont {Lin}}, \ and\
  \bibinfo {author} {\bibfnamefont {R.~J.}\ \bibnamefont {Rivers}},\
  }\href@noop {} {\bibfield  {journal} {\bibinfo  {journal} {Phys. Rev. Lett.}\
  }\textbf {\bibinfo {volume} {98}},\ \bibinfo {pages} {020603} (\bibinfo {year}
  {2007})}\BibitemShut {NoStop}%
\bibitem [{\citenamefont {Yi}\ and\ \citenamefont {Duan}(2006)}]{Yi06}%
  \BibitemOpen
  \bibfield  {author} {\bibinfo {author} {\bibfnamefont {W.}~\bibnamefont
  {Yi}}\ and\ \bibinfo {author} {\bibfnamefont {L.-M.}\ \bibnamefont {Duan}},\
  }\href {\doibase 10.1103/PhysRevA.73.063607} {\bibfield  {journal} {\bibinfo
  {journal} {Phys. Rev. A}\ }\textbf {\bibinfo {volume} {73}},\ \bibinfo
  {pages} {063607} (\bibinfo {year} {2006})}\BibitemShut {NoStop}%
\bibitem [{\citenamefont {Breid}\ and\ \citenamefont {Anglin}(2008)}]{Brei08}%
  \BibitemOpen
  \bibfield  {author} {\bibinfo {author} {\bibfnamefont {B.~M.}\ \bibnamefont
  {Breid}}\ and\ \bibinfo {author} {\bibfnamefont {J.~R.}\ \bibnamefont
  {Anglin}},\ }\href {\doibase 10.1098/rsta.2008.0082} {\bibfield  {journal}
  {\bibinfo  {journal} {Phil. Trans. R. Soc. A}\ }\textbf {\bibinfo {volume}
  {366}},\ \bibinfo {pages} {2813} (\bibinfo {year} {2008})}\BibitemShut
  {NoStop}%
\bibitem [{\citenamefont {Breid}\ and\ \citenamefont {Anglin}()}]{Brei14}%
  \BibitemOpen
  \bibfield  {author} {\bibinfo {author} {\bibfnamefont {B.~M.}\ \bibnamefont
  {Breid}}\ and\ \bibinfo {author} {\bibfnamefont {J.~R.}\ \bibnamefont
  {Anglin}},\ }\href@noop {} {}\bibinfo {note} {In preparation}\BibitemShut
  {NoStop}%
\bibitem [{\citenamefont {Walczak}\ and\ \citenamefont
  {Anglin}(2011)}]{Walc11}%
  \BibitemOpen
  \bibfield  {author} {\bibinfo {author} {\bibfnamefont {P.~B.}\ \bibnamefont
  {Walczak}}\ and\ \bibinfo {author} {\bibfnamefont {J.~R.}\ \bibnamefont
  {Anglin}},\ }\href {\doibase 10.1103/PhysRevA.84.013611} {\bibfield
  {journal} {\bibinfo  {journal} {Phys. Rev. A}\ }\textbf {\bibinfo {volume}
  {84}},\ \bibinfo {pages} {013611} (\bibinfo {year} {2011})}\BibitemShut
  {NoStop}%
\bibitem [{\citenamefont {Combescot}\ \emph {et~al.}(2006)\citenamefont
  {Combescot}, \citenamefont {{M. Yu. Kagan}},\ and\ \citenamefont
  {Stringari}}]{Comb06}%
  \BibitemOpen
  \bibfield  {author} {\bibinfo {author} {\bibfnamefont {R.}~\bibnamefont
  {Combescot}}, \bibinfo {author} {\bibnamefont {{M. Yu. Kagan}}}, \ and\
  \bibinfo {author} {\bibfnamefont {S.}~\bibnamefont {Stringari}},\ }\href
  {\doibase 10.1103/PhysRevA.74.042717} {\bibfield  {journal} {\bibinfo
  {journal} {Phys. Rev. A}\ }\textbf {\bibinfo {volume} {74}},\ \bibinfo
  {pages} {042717} (\bibinfo {year} {2006})}\BibitemShut {NoStop}%
\bibitem [{\citenamefont {Bender}\ and\ \citenamefont {Orszag}(1978)}]{Bend78}%
  \BibitemOpen
  \bibfield  {author} {\bibinfo {author} {\bibfnamefont {C.~M.}\ \bibnamefont
  {Bender}}\ and\ \bibinfo {author} {\bibfnamefont {S.~A.}\ \bibnamefont
  {Orszag}},\ }\href@noop {} {\emph {\bibinfo {title} {Advanced Mathematical
  Methods for Scientists and Engineers}}}\ (\bibinfo  {publisher}
  {McGraw-Hill},\ \bibinfo {address} {New York},\ \bibinfo {year}
  {1978})\BibitemShut {NoStop}%
\bibitem [{\citenamefont {Byrd}\ and\ \citenamefont {Friedman}(1971)}]{Byrd71}%
  \BibitemOpen
  \bibfield  {author} {\bibinfo {author} {\bibfnamefont {P.~F.}\ \bibnamefont
  {Byrd}}\ and\ \bibinfo {author} {\bibfnamefont {M.~D.}\ \bibnamefont
  {Friedman}},\ }\href@noop {} {\emph {\bibinfo {title} {Handbook of Elliptic
  Integrals for Engineers and Physicists}}}\ (\bibinfo  {publisher}
  {Springer},\ \bibinfo {address} {Berlin},\ \bibinfo {year}
  {1971})\BibitemShut {NoStop}%
\end{thebibliography}
\end{document}